\documentclass[a4paper]{cas-sc}
\usepackage[numbers]{natbib}
\usepackage{caption}
\usepackage{subcaption}
\usepackage{caption}
\usepackage{float}
\usepackage{changepage} 
\usepackage{hyperref} %
\usepackage[most]{tcolorbox}
\newcommand{\todo}[1]{}
\renewcommand{\todo}[1]{{\color{red} TODO: {#1}}}

\newcommand{\sectopic}[1]{\vspace{0.2em}\par\noindent{\textit{\bfseries #1}}}

\def\tsc#1{\csdef{#1}{\textsc{\lowercase{#1}}\xspace}}
\tsc{WGM}
\tsc{QE}
\tsc{EP}
\tsc{PMS}
\tsc{BEC}
\tsc{DE}
\newcommand\Tstrut{\rule{0pt}{2.6ex}}       
\newcommand\Bstrut{\rule[-0.9ex]{0pt}{0pt}} 
\newcommand{\TBstrut}{\Tstrut\Bstrut} 

\newtcolorbox{myframe}[1][]{
  enhanced,
  arc=0pt,
  outer arc=0pt,
  colback=white,
  boxrule=0.8pt,
  #1
}

\begin{document}
\let\WriteBookmarks\relax
\def\floatpagepagefraction{1}
\def\textpagefraction{.001}

\shorttitle{Model Driven Engineering for Machine Learning Components}

\shortauthors{Naveed et~al.}

\title [mode = title]{Model Driven Engineering for Machine Learning Components: A Systematic Literature Review}     



\author[1]{Hira Naveed}


\ead{hira.naveed@monash.edu}
\affiliation[1]{organization={Monash University}, 
    city={Clayton},
    state={VIC},
    country={Australia}}

\author[1]{Chetan Arora}
\ead{chetan.arora@monash.edu}

\author[2]{Hourieh Khalajzadeh}
\ead{hourieh.khalajzadeh@deakin.edu.au}
\affiliation[2]{organization={Deakin University}, 
    city={Burwood},
    state={VIC},
    country={Australia}}

\author[1]{John Grundy}
\ead{john.grundy@monash.edu}

\author[1]{Omar Haggag}
\ead{omar.haggag@monash.edu}
\begin{abstract}
\textbf{\textit{\textbf{Context}: }} Machine Learning (ML) has become widely adopted as a component in many modern software applications. Due to the large volumes of data available, organizations want to increasingly leverage their data to extract meaningful insights and enhance business profitability. ML components enable predictive capabilities, anomaly detection, recommendation, accurate image and text processing, and informed decision-making. However, developing systems with ML components is not trivial; it requires time, effort, knowledge, and expertise in ML, data processing, and software engineering. There have been several studies on the use of model-driven engineering (MDE) techniques to address these challenges when developing traditional software and cyber-physical systems. Recently, there has been a growing interest in applying MDE for systems with ML components. \\
\textbf{\textit{Objective: }} The goal of this study is to further explore the promising intersection of MDE with ML (MDE4ML) through a systematic literature review (SLR). Through this SLR, we wanted to analyze existing studies, including their motivations, MDE solutions, evaluation techniques, key benefits and limitations.\\
\textbf{\textit{Method: }} Our SLR is conducted following the well-established guidelines by Kitchenham. We started by devising a protocol and systematically searching seven databases, which resulted in 3,934 papers. After iterative filtering, we selected 46 highly relevant primary studies for data extraction, synthesis, and reporting. \\
\textbf{\textit{Results: }} We analyzed selected studies with respect to several areas of interest and identified the following: 1) the key motivations behind using MDE4ML; 2) a variety of MDE solutions applied, such as modeling languages, model transformations, tool support, targeted ML aspects, contributions and more; 3) the evaluation techniques and metrics used; and 4) the limitations and directions for future work. We also discuss the gaps in existing literature and provide recommendations for future research.  \\
\textbf{\textit{Conclusion: }} This SLR highlights current trends, gaps and future research directions in the field of MDE4ML, benefiting both researchers and practitioners.
\end{abstract}




\maketitle

\section{Introduction}\label{sec:introduction}

The ability of Machine Learning (ML) to autonomously learn data patterns and predict outcomes has tremendous potential to solve complex problems~\cite{zhang2003machine}. The proliferation of data and advances in hardware processing capabilities have contributed to the rapid growth and adoption of ML in recent years. ML components have now becoming integral to software systems with application domains including healthcare~\cite{ghassemi2020review,beam2018big}, finance~\cite{goodell2021artificial,dixon2020machine}, transport~\cite{zantalis2019review}, entertainment~\cite{galway2008machine, bennett2007netflix}, and many more. However, the development, integration, and maintenance processes of traditional software components and ML components  differ significantly~\cite{ahmad2023requirements}. Traditional software components are deterministic and developed through a {deductive} development process by explicitly coding the system's required behaviour. In contrast, ML component development follows an {inductive} process by exploring data and recognizing patterns~\cite{khomh2018software}. ML components are dynamic and it is extremely difficult to specify their `well-defined' behavior~\cite{ahmad2023requirements,arora2023advancing}. Unlike software engineers, ML engineers need to perform exploratory steps to identify and curate datasets, select relevant features, select the most suitable ML model, tune hyperparameters, monitor the ML component, and re-train in case of performance degradation~\cite{lwakatare2019taxonomy}. Hence, it is challenging to build and integrate ML components into software systems ~\cite{khomh2018software}. 

\begin{figure*}[htbp]
    \centering
    \includegraphics[width=1\textwidth]{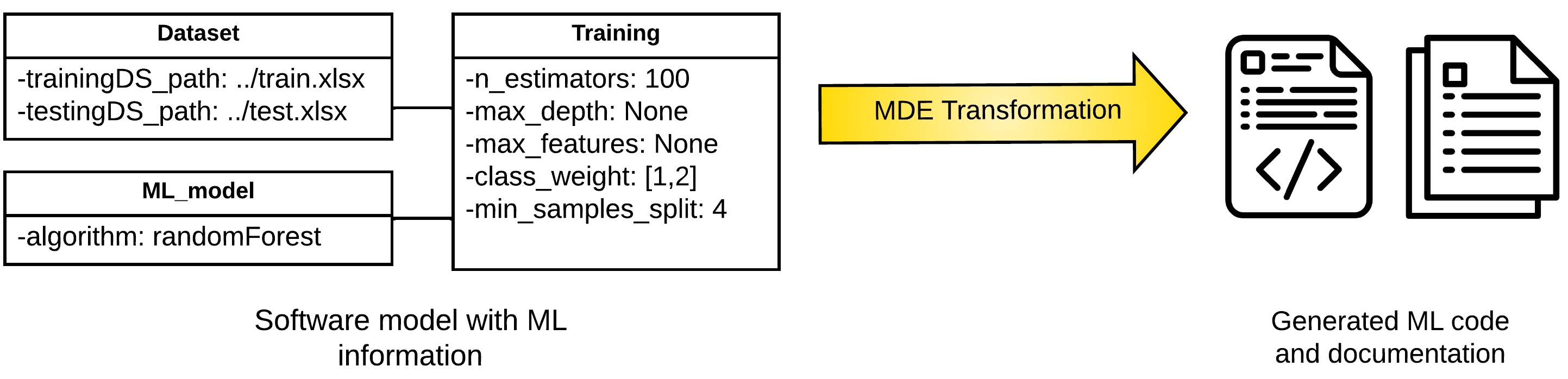}
    \caption{Model-driven Engineering for Machine Learning}
    \label{fig:MDE4ML}
\end{figure*}

The Model-driven Engineering (MDE) paradigm offers a potential solution to reduce the aforementioned complexities through abstraction~\cite{hutchinson2011model,moin2022model}. MDE advocates using software models at different abstraction levels to (semi-)automatically build software systems~\cite{brambilla2017model, moin2022model}. It has been effectively applied over several years in aerospace, automotive, telecommunication, business information systems, and mobile apps~\cite{mussbacher2014relevance, hutchinson2011model, hutchinson2011empirical, shamsujjoha2021developing}. Similar MDE techniques apply to software systems with ML components provided that existing modeling techniques are customized for ML-specific information, e.g., software architecture models describe classical software components and ML components, and the generated artifacts are tailored to ML, e.g., ML models or training code~\cite{moin2022model}. MDE has the potential to significantly enhance the development of ML-based systems~\cite{bucchiarone2020grand} by hiding complexities, increasing productivity, and improving system quality~\cite{moin2022model, mohagheghi2008proof}. MDE approaches, with their capability to model ML-based systems at a high level of abstraction, facilitate ML novices and experts~\cite{bucchiarone2020grand} in the development, integration, and maintenance of the systems. The effort and time required to develop and maintain an ML-based system may also be reduced through automated artifact generation~\cite{volter2013model}. 
Figure \ref{fig:MDE4ML} models an example ML component by specifying the ML algorithm (Random Forest), training parameters, and the training and test datasets. By applying MDE, this model is automatically transformed into code and documentation for the ML component with lower technical barriers and higher efficiency. Additional benefits of MDE include easier system management~\cite{bhattacharjee2019stratum}, early detection of bugs~\cite{mohagheghi2008proof}, lower costs~\cite{fleurey2007model, mohagheghi2013empirical}, and improved understanding and collaboration among diverse stakeholders~\cite{khalajzadeh2020end}.

To explore this promising and relatively new research area, we conducted a systematic literature review (SLR) focusing on MDE approaches used to develop systems with ML components (that we term `MDE4ML'). Through this study, we aim to collate, summarize, and report interesting findings in the literature on MDE4ML. We identify the goals of existing studies, key MDE approaches used, and the modeling languages, frameworks, and model transformation tools applied to develop ML-based systems. We also analyze the ML aspects addressed in the studies, evaluation methods, existing limitations, and future opportunities. Our analysis reveals that most MDE solutions for ML lack maturity and good tooling, often ignore data pre-processing steps, responsible ML development practices and scalability considerations, and have limited emphasis on ML aspects other than design, development, and training. Our findings can help future researchers efficiently identify current research trends and limitations in studies on MDE for ML components and guide future research. We followed the well-known and widely accepted SLR guidelines by Kitchenham et al.~\cite{kitchenham2009systematic, kitchenham2007guidelines}. The main contributions of this SLR are as follows:
\begin{itemize}
    \item Identification, analysis, data extraction, and synthesis of 46 primary studies highly relevant to MDE4ML;
    \item Insights into current trends in MDE for ML components, e.g., most MDE approaches focus on supervised learning;.
    \item Key limitations in existing studies on MDE for ML components, e.g., scalability -- one of the most important concerns in ML development -- is seldom considered in MDE4ML; and
    \item Key future research directions and recommendations for further studies on MDE for ML.
\end{itemize}

The rest of the paper is organised as follows: Section~\ref{sec:background} provides an overview of the background and related work on MDE4ML. Section~\ref{sec:Method} presents details of our research methodology. Section~\ref{sec:Results}
reports our findings from the selected primary studies. Section~\ref{sec:Threats} addresses the threats to validity. Section~\ref{sec:Discussion}  discusses the interesting results of our SLR and recommendations for future research, and Section~\ref{sec:Conclusion}  concludes the paper.

\section{Background and Related Work}~\label{sec:background}
\subsection{Model-driven Engineering}~\label{subsec:MDEBackground}
Model-driven Engineering (MDE) is a software development methodology that relies on models as the primary artifacts that drive the development process~\cite{ciccozzi2019execution, almonte2021recommender,hutchinson2011model}. This differs from traditional software development processes such as waterfall and agile, where the focus is on development phases like requirements engineering, design, and implementation, and models are only used as auxiliary artifacts to support these activities and serve as documentation~\cite{ciccozzi2019execution}. 
The focus of MDE is on the continual refinement and transformation of models, beginning with computation-independent models (CIMs), to platform-independent models (PIMs) and then platform-specific models (PSMs)~\cite{brambilla2017model}. Finally, these models are transformed into code, documentation, configurations, and tests for the software system.

MDE relies on two key aspects: abstraction and automation~\cite{mohagheghi2009mde}. Models are abstractions of complex entities; they hide unwanted information so modelers can easily focus on areas of interest~\cite{schmidt2006model, brambilla2017model}. 
In MDE, models are automatically transformed into artifacts such as code, documentation, and other models to achieve various goals such as merging, translation, refinement, refactoring, or alignment~\cite{brambilla2017model}. These transformations help reduce developers' manual effort and production time by generating executable artifacts -- leading to improved software quality, reduced complexity, and decreased development time and effort~\cite{kelly2008domain}. There are two types of transformations in MDE: 1) Model-to-Text (M2T) transformations, for a given input model a M2T transformation produces a textual artifact such as code or documentation as output; and ) Model-to-model (M2M) transformations, for a given input model an M2M transformation produces a different kind of model, for example translating a model from one language to another~\cite{brambilla2017model}.

A model is created in a modeling language, conforming to a meta-model that defines the syntax and semantics of that language. There are two types of modeling languages: general-purpose languages (GPL) and domain-specific languages (DSL). GPLs are intended for modeling generic concepts applicable to multiple domains; some examples include the Unified Modeling Language (UML)~\cite{eriksson2003uml}, Petri-nets~\cite{peterson1977petri} and finite state machines~\cite{wagner2006modeling}. On the other hand, a DSL has modeling concepts tailored to a specific domain or context, like SysML for embedded systems, HTML for web page development, and SQL for database queries~\cite{brambilla2017model}.

While exploring the literature, one encounters terms similar to MDE: examples include model-driven architecture (MDA), model-driven development (MDD), and model-based engineering (MBE). MDA is an architectural standard~\cite{mda} developed by the Object Management Group (OMG) \cite{omg} for MDD. MDD refers to automatically generating artifacts from models, whereas MDE has a broader scope and includes analysis, validation~\cite{almonte2021recommender}, interoperability of artifacts and reverse engineering \cite{brambilla2017model}. MBE is a lighter version of MDE, where models are not necessarily the central focus of the engineering process; however, they provide critical support~\cite{brambilla2017model}. This SLR primarily focuses on MDE.

\subsection{Machine Learning}
Machine Learning (ML) is a branch of Artificial Intelligence (AI) that enables machines to learn patterns from data without being explicitly programmed~\cite{samuel1959machine}. ML algorithms are fed with existing data to \textit{train} them and produce an ML model. This trained ML model then has the capability to \textit{infer}, i.e., predict outcomes for new data inputs or also commonly known as \emph{ML model inference}~\cite{mueller2021machine}. For example, an ML model trained on stock prices for a company till September 2023 can predict stock prices in the following months. ML is preferable when solving problems that would require very complex and difficult-to-maintain traditional algorithms~\cite{geron2022hands}. Since ML algorithms can learn autonomously, they reduce complexity and facilitate easier maintenance~\cite{geron2022hands}. This ability of ML to minimise complexity, learn from changing data, and make future predictions is immensely valuable for businesses~\cite{lee2020machine}. According to a recent survey~\cite{rackspace2023report}, organizations report that applying ML increases employee efficiency by 20\%, innovation by 17\%, and lowers costs by 16\% -- leading to increased adoption of ML in practical settings~\cite{rackspace2023report}.

ML can further be divided into three broad categories: supervised learning, unsupervised learning, and reinforcement learning. The most suitable ML approach depends on the specific problem and data.
Supervised learning is when an ML algorithm is trained on a labeled dataset that has labels to define the meaning of data~\cite{mueller2021machine}. For example, a dataset with images labeled as ``cat'' or ``not cat'' images. Supervised learning algorithms learn to make classifications or predictions by learning patterns and relationships in labeled data~\cite{lee2020machine,mueller2021machine}. When the labels are discrete, this is known as \textit{classification} and when labels are continuous, this is known as \textit{regression}~\cite{mueller2021machine}. Once the algorithm is trained, the performance is evaluated on unseen or test data. Some popular supervised learning algorithms include linear regression, decision trees, naive Bayes classifier, support vector machines (SVM), random forest, and artificial neural networks (ANNs)~\cite{lee2020machine}. Supervised model applications include fraud detection and recommender systems~\cite{mueller2021machine}. 

Unsupervised learning is when an ML algorithm is trained on an unlabeled dataset with few or no labels to define the meaning of data~\cite{mueller2021machine,lee2020machine}. Unsupervised learning algorithms attempt to understand hidden patterns in data and group similar data together creating a classification of the data~\cite{mueller2021machine}. Unsupervised learning works without any guidance, hence it is most suitable for large volumes of data when classifications are unknown and data cannot be labeled~\cite{mueller2021machine}. Evaluating the performance of such algorithms can be challenging due to the lack of ground truth. Some popular unsupervised techniques include clustering, k-means, principal component analysis, and association rules~\cite{lee2020machine}. Applications of unsupervised models include customer segmentation and clustering user reviews~\cite{mueller2021machine}.

 Reinforcement learning is when an ML algorithm receives feedback on actions to guide the behavior toward an optimal outcome~\cite{mueller2021machine, lee2020machine}. Reinforcement learning algorithms are not trained with datasets; instead, they learn from trial and error in a simulated environment or a real-world environment~\cite{mueller2021machine}. Desired behaviors are rewarded and reinforcement learning algorithms attempt to maximize rewards through successful decisions~\cite{lee2020machine,mueller2021machine}. These algorithms are most suitable when sequential decision-making is required, interaction with an environment is possible and feedback is available. 
 Some popular reinforcement learning algorithms are Q-learning, temporal difference learning, hierarchal reinforcement learning, and policy gradient~\cite{lee2020machine}. Applications of reinforcement learning include robotics, self-driving cars, and game playing~\cite{lee2020machine}.
 
\subsection{Model-driven Engineering for Machine Learning (MDE4ML)}
Developing and managing systems with ML models and components is challenging. 
Some aspects of this complexity are immature requirements specification~\cite{kuwajima2020engineering, ahmad2023requirements}, constantly evolving data~\cite{baumann2022dynamic}, lack of ML domain knowledge~\cite{yohannis2022towards}, integration with traditional software \cite{atouani2021artifact}, responsible use of ML~\cite{yohannis2022towards}, and deployment and maintenance of ML models~\cite{kourouklidis2021model, langford2021modalas}. 

These complexities introduce several challenges. For example, Nils Baumann et al.~\cite{baumann2022dynamic} describe how challenging it is to handle changing datasets; ML engineers have to manually merge new and old datasets and re-train the entire ML model;  Benjamin Jahi et al. \cite{jahic2023semkis} point out how challenging it is to describe the dataset and neural network requirements to satisfy customer expectations;  Benjamin Benni et al. \cite{benni2019devops} state how the development of a correct ML pipeline is a highly demanding task, data scientists must have knowledge and experience to go through numerous data pre-processing and ML models to select the best one; and Kaan Koseler et al. \cite{koseler2019realization} mention the difficulties developers face when attempting to use ML techniques with big data, developers need to acquire knowledge of the problem space, domain and ML concepts. There is a need for solutions to efficiently and effectively address these challenges~\cite{raedler2023model}.

A synergy between MDE and ML development exists, where software models are leveraged to drive the development and management of ML components~\cite{safdar2022modlf, yohannis2022towards, kourouklidis2021model}. This should not to be confused with AI or ML for MDE (AI4MDE), where intelligent agents and recommenders support users in modeling and related activities \cite{almonte2021recommender, gil2021artificial, boubekeur2020towards, saini2019teaching}. The application of MDE for ML-based systems (MDE4ML) offers many potential benefits to developers, such as reduced complexity~\cite{kourouklidis2021model, bucchiarone2020grand}, development effort, and time~\cite{yohannis2022towards,gatto2019modeling}. Domain experts, software engineers and ML novices can also take advantage of ML through the abstraction and automation of MDE \cite{shi2022feature,moin2022supporting, bucchiarone2020grand}. Additionally, MDE can also improve the quality of the ML-based system through easier maintainability, scalability~\cite{selic2003pragmatics}, reusability, and interoperability~\cite{brambilla2017model}.

\subsection{Key MDE4ML Related Work}
MDE4ML has received growing attention from researchers in recent years. We found six relevant secondary studies comprising SLRs, scoping reviews, and surveys. In their SLR \cite{raedler2023model}, the authors identify 15 primary studies on MDE for AI and analyze them with respect to MDE practices for the development of AI systems and the stages of AI development aligned with CRoss Industry Standard Process for Data Mining (CRISP-DM) \cite{wirth2000crisp} methodology. However, this study only considers a small subset of studies and performs a shallow analysis with no details about goals, end-users, types of models, implemented tools, and evaluation. A second SLR \cite{zafar2017systematic} reviews 24 papers on MDE for ML in the context of Big data analytics. This study has a narrower scope compared to ours and provides only a brief overview of the models, approaches, tools, and frameworks in the studies. In a third SLR \cite{li2022can}, 31 studies on no/low code platforms for ML applications are reviewed. This study is limited to no/low code approaches and therefore misses out on many other MDE for ML studies. A scoping review is presented in \cite{mardani2023model} on MDE for ML in IoT applications. The study examines 68 studies in depth; however, the review focuses more on MDE for IoT applications and only four of the selected studies apply ML techniques. A preliminary survey on DSLs for ML in Big data is presented in \cite{portugal2016preliminary}, with an extended version in \cite{portugal2016survey}. These surveys do not follow a systematic review process, include studies only for big data, and briefly highlight the DSLs and frameworks in the studies. From the analysis of existing literature, we found that the available secondary studies consist of limited subsets of papers on MDE for ML, lack analysis of key areas like goals, end-users, ML aspects, MDE approach details, evaluation methods, and limitations, and often do not follow a systematic and rigorous review process. Therefore, we aim to address these gaps in this SLR.

\section{Research Methodology}~\label{sec:Method}
We conducted a Systematic Literature Review (SLR) on Model-driven Engineering (MDE) approaches for systems with machine learning (ML) components (MDE4ML). This review aims to analyze existing primary studies and synthesize significant findings to guide future research and practice. This literature review has been conducted in conformance with the systematic literature review guidelines for SE presented by Kitchenham et al. \cite{kitchenham2009systematic, kitchenham2007guidelines}. 

\begin{figure*}[ht]
    \centering
    \includegraphics[width=0.875\textwidth]{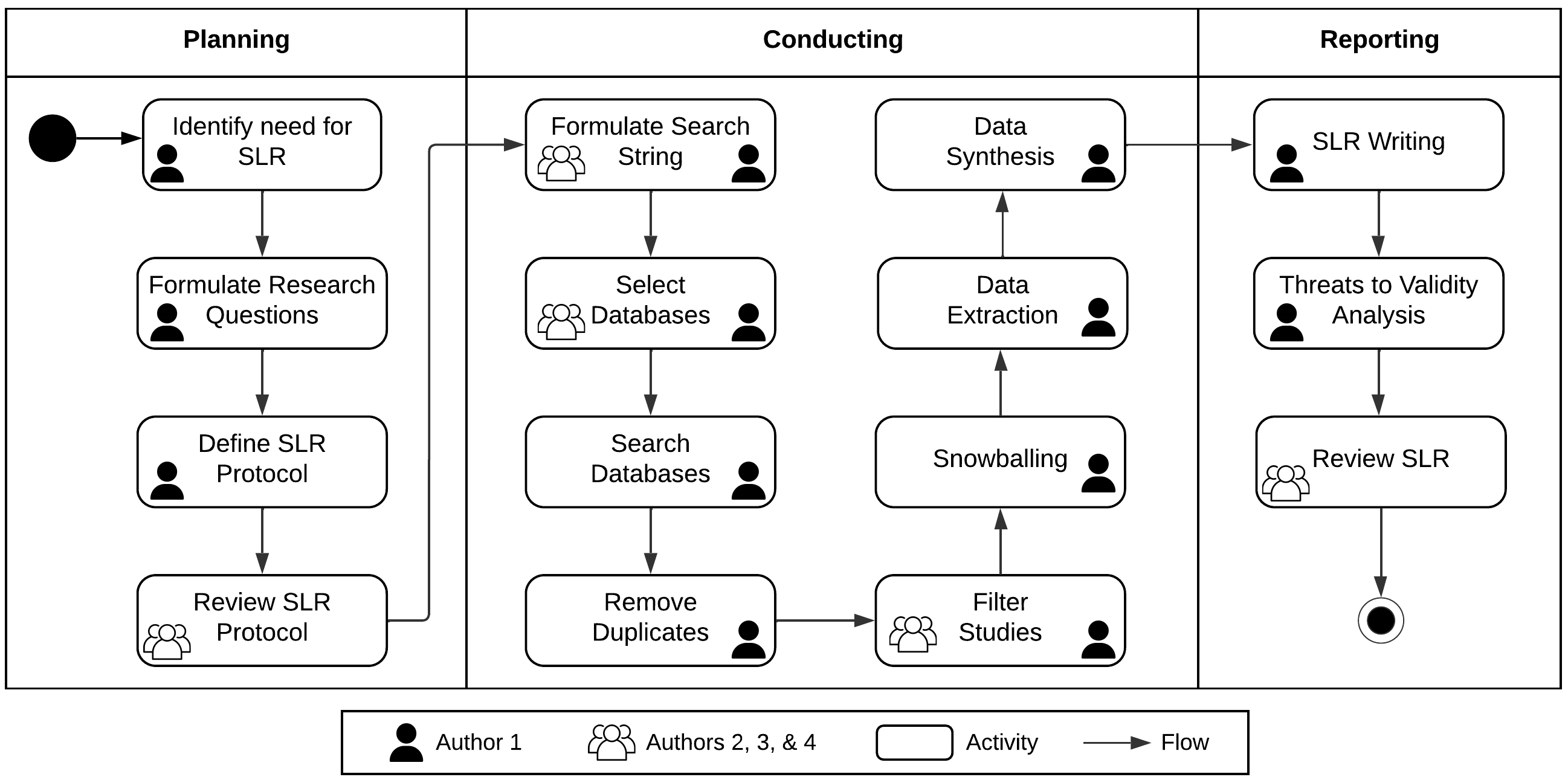}
    \caption{Systematic Literature Review Process}
    \label{fig:SLR process}
    \vspace*{-1em}
\end{figure*}

Figure~\ref{fig:SLR process} provides an overview of our review process. We divide the work into three stages: planning, conducting, and reporting. During the \textit{planning} stage, we identified the need for this SLR, formulated research questions (RQs), and defined the SLR protocol. 
In the \textit{conducting} stage, all authors collaborated to formulate search strings and select databases. The first author searched all selected databases and removed duplicate studies to create an initial pool of papers. Over multiple iterations, the first author filtered studies based on predefined criteria. Cross-validation was used to check searches and selection and to resolve ambiguities with the other authors. Forward and backward snowballing was used to identify other highly relevant primary studies. This resulted in 46 primary studies for analysis. Data extraction and synthesis activities for the final 46 papers, listed in Appendix A, were performed by the first author under the close supervision of the other authors. In the final stage \textit{reporting}, we documented all the significant findings and analyzed threats to the validity of the SLR.

\subsection{Research Questions}
The objectives of this study were to identify the motivations, solutions, evaluation techniques, and limitations of MDE4ML. Following the PICOC (population, intervention, comparison, outcomes, and context) approach, we developed four research questions (RQs)~\cite{petticrew2008systematic}. The PICOC for RQs in this SLR is shown in Table~\ref{table:picoc}.

\begin{table}[htbp]
\centering
\caption{PICOC for Research Questions}
\label{table:picoc}
\footnotesize
\begin{tabular}{ p{2cm} p{13cm}} 
 \hline
 \textbf{Population} & Systems with machine learning (ML) components \TBstrut \\ 
 \textbf{Intervention} & Model-driven engineering (MDE) approaches for systems with ML components  \TBstrut \\
 \textbf{Comparison} & Not applicable  \TBstrut \\
 \textbf{Outcomes} & The consequence of using MDE for systems with ML components \TBstrut \\
 \textbf{Context} & Include: MDE approaches for systems with ML components \TBstrut \\ & Exclude: AI approaches automating, recommending, or enhancing the MDE process, MDE approaches for systems with AI components
other than ML, pre-deployment Model-based testing approaches for ML systems \TBstrut \\ 
\hline
\end{tabular}
\end{table}

\par\textbf{RQ1. What is the motivation behind applying MDE approaches to systems with ML components?} -- RQ1 explores the key motivations and goals for applying MDE for ML-based systems in our selected primary studies. We further looked into other relevant aspects, such as the ML techniques, application domain, end users, and outcomes. 

\par\textbf{RQ2. Which MDE approaches and tools are presented in the literature for systems with ML components?} -- RQ2 examines the MDE solutions presented in the literature for ML-based systems. We identified and classified the modeling languages applied, transformations and automation levels supported, and the artifacts generated. We also explore the MDE tools available and the underlying meta-tools and frameworks. From the ML perspective, we looked at the ML phases, training data, and ML libraries considered in the primary studies.

\par\textbf{RQ3. How are existing studies on MDE4ML evaluated?} -- RQ3 investigates the evaluation techniques, metrics, datasets, and settings applied in the selected primary studies. 

\par\textbf{RQ4. What are the limitations and future work of existing studies on MDE4ML?} -- RQ4 identifies the limitations of current studies and key future research challenges of MDE4ML. In this context, we further examine the studies in terms of limitations in their approach, evaluation, and solution quality. 

\subsection{Study Selection}
Our database search results contain a number of irrelevant studies. Therefore, we developed detailed selection criteria as part of our SLR protocol. These criteria are divided into inclusion and exclusion criteria (shown in Table \ref{table:inclusion}) that we used to filter primary studies. For this review, we only consider primary English-language studies that focus on MDE for ML-based systems. The studies must be academic and have their full text available online. We excluded any irrelevant papers or did not have enough information to extract, such as vision papers, posters, magazine articles, keynotes, opinion papers, and experience papers. This SLR focuses on MDE4ML approaches, so we did not include any AI4MDE papers. These criteria were applied to all papers to select the most relevant ones, and discussions between all authors resolved any ambiguities in the study filtering process.

\begin{table}[htbp]
\centering
\caption{Inclusion and Exclusion Criteria}
\label{table:inclusion}
\footnotesize
\begin{tabular}{ p{2cm} p{13cm}} 
\hline
\textbf{Criteria ID} & \textbf{Criterion} \TBstrut \\
\hline
I01 & Papers about MDE for systems with ML components \TBstrut \\ 
I02 & Full text of the article is available  \TBstrut \\
I03 & Peer-reviewed studies that have been used in academia with references from literature  \TBstrut \\
I04 & Papers written in English language  \TBstrut \\
\hline
\hline

E01 & Papers about ML (or its subsets) that do not use MDE \TBstrut \\ 
E02 & Papers about MDE for any subset of AI other than ML  \TBstrut \\
E03 & Papers about pre-deployment model-based testing of systems with ML components.  \TBstrut \\
E04 & Studies leveraging AI to improve, enhance, or automate MDE \TBstrut \\
E05 & Short papers that are less than four pages \TBstrut \\
E06 & Conference or workshop papers if an extended journal version of the same paper exists \TBstrut \\
E07 & Papers with inadequate information to extract \TBstrut \\
E08 & Non-primary studies (Secondary or Tertiary Studies) \TBstrut \\
E09 & Vision papers and grey literature (unpublished work), books (chapters), posters, discussions, opinions, keynotes, magazine articles, opinion, experience and comparison papers \TBstrut \\
\hline
\end{tabular}
\end{table}

\subsection{Search Strategy}
\begin{figure*}[htbp]
    \centering
    \includegraphics[width=\textwidth]{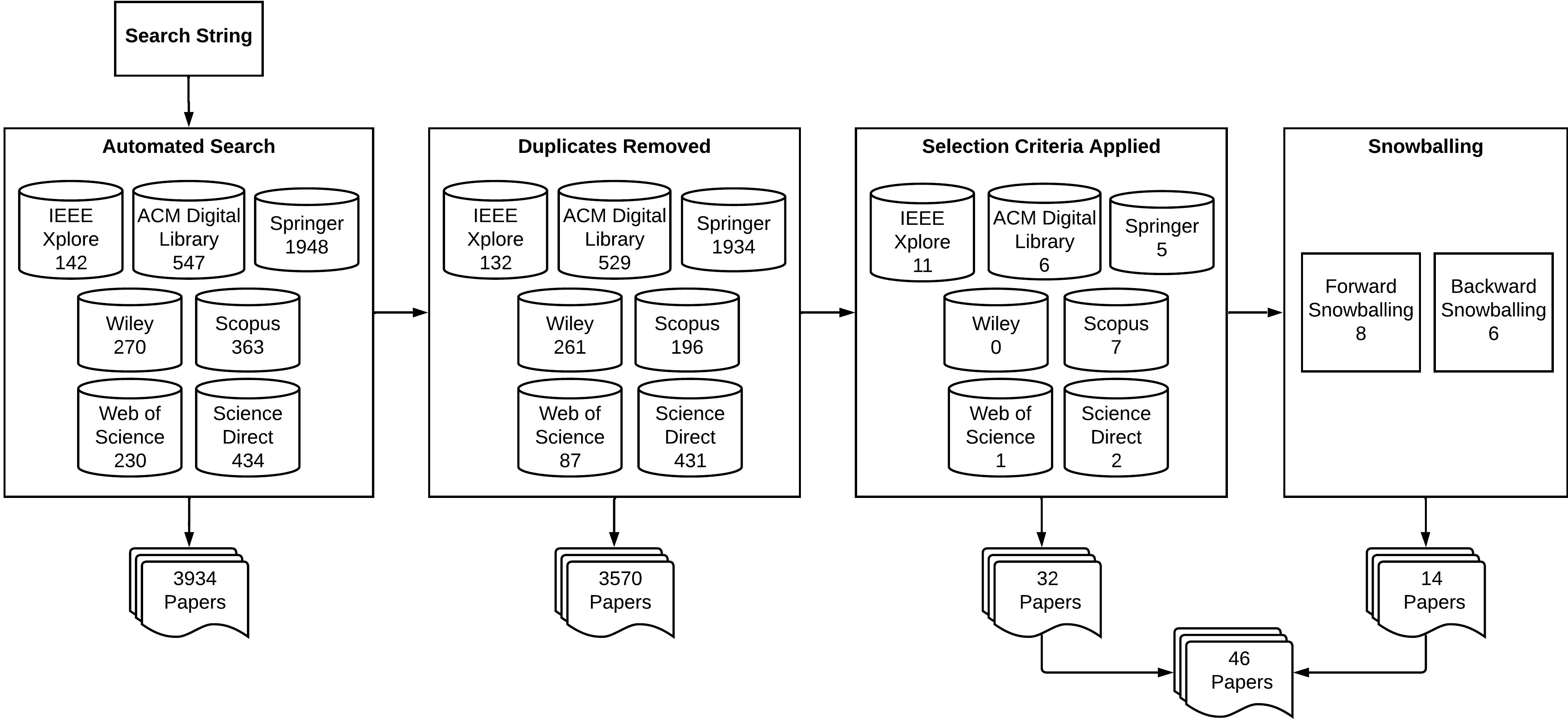}
    \caption{Study Search Process}
    \label{fig:searchProcess}
    \vspace*{-1em}
\end{figure*}

Figure \ref{fig:searchProcess} provides an overview of our search process, with all the steps detailed below.

\subsubsection{Search String Formulation}
For automated search in online databases, a search string was defined using key terms from the PICOC in Table \ref{table:picoc} and alternative terms. We realize that model-based software engineering and low code/no code are not the same as MDE; however, we included them in the search string since they are sometimes used interchangeably in the literature. This string was tested on seven online databases, IEEE Xplore, ACM Digital Library, Springer, Wiley, Scopus, Web of Science, and Science Direct, and refined over several iterations to get the most relevant results. Each time the search was executed, the first author randomly selected a sample of 5-7 papers from the search results and skimmed through them to ensure relevance and further refine the string. We combine search string keywords in the same category using the OR operator and keywords in different categories using the AND operator. For example, all ML keywords are joined by the OR operator, whereas the AND operator joins the ML and MDE keywords. The final search string below was slightly modified for some databases to get the best results. 
 
\subsubsection{Automated Search and Duplicate Removal}
The automated search was conducted in March 2023. The final search string was executed on the selected online database search engines to extract an initial pool of 3,934 papers. The search was restricted to academic articles, including journals, conference papers, and workshop papers, with no time range specified. We used the default search for all databases except Scopus, where we performed an advanced search on titles, abstracts, and keywords. The list of resultant papers was downloaded, and a Python script was developed to remove duplicates. The script also removed conference or workshop papers from the list if a journal version was found with the same title and authors. After duplicate removal, we were left with 3,570 papers for further screening.

\subsubsection{Filtering Studies}
We filtered the 3,570 potentially relevant papers in three iterations by applying the inclusion and exclusion criteria. The first screening based on title and abstract yielded 72 potentially relevant papers. In the second screening, we skimmed through the entire paper and further narrowed it down to 55 papers. The third and final screening was done during data extraction, resulting in 32 highly relevant papers. During the filtering process, we maintained a list of all papers on \textit{Google Sheets} and color-coded them on relevance, i.e., green for `relevant', yellow for `might be relevant', and red for `irrelevant'. The color codes helped authors focus and classify the papers that needed further attention. 

\begin{center}
\begin{myframe}[width=45em,top=10pt,bottom=10pt,left=10pt,right=10pt,arc=10pt,auto outer arc,title=\centering\textbf{Search String}]
\footnotesize
\begin{center}
("machine learning" OR "supervised learning" OR "unsupervised learning" OR "semi-supervised
learning" OR "reinforcement learning" OR "deep learning" OR "ensemble learning" OR "neural network" OR
"naive bayes" OR "decision tree" OR "deep boltzmann machine" OR "deep belief network" OR "convolutional
neural network" OR "recurrent neural network" OR "generative adversarial network" OR "auto encoder"
OR "gradient boosted regression trees" OR "adaboost" OR "gradient boosting machine" OR "random forest"
OR "perceptron" OR "back propagation" OR "hopfield network" OR "radial basis function network" OR
"linear regression" OR "stepwise regression" OR "logistic regression" OR "least square regression" OR
"adaptive regression" OR "locally estimated scatterplot" OR "k means clustering" OR "k medians clustering"
OR "hierarchical clustering" OR "mean shift clustering" OR "expectation maximization" OR "gaussian naïve
bayes" OR "multinomial naïve bayes" OR "bayesian network" OR "bayesian belief network" OR "k nearest
neighbour" OR "learning vector quantization" OR "self organizing map" OR "locally weighted learning" OR
"transfer learning" OR "support vector machines" OR "classification and regression tree" OR "CHAID" OR
"conditional decision tree" OR "decision stump" OR "long short term memory network" OR "gaussian
mixture" OR "hidden markov model" OR "Q learning" OR "temporal difference learning" OR "dimensionality
reduction") \\ AND \\ ("model driven development" OR "model driven engineering" OR "model based software
engineering" OR "model driven software engineering" OR "model transformation" OR "model driven
architecture" OR "low code solution" OR "low code application" OR "low code applications" OR "low code
paradigm" OR "low code approach" OR "low code platform" OR "low code development" OR "low code/no
code" OR "no code/low code" OR "no code development platform")
        \end{center}
    \end{myframe}
\end{center}

\subsubsection{Snowballing}
To account for any missed papers in the automated search process, we performed a manual search following the snowballing guidelines by Wohlin~\cite{wohlin2014guidelines}. The 32 selected papers and related work papers were snowballed in forward and backward directions over three iterations until no new suitable papers were found. We gathered eight (8) papers through our forward snowballing and six (6) papers through our backward snowballing.

\subsection{Data Extraction and Synthesis}
We created a Google Form with 40 questions corresponding to our four RQs to ensure all required data was extracted from the papers. This form was divided into five sections: the first section for general information and publication trends, including title, authors, publication venue, and citation count; the second section for motivations, goals, application domain, and users; the third section for MDE approaches for ML-based systems; the fourth section for evaluation techniques and tools and the last section for limitation and future challenges. The answer options for the questions in the form consisted of 23 short answers, ten long answers, two checkboxes and 14 radio buttons. For quality assessment, we ran pilot tests; the first author extracted data for six papers and compared it with data extracted by the other authors for the same papers. A close match was found between both, after which the first author extracted data for all the remaining papers. During pilot tests, small updates were made to improve the Google Form, for example, adding check boxes for common answers and improving the structure of the question. From the data extraction, we gathered qualitative and quantitative data for the next stage of data synthesis. The first author performed synthesis using various graphs, figures, and tables under the guidance of the other authors. 

\subsection{Quality Assessment}
We devised a five-point scoring system to assess the quality of the selected primary studies to answer five predefined questions. The scores range from very poor~(1) and inadequate~(2) to moderate~(3), good~(4), and excellent~(5). Such QA scoring for evaluating the quality of primary studies is commonplace in SLRs~\cite{shamsujjoha2021developing, hidellaarachchi2021effects}. The quality assessment questions we used are based on Kitchenham's guidelines ~\cite{kitchenham2007guidelines} and are shown below: \\
\textbf{QA1:} Are the aims clearly stated? \\
\textbf{QA2:} Is the solution clearly defined?\\
\textbf{QA3:} Are the measures used clearly defined? \\
\textbf{QA4:} Does the report have implications for practice? \\
\textbf{QA5:} Does the report discuss how the results add to the literature? \\

\begin{table}[htb]
    \centering
        \caption{Quality Assessment of Selected Primary Studies}
\footnotesize
    \begin{tabular}{cccccc|cccccc}
    \hline
      \textbf{ID}&\textbf{QA1} &\textbf{QA2} & \textbf{QA3}  & \textbf{QA4}  & \textbf{QA5}  &\textbf{ID}  & \textbf{QA1}  & \textbf{QA2} & \textbf{QA3}  & \textbf{QA4}  & \textbf{QA5}  \\
    \hline
    
       P1  & 5 & 5 & 3 & 4 & 5 & P24 & 4 & 2 & 1 & 2 & 2 \\
       P2  & 5 & 5 & 3 & 5 & 5 & P25 & 5 & 5 & 1 & 5 & 5 \\
       P3  & 5 & 4 & 5 & 3 & 3 & P26 & 2 & 3 & NA & NA & NA \\
       P4  & 5 & 5 & 4 & 3 & 5 & P27 & 3 & 4 & 5 & 3 & 1 \\
       P5  & 5 & 5 & NA & NA & NA & P28 & 4 & 4 & 2 & 3 & 1\\
       P6  & 5 & 3 & NA & NA & NA & P29 & 3 & 2 & 1 & 2 & 1\\
       P7  & 3 & 3 & 1 & 2 & 1 & P30 & 5 & 5 & 5 & 4 & 3 \\
       P8  & 5 & 5 & NA & NA & NA & P31 & 5 & 5 & 5 & 5 & 5\\
       P9  & 4 & 4 & 1 & 4 & 4 & P32 & 4 & 3 & 1 & 1 & 1 \\
       P10  & 5 & 5 & 2 & 5 & 5 & P33 & 5 & 4 & 4 & 2 & 2 \\
       P11  & 5 & 4 & 1 & 4 & 2 & P34 & 5 & 4 & 3 & 4 & 1\\
       P12  & 5 & 4 & 1 & 4 & 2 & P35 & 4 & 5 & 2 & 5 & 5\\
       P13  & 5 & 3 & 1 & 4 & 3 & P36 & 4 & 3 & 1 & 2 & 1\\
       P14  & 5 & 3 & 1 & 3 & 3 & P37 & 5 & 4 & 1 & 3 & 1 \\
       P15  & 3 & 3 & 1 & 1 & 2 & P38 & 4 & 3 & 1 & 2 & 1 \\
       P16  & 4 & 4 & 1 & 4 & 5 & P39 & 5 & 5 & 5 & 1 & 4\\
       P17  & 4 & 3 & 1 & 3 & 5 & P40 & 4 & 3 & 1 & 3 & 1 \\
       P18  & 4 & 3 & NA & NA & NA & P41 & 4 & 3 & 1 & 4 & 4 \\
       P19  & 4 & 3 & 4 & 3 & 3 & P42 & 5 & 3 & 1 & 3 & 4\\
       P20  & 5 & 4 & 1 & 4 & 4 & P43 & 2 & 2 & NA  & NA & NA\\
       P21  & 3 & 3 & NA & NA & NA & P44 & 5 & 4 & 1 & 4 & 4 \\
       P22  & 5 & 5 & 4 & 5 & 5 & P45 & 4 & 4 & 1 & 3 & 3 \\
       P23  & 5 & 5 & 1 & 5 & 4 & P46 & 3 & 4 & 1 & 3 & 1 \\
    \hline
    \end{tabular}
    \label{tab:quality}
\end{table}

The results of our quality assessment are captured in Table \ref{tab:quality}. Questions QA3-QA5 are only applicable to studies that provide an evaluation and are marked as not applicable (NA) for all other studies without an explicit evaluation component. From our quality assessment, we found that 19/46 included papers were of good quality, 15/46 were of average quality, and 12/46 were poor quality. We did not exclude any papers since MDE4ML is an emerging research area and our selection of 46 studies was already limited. Furthermore, we aimed to minimize any publication bias in our research.

\section{Results}~\label{sec:Results}
This section presents the results of our SLR on MDE approaches for systems with ML components. We organize these findings based on the four previously mentioned research questions.
\subsection{Publication Trends}

\begin{figure}[htbp]
    \centering
    \begin{subfigure}{0.5\textwidth}
    \includegraphics[width=\textwidth]{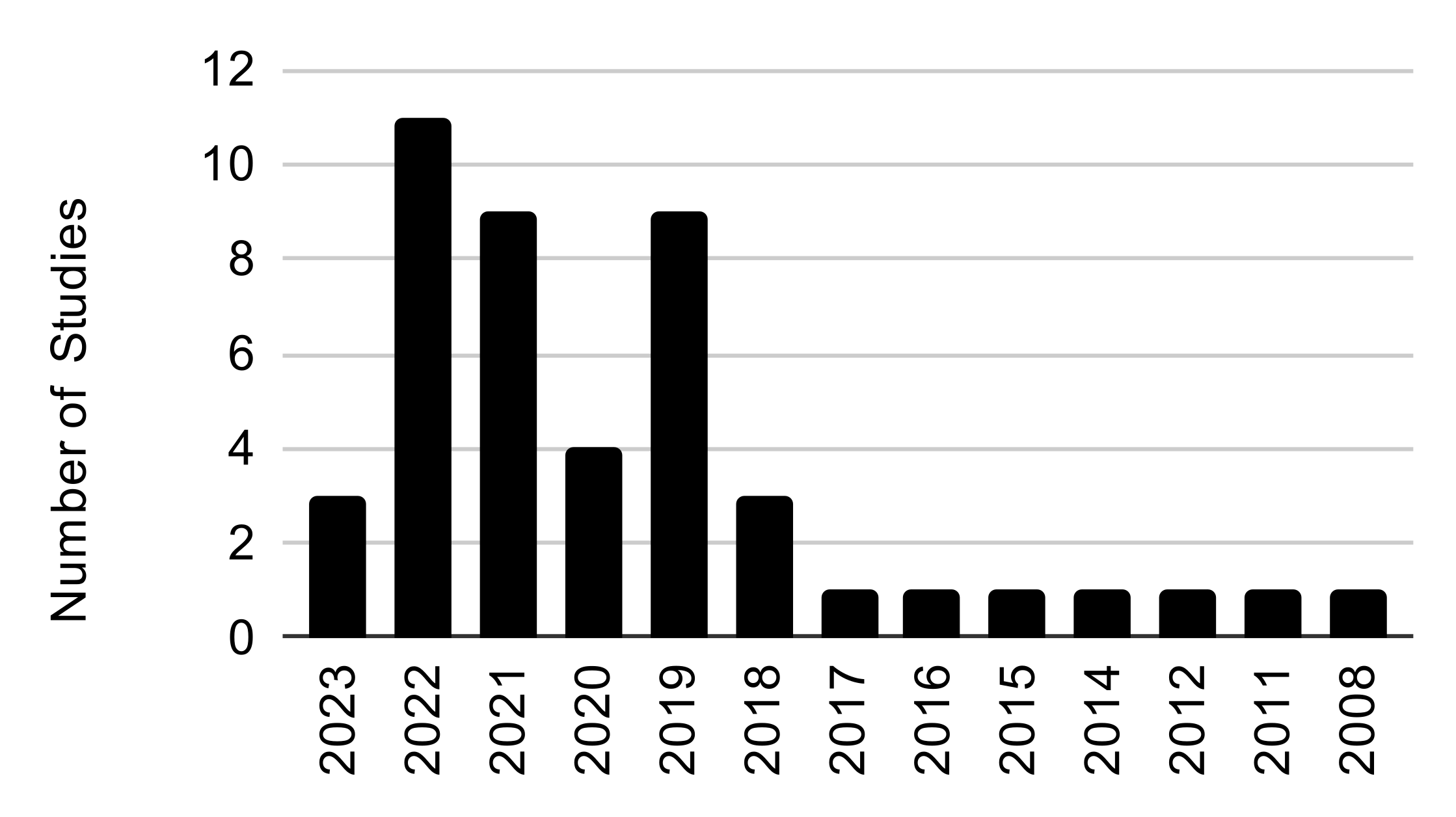}
    \caption{Study Distribution by Year}
    \label{fig:yearDist}
    \vspace*{-1em}
    \end{subfigure}
\hfill
    \begin{subfigure}{0.45\textwidth}
    \centering
    \includegraphics[width=\textwidth]{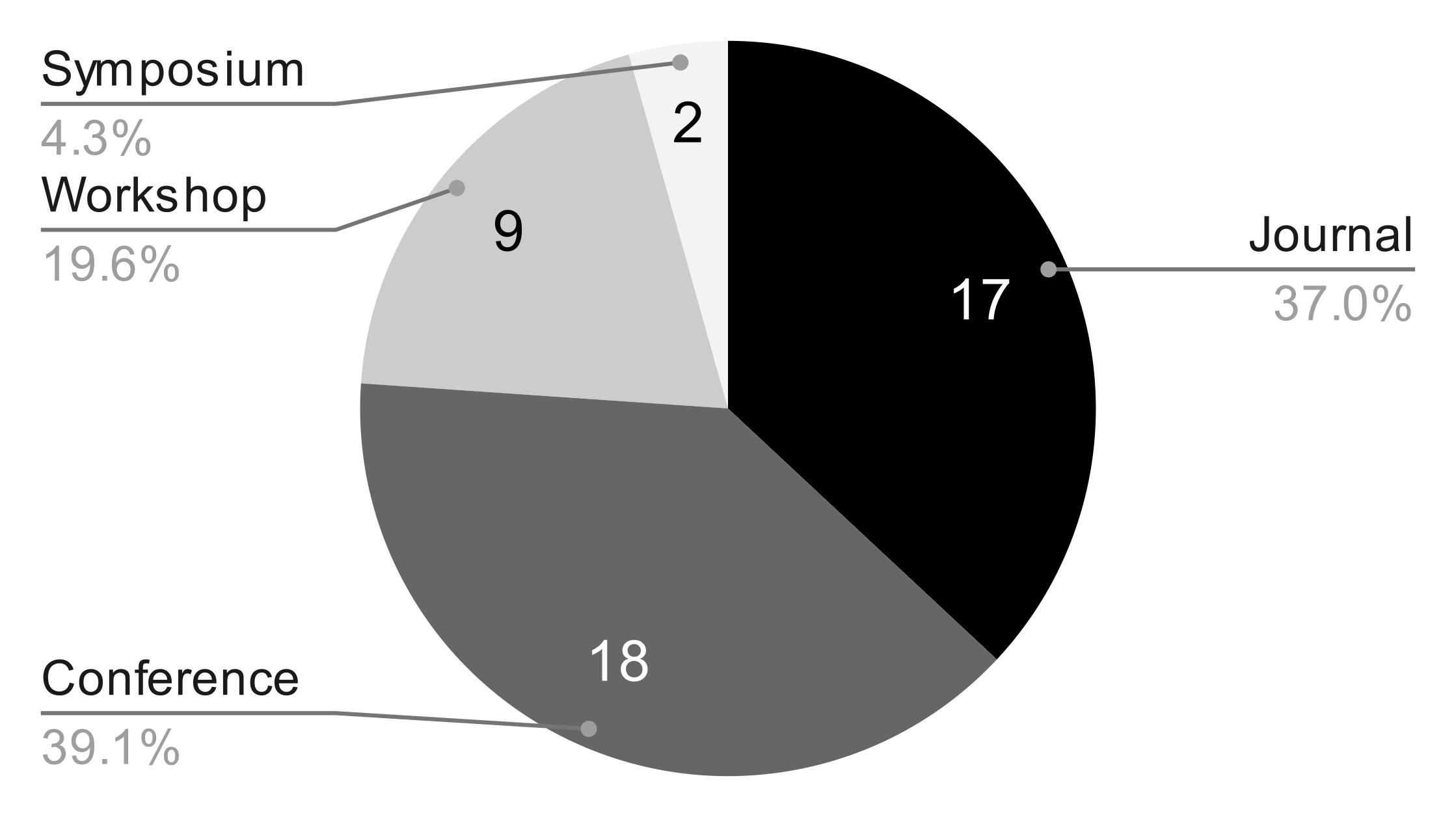}
    \caption{Study Distribution by Publication Type}
    \label{fig:studyType}
    \vspace*{-1em}
\end{subfigure}
    \vspace*{1em}
 \caption{Publication Trends}
\end{figure}

We selected 46 primary studies on MDE4ML, published over a span of 16 years, as shown in Figure~\ref{fig:yearDist}. Appendix A lists the full citation details of these primary studies. There are no studies included from 2009, 2010, and 2013. The number of papers published from 2011 to 2018 was low compared to 2019 and onward. There was a drop in 2020, which could have been due to the COVID-19 outbreak. However, we are not sure. The high number of studies in 2021 and 2022 show the increasing research interest in MDE for ML and we hope to see the same trend in the future. Our search process was conducted in March 2023 and could thus be the reason for the low numbers in 2023.  

Most of the studies were published in conferences and journals, contributing to ~39\% and 37\% of the total paper count, followed by workshops and symposiums, making up for the remaining ~20\% and ~4\%, respectively. Figure \ref{fig:studyType} provides more detail about the type of publications included in this study. The selected 46 studies belonged to 34 different venues; the venues with two or more published studies are shown in Table \ref{table:pubTrends}. Model Driven Engineering Languages and Systems (MODELS) conference, co-located workshops, and Software and Systems Modeling (SoSym) journal have the highest number of papers in each publication type category. This is unsurprising as these are all highly reputed publication venues dedicated to MDE. 

\begin{table}[htbp]
\centering
\caption{Publication Venues with two or more research studies}
\footnotesize
\begin{tabular}{ p{11cm} p{2cm} p{2cm}} 
\hline
\textbf{Publication Venue} & \textbf{Type} & \textbf{No. of Studies}\TBstrut \\
\hline
Model Driven Engineering Languages and Systems (MODELS) & Conference & 4 \TBstrut \\ 
Model-Based Software and Systems Engineering (MODELSWARD) & Conference & 2 \TBstrut \\
Model Driven Engineering Languages and Systems (MODELS) - Companion  & Workshop  & 6 \TBstrut \\
Computer Languages & Journal & 2  \TBstrut \\
Software and System Modeling (SoSym) & Journal & 3  \TBstrut \\
\hline
\end{tabular}
\label{table:pubTrends}
\end{table}

\begin{figure*}[htbp]
    \centering
    \includegraphics[width=1\textwidth]{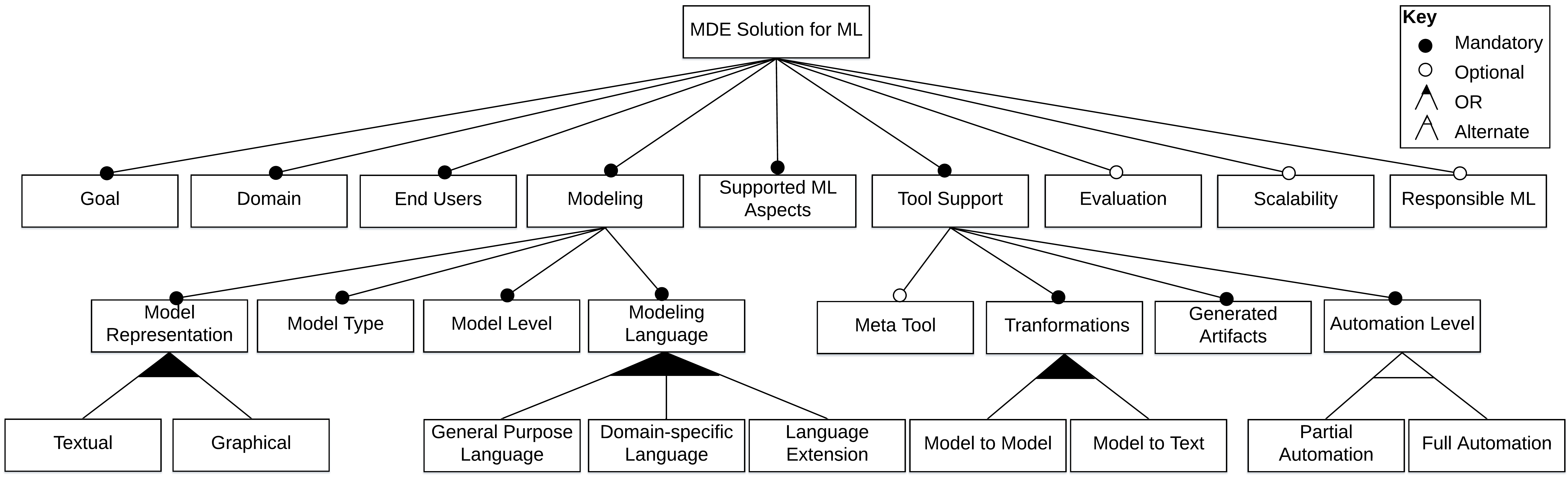}
    \caption{Features of Selected Primary Studies}
    \label{fig:features}
    \vspace*{-1em}
\end{figure*}

In the following sections, we describe the features of MDE solutions for ML-based systems with respect to our RQs. Figure~\ref{fig:features} provides an overview of these features, including the goal, end users, modeling, supported ML aspects and more. We extracted data corresponding to these features from our selected primary studies and reported our findings.

\subsection{RQ1 - Motivation for MDE4ML approaches}

\begin{figure}[htbp]
    \centering
    \includegraphics[width=0.35\textwidth]{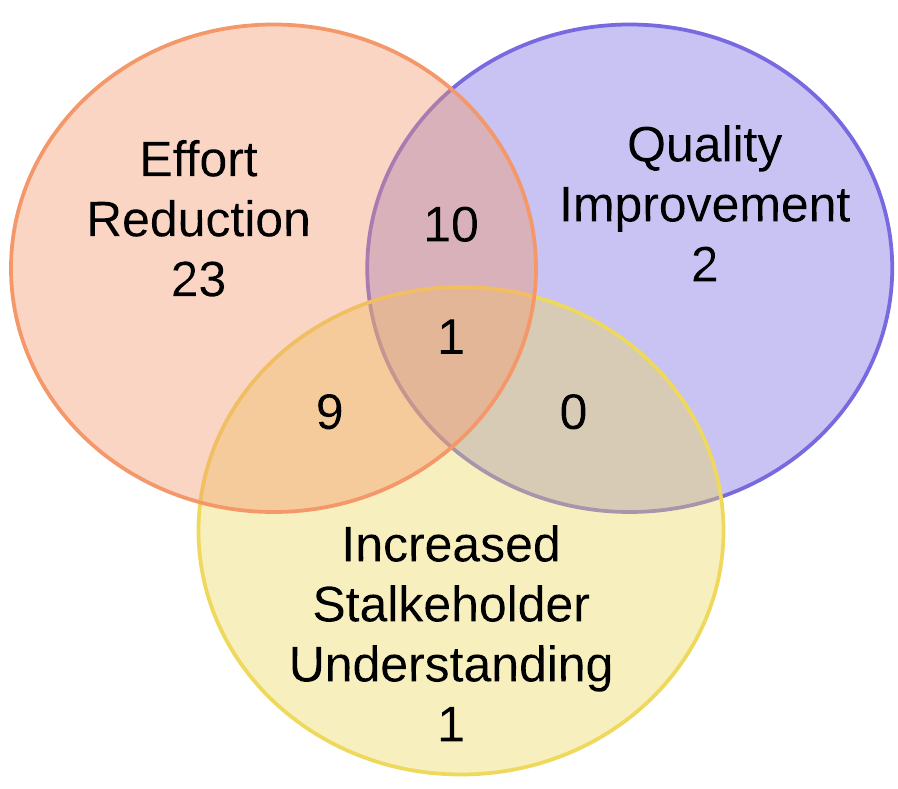}
    \caption{Goal Distribution in Studies}
    \label{fig:goals}
\end{figure}

\subsubsection{Motivation, Goals, and Objectives}
All the analyzed studies described their goals, objectives, and motivations for applying MDE techniques to systems with ML components. We divided these goals into three high-level categories \textit{effort reduction}, \textit{increased stakeholder understanding}, and \textit{quality improvement}. These categories are not mutually exclusive and many studies fall under more than one category, as shown in the Venn diagram in Figure \ref{fig:goals}. The full breakdown of goals, sub-goals, and relevant studies can be seen in Table \ref{table:goals}. \\

\begin{table}[htbp]
\centering
\caption{Goals of Primary Studies}
    \label{table:goals}
    \footnotesize
        \begin{tabular}{p{3cm} p{3.5cm} p{8.5cm}}
        \hline
        \textbf{Goal} & \textbf{Sub-goal} & \textbf{Studies} \TBstrut \\
            \hline
            \multirow{10}{*}{Effort Reduction}  & \multicolumn{1}{p{2.5cm}}{Abstraction} & \multicolumn{1}{p{8.5cm}}{P1, P4, P5, P6, P7, P8, P9, P10, P14, P16, P19, P21, P22, P23, P25, P27, P28, P29, P30, P33, P35, P36, P40, P41, P42, P43, P44, P46} \TBstrut\\
                                                & \multicolumn{1}{p{2.5cm}}{Automation} & \multicolumn{1}{p{8.5cm}}{P2, P4, P5, P9, P11, P12, P13, P16, P17, P18, P19, P21, P23, P24, P25, P27, P28, P30, P31, P32, P33, P34, P35, P36, P37, P38, P39, P41, P42, P43, P44, P46} \TBstrut\\
                                                & \multicolumn{1}{p{2.5cm}}{Integration} & \multicolumn{1}{p{8.5cm}}{P1, P5, P8, P11, P20, P22} \TBstrut\\
                                                & \multicolumn{1}{p{2.5cm}}{Monitoring} & \multicolumn{1}{p{8.5cm}}{P3, P6, P13} \TBstrut\\
                                                & \multicolumn{1}{p{3.5cm}}{System Management} & \multicolumn{1}{p{8.5cm}}{P3, P13} \TBstrut\\
                                                & \multicolumn{1}{p{2.5cm}}{Data management} & \multicolumn{1}{p{8.5cm}}{P11, P12} \TBstrut\\

            \hline
            \multirow{9}{*}{Quality Improvement}  & \multicolumn{1}{p{2.5cm}}{Reusability} & \multicolumn{1}{p{8.5cm}}{P8, P19, P23, P25} \TBstrut\\&{Extensibility} & \multicolumn{1}{p{8.5cm}}{P1, P8, P25, P26} \TBstrut\\&
            {Standardisation} & \multicolumn{1}{p{8.5cm}}{P1, P7, P10} \TBstrut\\
                                                     & \multicolumn{1}{p{3.5cm}}{Responsible ML} & \multicolumn{1}{p{8.5cm}}{P2, P3, P10} \TBstrut\\
                                                     & \multicolumn{1}{p{2.5cm}}{Interoperability} & \multicolumn{1}{p{8.5cm}}{P7, P45} \TBstrut\\

                                                     &\multicolumn{1}{p{2.5cm}}{Maintainability} & \multicolumn{1}{p{8.5cm}}{P11} \TBstrut\\
                                                     & \multicolumn{1}{p{2.5cm}}{Scalability} & \multicolumn{1}{p{8.5cm}}{P16} \TBstrut\\
                                                     & \multicolumn{1}{p{2.5cm}}{Reliability} & \multicolumn{1}{p{8.5cm}}{P16} \TBstrut\\
                                                     
             \hline
             \multirow{2}{3cm}{Increased Stakeholder Understanding}  & \multicolumn{1}{p{3.5cm}}{Support non-ML Experts} & \multicolumn{1}{p{8.5cm}}{P2, P17, P24, P28, P34, P39} \TBstrut\\
                                                     & \multicolumn{1}{p{3.5cm}}{Common Language} & \multicolumn{1}{p{8.5cm}}{P14, P15, P32, P35, P36} \TBstrut\\
            \hline
        \end{tabular}
\end{table} 

\sectopic{Effort Reduction}
was the most common aim mentioned in (43 out of 46 studies) MDE4ML papers, and focuses on effort reduction in development (e.g., P4, P9, P13), integration (e.g., P1, P8, P20), monitoring (e.g., P3, P6, P13) and managing systems (e.g., P3, P11, P12) with ML components. \textit{Abstraction} is one way to achieve effort reduction by creating models of complex systems that hide unnecessary details to focus only on the required aspects. For example, the goal of P5 and P9 is to use models to reduce complexity when developing ML solutions for cyber-physical systems. \textit{Automation} is another way to reduce development effort, artifacts such as code, configurations, and documentation can be generated from the models without any manual effort. For example, the goal of P10 is to generate dataset description documents from models with dataset details such as structure, provenance, and social concerns. P28 aims to automatically generate code for neural networks by transforming MDE models into code. \textit{Integration} of ML components into the rest of the system is not a trivial task; the objective of P11 is to ease this process through an ML artifact model. Another example is P8, which aims to simplify integration between neural network components and non-ML components to create reusable neural networks. \textit{Monitoring} refers to observing a system at runtime to ensure desired behavior. Despite being important, setting up monitoring mechanisms is a complex task. P6 aims to support ML experts through MDE by setting up a monitoring solution to detect performance drops. \textit{System management} and \textit{data management} are two other goals mentioned in the studies that correspond to less effort required for management. P12 identifies the challenges of managing dynamic datasets and presents a model-driven approach with the goal of automating data retraining and version management. For system management, a model-driven approach is described in P13 with the objective of managing ML-based analytics services on IoT devices.\\

\sectopic{Quality Improvement}
is the second common category (13 out of 46 studies) that contains studies that intend to improve the quality of ML-based systems using MDE. These qualities include \textit{reusability}, the ability to be used in different contexts without needing significant effort  (e.g., P8, P23); \textit{extensibility}, the ability to be extended with new features or functionalities (e.g., P25, P26); \textit{standardization}, the adherence to standards or best practice guidelines (e.g., P1, P10); \textit{responsible ML}, developing ML systems in a manner that maximize benefit and minimize risk  (e.g., P2, P3); \textit{interoperability}, the ability to easily communicate with other systems (e.g., P7);  \textit{maintainability}, ease of incorporating changes (e.g., P11); \textit{scalability}, the capability of handling increased workloads  (e.g., P16); and \textit{reliability}, the ability to perform consistently as required without failures (e.g., P16). In this category, we found the largest number of papers for reusability and extensibility, followed by standardization and responsible ML. P2 is an example of a responsible ML study and aims to measure and mitigate biases in ML models using a DSL. P45 is an example of an interoperability study and aims to create models for manufacturing intelligence that can seamlessly operate across various devices and systems. Through modeling various aspects, study P10 intends to standardize machine learning datasets. \\

\sectopic{Increased Stakeholder Understanding} 
is the least common category (11 out of 46 studies) with papers that aim to support collaboration (e.g., P35, P36) and improve system understanding in stakeholders other than ML experts (e.g., P2, P17, P34). We found six studies that \textit{support non-ML experts} in developing ML-based systems; these studies use domain-specific terminology and hide complex technical details. In P39, the goal is to provide a DSL to support software engineers in specifying requirements for neural networks. This is, otherwise, a challenging task since most software engineers are not experienced in deep learning~\cite{ahmad2023requirements}. We also found five studies focused on providing a \textit{common language} for easier communication and stakeholder collaboration. For example, P35 aims to facilitate multi-disciplinary teams developing data analytics and machine learning solutions using DSLs.  \\

\subsubsection{Machine Learning Techniques}

\begin{table}[htbp]
\centering
    \caption{Machine Learning Techniques Used in Studies}
    \label{table:MLtypes}
        \begin{tabular}{p{4.5cm} p{3.5cm} p{7cm}}
        \hline
        \textbf{Machine Learning Technique} & \textbf{Sub-type} & \textbf{Studies} \TBstrut \\
            \hline
            Generic Machine Learning \TBstrut	& - & P9, P10, P13, P15, P21, P22, P23, P32, P33, P41, P43 \TBstrut \\ 
            \hline
            \multirow{5}{*}{Supervised Machine Learning	}  & \multicolumn{1}{p{2.5cm}}          {Traditional} & \multicolumn{1}{p{7cm}}{P2, P6, P16, P17, P19, P35, P38} \TBstrut\\
                                                & \multicolumn{1}{p{2.5cm}}{Neural Networks} & \multicolumn{1}{p{7cm}}{P1, P3, P4, P7, P8, P11, P12, P14, P20, P24, P25, P26, P28, P29, P36, P39, P42, P44, P46} \TBstrut\\
                                                & \multicolumn{1}{p{3.5cm}}{Traditional and Neural Networks} & \multicolumn{1}{p{7cm}}{P27, P31, P34, P37, P45} \TBstrut\\
            
            \hline
                        Unsupervised Machine Learning \TBstrut	& - & - \TBstrut \\ 
            \hline
            Reinforcement Learning & - & P5, P18, P30, P40 \TBstrut\\
            \hline
        \end{tabular}
\end{table}

Table~\ref{table:MLtypes} shows the different subsets of ML techniques considered in the included primary studies. Of the 46 primary studies, 31 studies (67\%) explicitly focused on supervised ML, and 4 studies (9\%) on reinforcement learning. None of the studies focused explicitly on unsupervised learning or clustering. The remaining 11 studies (24\%) are termed as \emph{generic} ML studies, i.e., (i) the studies did not explicitly specify the type of ML their solution targets (e.g., P41, P43); (ii) the studies target both supervised and unsupervised learning (e.g., P22, P32), and (iii) the studies target supervised, unsupervised and reinforcement learning (e.g., P13). For instance, P22 offers an MDE solution for IoT devices using \emph{generic} ML and supports various supervised and unsupervised ML algorithms. \textit{Supervised machine learning} papers are further categorized into \textit{traditional supervised learning}, where studies use models such as decision trees, linear regression and naive Bayes (e.g., P19, P35, P38); \textit{neural networks}, where studies use neural networks and deep learning (e.g., P29, P44, P46) and \textit{traditional and neural networks}, where the studies support traditional supervised learning models and neural networks (e.g. P27, P34, P45). An example of a study exclusively for reinforcement learning is P40, which presents a DSL to model multi-agent reinforcement learning in distributed systems. 


\subsubsection{Application Domain}
\begin{figure}[htbp]
    \centering

    \centering
    \includegraphics[width=0.5\textwidth]{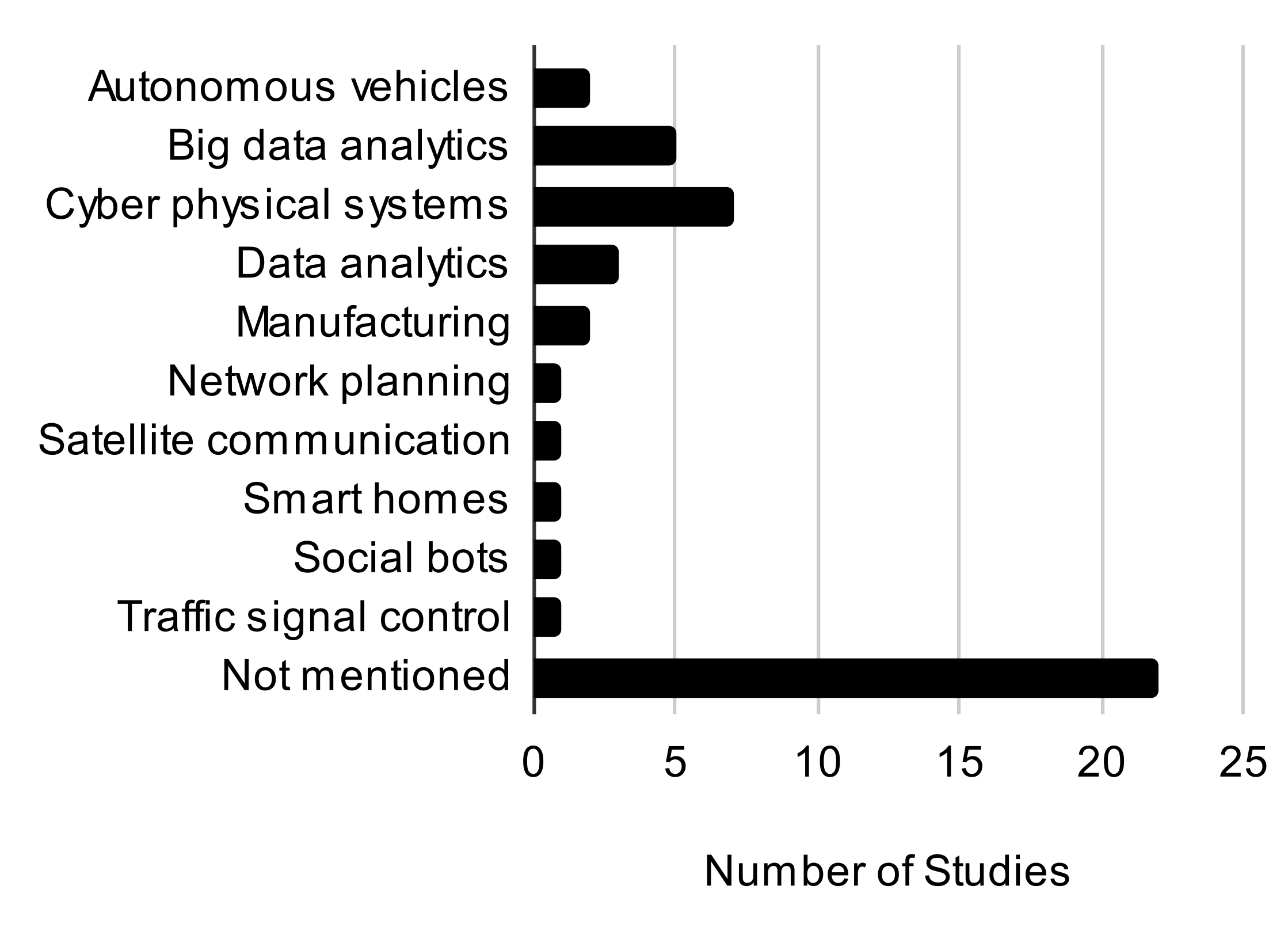}
    \caption{Application Domain Distribution}
    \label{fig:domain}
\end{figure}

\begin{table}[htbp]
\centering
    \caption{End Users Mentioned in Studies}
    \label{table:users}
    \footnotesize
        \begin{tabular}{p{3.5cm} p{5cm} p{6.5cm}}
        \hline
        \textbf{User Category} & \textbf{End User} & \textbf{Studies} \TBstrut \\
            \hline
            \multirow{3}{*}{ML-related Roles}  & \multicolumn{1}{p{4.5cm}}{ML Engineer} & \multicolumn{1}{p{6.5cm}}{P1, P5, P6, P7, P8, P10, P11, P12, P13, P15, P25, P37, P40, P41, P42, P43, P44, P46} \TBstrut\\
                                                & \multicolumn{1}{p{4.5cm}}{\mbox{Data Analyst/ Engineer/ Scientist}} & \multicolumn{1}{p{6.5cm}}{P2, P10, P12, P14, P16, P18, P22, P24, P32, P34, P35, P36, P38} \TBstrut\\
            \hline
            
            \multirow{6}{*}{Software \& Systems Roles}  & \multicolumn{1}{p{4.5cm}}{Software Engineer} & \multicolumn{1}{p{6.5cm}}{P1, P2, P3, P4, P9, P11, P13, P19, P20, P22, P23, P27, P28, P29, P30, P39} \TBstrut\\
                                                & \multicolumn{1}{p{4.5cm}}{Systems Engineer} & \multicolumn{1}{p{6.5cm}}{P9, P31, P33} \TBstrut\\
                                                & \multicolumn{1}{p{4.5cm}}{Business Analyst} & \multicolumn{1}{p{6.5cm}}{P35} \TBstrut\\
                                                & \multicolumn{1}{p{4.5cm}}{Formal Methods Analyst} & \multicolumn{1}{p{6.5cm}}{P14} \TBstrut\\
            \hline
            \multirow{2}{*}{Other Roles}  & \multicolumn{1}{p{4.5cm}}{Domain Expert} & \multicolumn{1}{p{6.5cm}}{P17, P21, P24, P25, P26, P28, P30, P32, P35, P36, P45} \TBstrut\\
            \hline
        \end{tabular}
\end{table}

Figure \ref{fig:domain} highlights the eleven application domains found in the selected studies. 52\% of studies mention an application domain with \textit{cyber physical systems (CPS)} and subsets of it being the most common. We found seven studies for generic \textit{CPS} (P5, P8, P9, P12, P22, P23, and P31), two for \textit{manufacturing systems} (P36 and P45), two for \textit{autonomous vehicles} (P1 and P3), one for \textit{smart homes} (P20), one for \textit{traffic signal control} (P30), one for \textit{satellite communication system} (P33) and one for \textit{network planning} (P25). The second most common application domain found in the primary studies was \textit{big data analytics} in five studies (P13, P17, P21, P35, and P43) and \textit{data analytics} in three studies (P32, P34, and P38). We also found one study (P24) for developing \textit{social bots}. The MDE solutions proposed in almost half the studies (P2, P4, P6, P7, P10, P11, P14-P16, P18, P19, P26-P29, P37, P39-P42, P44, and P46) were generic and could be applied in any domain.

\subsubsection{End Users}

\begin{figure}[htbp]
    \centering
    \includegraphics[width=0.35\textwidth]{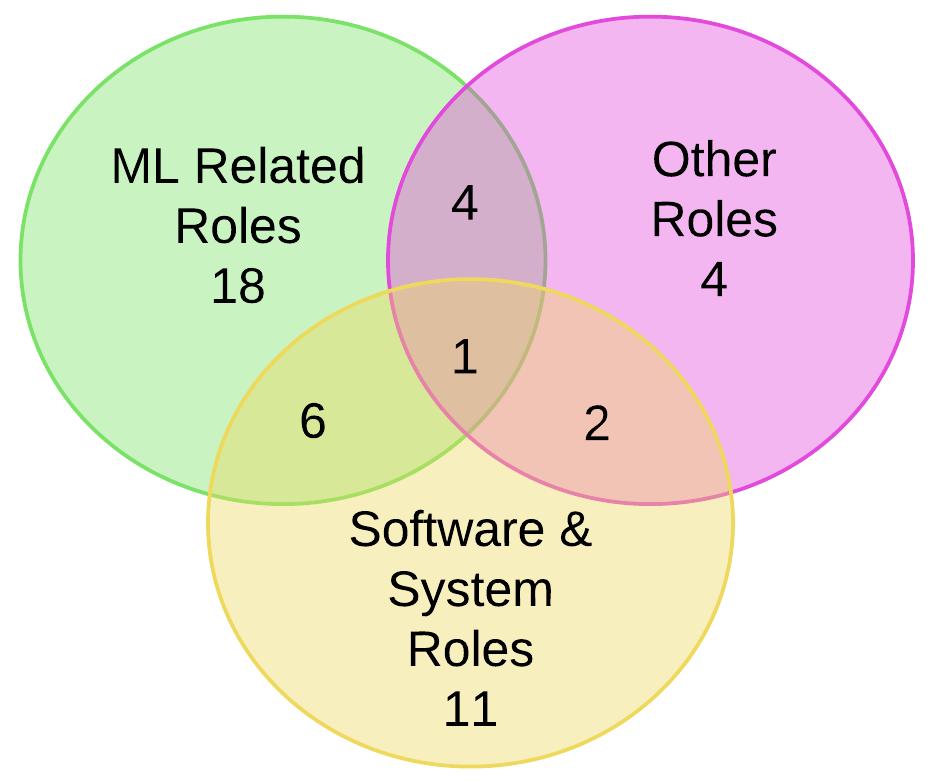}
    \caption{End Users Distribution in Studies}
    \label{fig:users}
\end{figure}

Figure \ref{fig:users} provides an overview of the end user distribution in the primary studies. The proposed MDE solutions for ML-based systems are intended for three different categories of users, as shown in Table \ref{table:users}. The \textit{ML-related roles} category includes \textit{ML engineer} and \textit{data analyst/ engineer/scientist}; the \textit{software and systems roles} category includes \textit{software engineer}, \textit{system engineer}, \textit{business analyst}, and \textit{formal methods analyst}; and the last category \textit{other roles} includes \textit{domain expert}. Of the 46 selected studies, 13 presented approaches were for more than one user category, while the remaining 33 focused on a single end-user category. Of all the studies (including overlaps), 18 focused on MDE solutions for ML engineers, followed by solutions for software engineers in 16 studies, and 11 focused on approaches for domain experts. An example of each of these three main categories is: P1 proposes an approach for ML engineers to model deep neural network architectures for autonomous vehicle perception; P9 presents a modeling approach for software engineers to model ML components integrated on edge IoT devices for data analytics; 
and P17 enables domain experts to represent ML problems as models and derive code from them.

\begin{table}[htbp]
\centering
    \caption{Contributions of Studies}
    \label{table:contributions}
    \footnotesize
        \begin{tabular}{p{4.5cm} p{2.5cm} p{2.5cm} p{5.5cm}}
        \hline
        \textbf{Contribution} &  & & \textbf{Studies} \TBstrut \\

            \hline
            \multirow{8}{*}{Tool}  & \multirow{6}{2.5cm}{Model-to-Text Transformer}  & \multicolumn{1}{p{2.5cm}}{Code Generator}& \multicolumn{1}{p{5.5cm}}{P1, P2, P4, P5, P7, P8, P9, P11, P12, P13, P14, P16, P17, P19, P20, P22, P23, P24, P25, P27, P28, P30, P31, P32, P34, P35, P37, P38, P39, P41, P42, P43, P44, P45, P46} \TBstrut\\
                                                & &\multicolumn{1}{p{2.5cm}}{Text Generator} & \multicolumn{1}{p{5.5cm}}{P3, P6, P10, P15, P21, P26, P35, P36} \TBstrut\\ \cline{2-4}
                                    & \multirow{1}{2.5cm}{Model-to-Model Transformer} & \multicolumn{1}{p{2.5cm}}{Model Generator} & \multicolumn{1}{p{5.5cm}}{P1, P3, P4, P18, P25, P29, P33, P40, P41, P42} \TBstrut\\
            \hline
            \TBstrut Domain-specific Language (DSL) \TBstrut	& & & P1, P2, P6, P7, P10, P11, P13, P14, P15, P16, P19, P21, P25, P26, P27, P28, P29, P30, P32, P35, P36, P37, P38, P39, P40, P41, P42, P43, P44, P46 \TBstrut \\ 
            \hline   
            Framework \TBstrut	& & & P1, P3, P5, P7, P8, P12, P13, P18, P19, P23, P24, P25, P26, P27, P29, P31, P34, P36, P37, P38, P44 \TBstrut \\ 
            \hline
            Model \TBstrut & & & P4, P11, P13, P20, P31, P37 \TBstrut\\
            \hline
            Modeling Approach \TBstrut	& & & P9, P11, P14, P16, P20, P30 \TBstrut \\ 
            \hline
            Modeling Language Extension \TBstrut	& & & P5, P17, P22, P33, P45 \TBstrut \\ 
            \hline
            ML Knowledge Base \TBstrut & & & P15, P16, P20, P34 \TBstrut\\
            \hline     
            Data Synthesizer \TBstrut & & & P31, P39, P45 \TBstrut\\
            \hline
            OCL Constraints \TBstrut & & & P27 \TBstrut\\
            \hline
            API \TBstrut & & & P40 \TBstrut\\
            \hline
            Meta-modeling Language  \TBstrut	& & & P23 \TBstrut \\ 
            \hline
        \end{tabular}
\end{table}

\subsubsection{Contributions}
\begin{figure}[htbp]
    \centering
    \includegraphics[width=0.35\textwidth]{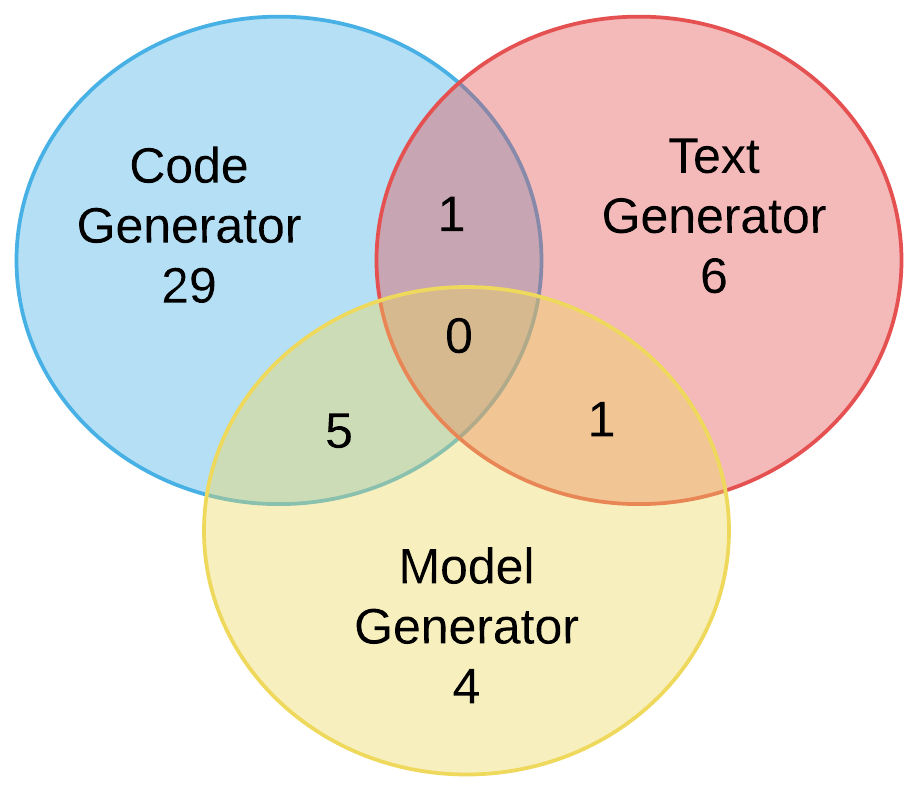}
    \caption{Tool Distribution in Studies}
    \label{fig:tools}
\end{figure}

\begin{figure*}[htbp]
    \centering
    \includegraphics[width=1\textwidth]{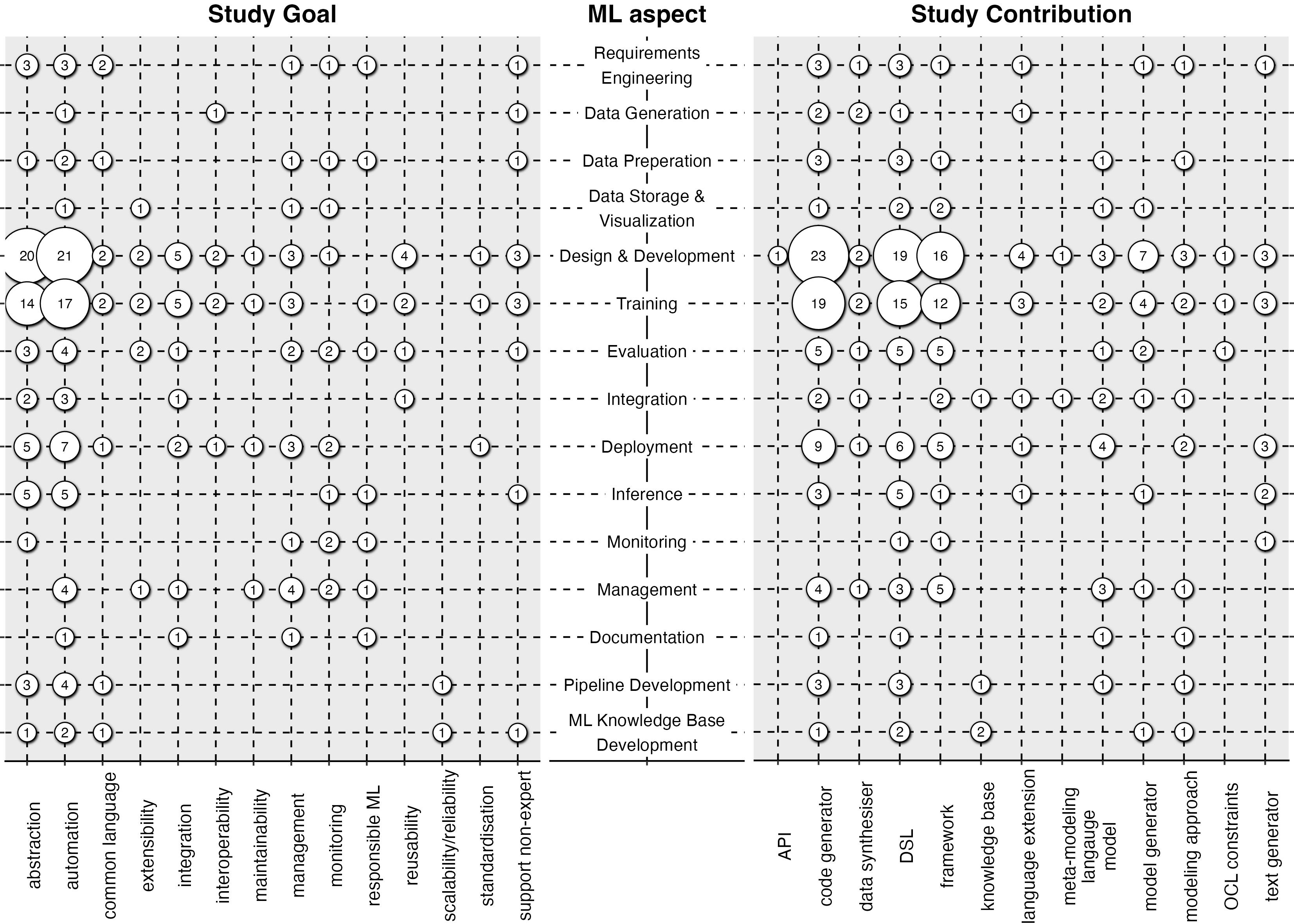}
    \caption{Bubble Chart for Study Goal, Study Contribution and Machine Learning Aspects}
    \label{fig:GoalvsContribution}
    \vspace*{-1em}
\end{figure*}

We identified eleven different contributions in the analysed MDE4ML approaches, tabulated in Table \ref{table:contributions}. We show tool-specific contributions in the Venn diagram in Figure~\ref{fig:tools}. The most common contribution (in 35 out of 46 studies) is a \textit{code generator} that transforms model(s) into code. For example, in P23, the code generator transforms CPS domain models to code for domain classes and then weaves ML code into it. An alternate approach in P17 reads a domain-specific model for baseball analytics using binary classification and produces ML classification code. Other kinds of generators presented in studies include \textit{model generator} (10 out of 46 studies) that transforms an input model(s) into an output model(s) and \textit{text generator} (8 out of 46 studies) that transforms model(s) into text other than code, e.g., documentation. The second common contribution (in 30 out of 46 studies) is a domain-specific language (\textit{DSL}), which includes graphical and textual modeling languages intended for a particular domain. A graphical DSL is proposed in P1 to model deep learning-based computer vision tasks in autonomous vehicles. A textual DSL is presented in P2 to mitigate biases in ML datasets, to model the dataset, training methods, bias metrics, and bias mitigation methods. Another common contribution (in 21 out of 46 studies) is an MDE \textit{framework} for systems with ML components. A framework to monitor ML components in CPS is introduced in P3. The study leverages goal models to evaluate the ML components at runtime and ensure correct behavior in uncertain environments. Several other outcomes were also identified such as \textit{models} (6 out of 46 studies), \textit{modeling approach} (6 out of 46 studies), 
\textit{modeling language extension} (5 out of 46 studies), 
\textit{ML knowledge base} (4 out of 46 studies),
\textit{data synthesizer} (3 out of 46 studies),
\textit{OCL constraints} (1 out of 46 studies study), \textit{APIs} (1 out of 46 studies study), and \textit{meta-modeling language} (1 out of 46 studies study). However, in comparison, these remaining contributions were significantly lower in number. Most papers make multiple contributions, e.g., P25 provides a framework, DSL, model generator, and code generator.

Figure~\ref{fig:GoalvsContribution} shows a bubble chart of the study goals and contributions against their relevant ML aspect. The size of the bubble depicts the frequency of studies in that category. For example, related to the \textit{requirements engineering} aspect of ML components, the goal of three studies is to achieve automation. In comparison, three studies provide a code generator as their contribution. While analyzing this figure, we found that in the selected studies, the preferred ML aspects are the \textit{design and development} of ML components, and \textit{training} of ML components. In contrast, \textit{monitoring} of ML components, and \textit{documentation} were the most neglected aspects.

\begin{center}
\begin{myframe}[width=45em,top=5pt,bottom=5pt,left=5pt,right=5pt,arc=10pt,auto outer arc,title=\centering\textbf{RQ1 Answer Summary}]
\footnotesize
The majority of studies aim to use MDE for ML-based systems to reduce effort through automation and abstraction, and provide code generators, DSLs, and frameworks. Considerably fewer studies have focused on quality improvement and increased stakeholder understanding as a goal of their research. Nearly half of the studies were not limited to any specific application domain, while most of the remaining studies focused on cyber-physical systems. The more significant part of the studies provides MDE solutions for users with ML-related roles like ML engineers and data scientists, whereas few are intended for domain experts. The most common type of ML technique found in the primary studies was deep learning, whereas the least common was unsupervised learning. We also found that the goals and contributions of most studies were related to the design, development, and training aspects of ML, whereas monitoring and documentation were ignored.
    \end{myframe}
\end{center}

\subsection{RQ2 - MDE4ML approaches and tools used} 

\subsubsection{Modeling Characteristics}

\begin{figure}[htbp]
    \centering
    \begin{subfigure}{0.3\textwidth}
    \includegraphics[width=\textwidth]{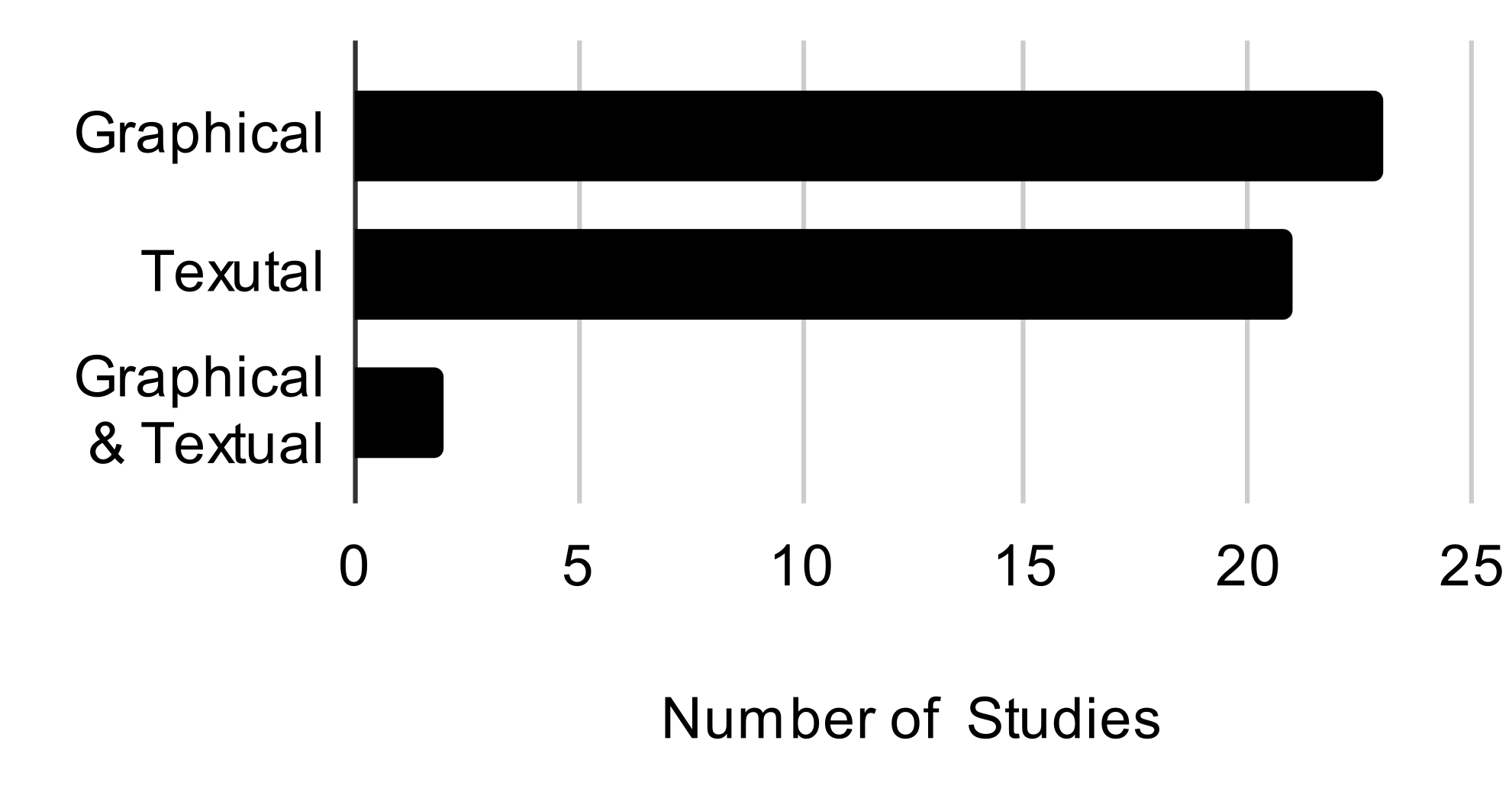}
    \caption{Model Representations in Studies}
    \label{fig:ModelRepresentation}
    \end{subfigure}
\hfill
    \begin{subfigure}{0.3\textwidth}
    \centering
\includegraphics[width=\textwidth]{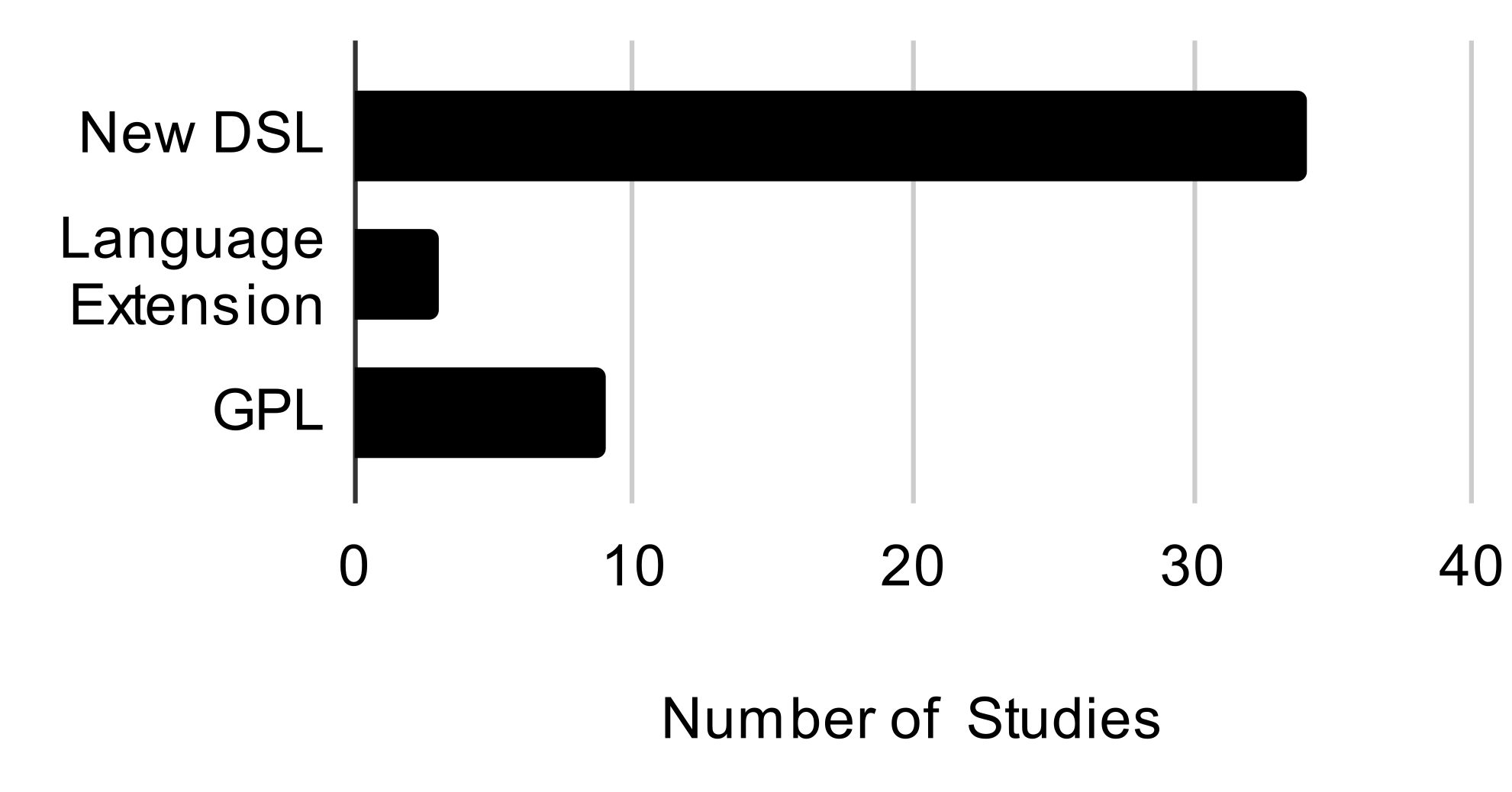}
    \caption{Language Types in Studies}
    \label{fig:ModelingLanguages}
\end{subfigure}
\hfill
    \begin{subfigure}{0.3\textwidth}
    \includegraphics[width=\textwidth]{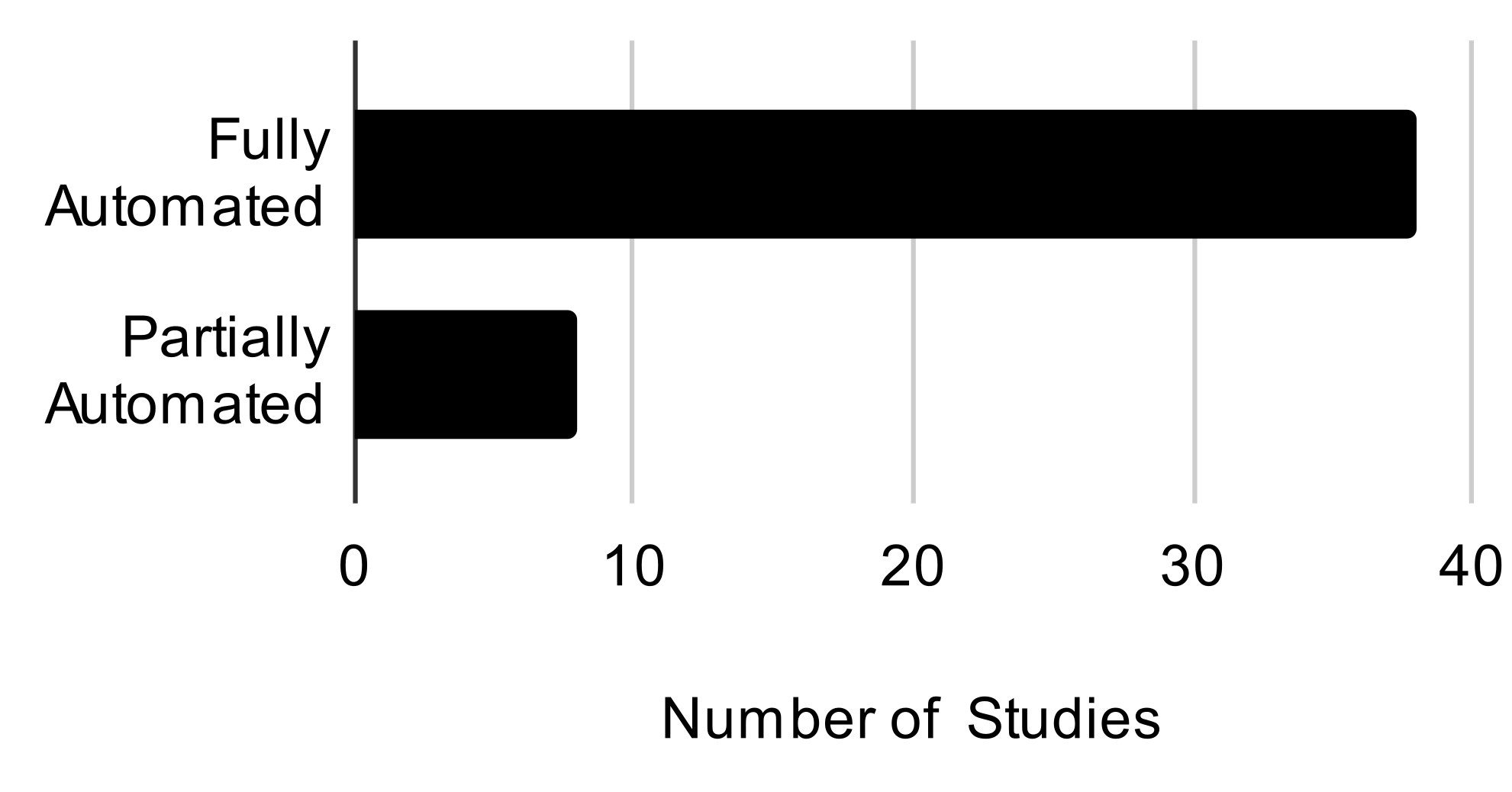}
    \caption{Automation Levels in Studies}
    \label{fig:AutomationLevels}
    \end{subfigure}
    \caption{MDE Solution Characteristics in Studies}
\end{figure}

\sectopic{Model Representation} refers to the \textit{graphical notations} or \textit{textual notations} used for expressing a model. Figure~\ref{fig:ModelRepresentation} shows our findings; half of the studies (23 out of 46 studies) use graphical models (P1, P3, P4, P13-P19, P21, P24-P27, P30, P32-P38), nearly half of the studies (21 out of 46 studies) use textual models (P2, P5-P12, P20, P23, P28, P31, P39-P46) and only a small number of studies (2 out of 46 studies) use both graphical and textual models (P22 and P29). An example of graphical representation is P33. It generates graphical SysML models of the architecture of autonomous systems with ML and non-ML components using restricted natural language requirements. An example of textual model representation is P41, wherein the authors build a textual domain-specific language (OptiML) to generate ML code for heterogeneous hardware platforms. P29 uses both graphical and textual representations to model artificial neural networks.\\

\sectopic{Modeling Languages} refer to specific languages or notations for creating abstract representations of systems in graphical, textual, or combined form. As discussed in Section~\ref{subsec:MDEBackground}, modeling languages can be classified as \textit{general purpose languages} (GPLs) such as UML and iStar, \textit{domain-specific languages} (DSLs), or \textit{extensions of an existing language}, such as UML profiles. As shown in Table~\ref{fig:ModelingLanguages}, we found the majority of studies (34 out of 46 studies) propose a new DSL (P1, P2, P6-P8, P10-P16, P19, P21, P23, P25-P30, P32, P34-P44, and P46), a significantly fewer fraction (9 out of 46 studies) use a GPL (P3, P4, P9, P17, P18, P20, P24, P31, and P45), and only 3 out of 46 studies extend an existing language (P5, P22, and P33). A DSL for MLOps is introduced in P37 to automate the ML pipeline. The authors use Kubernetes and a blockchain-based infrastructure. P6 presents a DSL for monitoring ML models and selects Kubernetes as a platform.  In contrast, P3 leverages GSN and KOAS general-purpose goal models to monitor ML components. An example of a study that extends an existing modeling language is P5. The authors extend the deep learning DSL family MontiAnna to support reinforcement learning, particularly for cyber-physical systems. \\

\sectopic{Information Modeled, Model Types and Levels.} Table~\ref{table:modelinfo} provides an overview of the models proposed in the selected studies. Models in MDE can be categorized into three levels: computation-independent models (CIMs), platform-independent models (PIMs), and platform-specific models (PSMs). CIMs contain high-level information about requirements or business processes without implementation details (e.g., P3, P18, P35). PIMs contain design or solution details without platform information (e.g., P5, P26, P39), and PSMs contain low-level platform-specific implementation details (e.g., P9, P35, P42). As shown in Table~\ref{table:modelinfo}, a significant fraction of studies model PIMs (42 out of 46 studies), of which 35 are only PIMs, six studies model PIMs and PSMs, and only one study (P35) models all three categories. Only two studies exclusively model CIMs (P3 and P18) and PSMs (P4 and P37).


{In terms of model types, 39 out of 46 studies were design-level models, followed by requirements-level models (6 out of 46 studies), and finally data-representation models (5 out of 46 studies). These results are not surprising since most studies focus on the design, development, and training of ML components. Other types of models, such as feature models, process models, and deployment models, were only found in a few studies.}

\begin{table}[htbp]
\centering
    \caption{Modeled Information, Model Level and Model Type in studies}

    \resizebox{\textwidth}{!}{ 
        \begin{tabular}{l l l l }
        \hline
       \TBstrut  \textbf{Paper ID} \TBstrut  & \textbf{Model Level(s)} & \textbf{Model Type(s)} & \textbf{Information Modeled}  \TBstrut\\
            \hline
            \TBstrut P1 \TBstrut	& PIM & Design model & DL framework application domains, CNN model, layers, datasets, training, hyper-parameters and evaluation \TBstrut \\
            \TBstrut P2 \TBstrut	& PIM & Design model & ML bias metrics, mitigation algorithms, datasets, training methods and hyper-parameters \TBstrut \\ 
            \TBstrut P3 \TBstrut	& CIM & Requirements model & Assurance cases for design and runtime, System goals, sub-goals, agents, objectives, and utility functions for goals\TBstrut \\
            \TBstrut P4 \TBstrut	& PSM   & Feature model, Design model & DL library features, API elements (i.e., classes, methods, and constructors) related to features, feature interactions\TBstrut \\
            \TBstrut P5 \TBstrut	& PIM    &  Design model & NN model, layers, training methods, and hyper-parameters\TBstrut \\
            \TBstrut P6 \TBstrut	& PIM    & Design model  & ML model to monitor, ML framework, data and concept drift detection algorithms, input features and output\TBstrut \\
            \TBstrut P7 \TBstrut	& PIM    & Design model &  ML problem, dataset, training method, NN model, neurons, layers, and connections\TBstrut \\
            \TBstrut P8 \TBstrut	& PIM    & Design model, Dataset model & NN architecture, layers, training method, hyper-parameters, input features, and output \TBstrut \\
            \TBstrut P9 \TBstrut	& PIM, PSM    &  Design model &  IoT service structure and behavior, data analytics model, data pre-processing, training methods, and prediction  \TBstrut \\
            \TBstrut P10 \TBstrut	& PIM    & Dataset model & Dataset description including metadata, composition, data instance, provenance, and social concerns \TBstrut \\
            \TBstrut P11 \TBstrut	& PIM    & Design model, Artifact model & Software, source code, dataset, pre-trained model and training environment archive, pipelines and hyper-parameters\TBstrut \\
            \TBstrut P12 \TBstrut	& PIM    & Design model, Dataset model  & NN architecture, layers, dataset, training methods, retraining, hyper-parameters, input features and output\TBstrut \\
            \TBstrut P13 \TBstrut	& PIM   & Design model  & ML models, frameworks, data pre-processing, training methods, hyper-parameters, deployment, and evaluation\TBstrut \\
            \TBstrut P14 \TBstrut	& PIM   & Requirements model & Dataset structure, properties, invariant properties, data, and equivalence classes\TBstrut \\
            \TBstrut P15 \TBstrut	& PIM    & Design model  & ML Knowledge source, version, relevance, reliability, decisions, ML algorithms and characteristics\TBstrut \\
            \TBstrut P16 \TBstrut	& PIM    & Design model  & ML pipeline, activities, ML model, ML component architecture, dataset, pipeline experiment, and justification \TBstrut \\
            \TBstrut P17 \TBstrut	& PIM    & Probabilistic Graphical model  & Classification problem, observed variables (input features), random variables, nodes, gates, plates and factors \TBstrut \\
            \TBstrut P18 \TBstrut	& CIM    & Requirements model & Goals, tasks, qualities, effects, preconditions, actor boundaries and links, \TBstrut \\
            \TBstrut P19 \TBstrut	& PIM    & Design model  & Fog layer, fog nodes, ML layer, ML algorithms, and rule-based algorithms \TBstrut\\
            \TBstrut P20 \TBstrut	& PIM    & Entity model & Physical entities, capabilities, states, location, users, activities, preferences, ML models\TBstrut \\
            \TBstrut P21 \TBstrut	& PIM    & Probabilistic Graphical model & Observed variables (input features), random variables, parameters, relationships, nodes, gates, plates and factors\TBstrut\\
            \TBstrut P22 \TBstrut	& PIM    & Design model  & ML features, ML models, data analytics libraries labels, results, datasets, timestamps, and AutoML support. \TBstrut \\
            \TBstrut P23 \TBstrut	& PIM    & Design model  & Specified, learned and derived properties, relations, parameters, ML algorithms, and features\TBstrut \\
            \TBstrut P24 \TBstrut	& PIM    & Design model  & Virtual learning environment instance, users, bots, user and bot actions, parameters, DL classifier, and triggers \TBstrut \\
            \TBstrut P25 \TBstrut	& PIM    & Design model  &  Wireless network plan properties, prediction problems, datasets, features, ML models, parameters, and evaluation \TBstrut \\
            \TBstrut P26 \TBstrut	& PIM    & Design model  & Data organized by Projects, project files (e.g. training data, testing data, validation data), runs and parameters\TBstrut \\
            \TBstrut P27 \TBstrut	& PIM    & Design model  & Entities, context, observations, notifications, properties, associations, ML models, inputs, output, and evaluation\TBstrut \\
            \TBstrut P28 \TBstrut	& PIM    & Design model  & NN architecture, layers, inputs, output, and training methods\TBstrut \\
            \TBstrut P29 \TBstrut	& PIM    & Design model  & ANN model, system, layers, links (weights), and bias\TBstrut \\
            \TBstrut P30 \TBstrut	& PIM    & Design model  & Agents, agent, decision and learning capabilities, entities, attributes, states, and decision options\TBstrut \\
            \TBstrut P31 \TBstrut	& PIM    & Design model  & NN architecture, layers, neurons in layers, inputs, weights, bias, and ML model meta data\TBstrut \\
            \TBstrut P32 \TBstrut	& PIM    & Process model & Dataflow process, sub-processes, function interface, inputs, output, constants, and ML algorithms\TBstrut \\
            \TBstrut P33 \TBstrut	& PIM    & Design model, Requirements model & System structure and behavior, properties, data, ML components, ML algorithms, and evaluation metrics \TBstrut \\
            \TBstrut P34 \TBstrut	& PIM    & Design model  & Dataset, dataset fields, domain, ML algorithms, parameters, data mining tasks, meta-features and predicted fields\TBstrut \\
            \TBstrut P35 \TBstrut	& CIM, PIM, PSM & Design model, Requirements model, & Big data analytics high-level tasks, sub-tasks, processes, stakeholders, operations, conditions, ML models, training  \TBstrut\\
            \TBstrut \TBstrut       &        & Process model, Artifact model, & methods, data processing techniques, data and artifacts, and deployment details\TBstrut \\
            \TBstrut \TBstrut       &        & Deployment model & \TBstrut \\
            \TBstrut P36 \TBstrut	& PIM    & Design model  & Manufacturing flows, processes, equipment and resources, NN model, layers, inputs, output, bias and edges\TBstrut \\
            \TBstrut P37 \TBstrut	& PSM    & Design model  & NN architecture, ML Pipeline, ML tasks, dataset import, training, evaluation, Kubernetes clusters, and nodes \TBstrut \\
            \TBstrut P38 \TBstrut	& PIM    & Design model, Dataset model & Classification problem, ML algorithm, features, evaluation metrics, labels, dataset, and hyper-parameters\TBstrut \\
            \TBstrut P39 \TBstrut	& PIM    & Design model, Requirements model & Requirements, NN behavior, inputs, output, datasets, equivalence classes, properties, and evaluation metrics \TBstrut \\
            \TBstrut P40 \TBstrut	& PIM    & Design model  & Goals, actors, actions, parameters, reward functions, properties, identifiers and messages \TBstrut \\
            \TBstrut P41 \TBstrut	& PIM, PSM  & Design model  & Vectors (vertices, edges, indices), Matrices, and Graphs to support operations in ML algorithms \TBstrut \\
            \TBstrut P42 \TBstrut	& PIM, PSM  & Design model  & NN model, layers, hyper-parameters, loops, tensor and scalar expressions, and functions\TBstrut \\
            \TBstrut P43 \TBstrut	& PIM    & Design model  & Inputs, output, functions, hyper-parameters, messages, graphs, nodes, expressions, and loops \TBstrut \\
            \TBstrut P44 \TBstrut	& PIM, PSM  & Design model  & NN model, layers, arrays and tensor operations, training, and evaluation\TBstrut \\
            \TBstrut P45 \TBstrut	& PIM, PSM  & Design model  & Regression model, NN model, inputs, layers, ML algorithms, functions, and bias \TBstrut \\
            \TBstrut P46 \TBstrut	& PIM, PSM  & Design model  & Sets, maps, iterations, functions, schedule, data dependencies, parameters, matrices, expressions and  statements
\TBstrut \\
            \hline
        \end{tabular}
        }
    \label{table:modelinfo}
\end{table}

\begin{figure}[htbp]
\centering
    \begin{subfigure}{0.49\textwidth}
    \includegraphics[width=\textwidth]{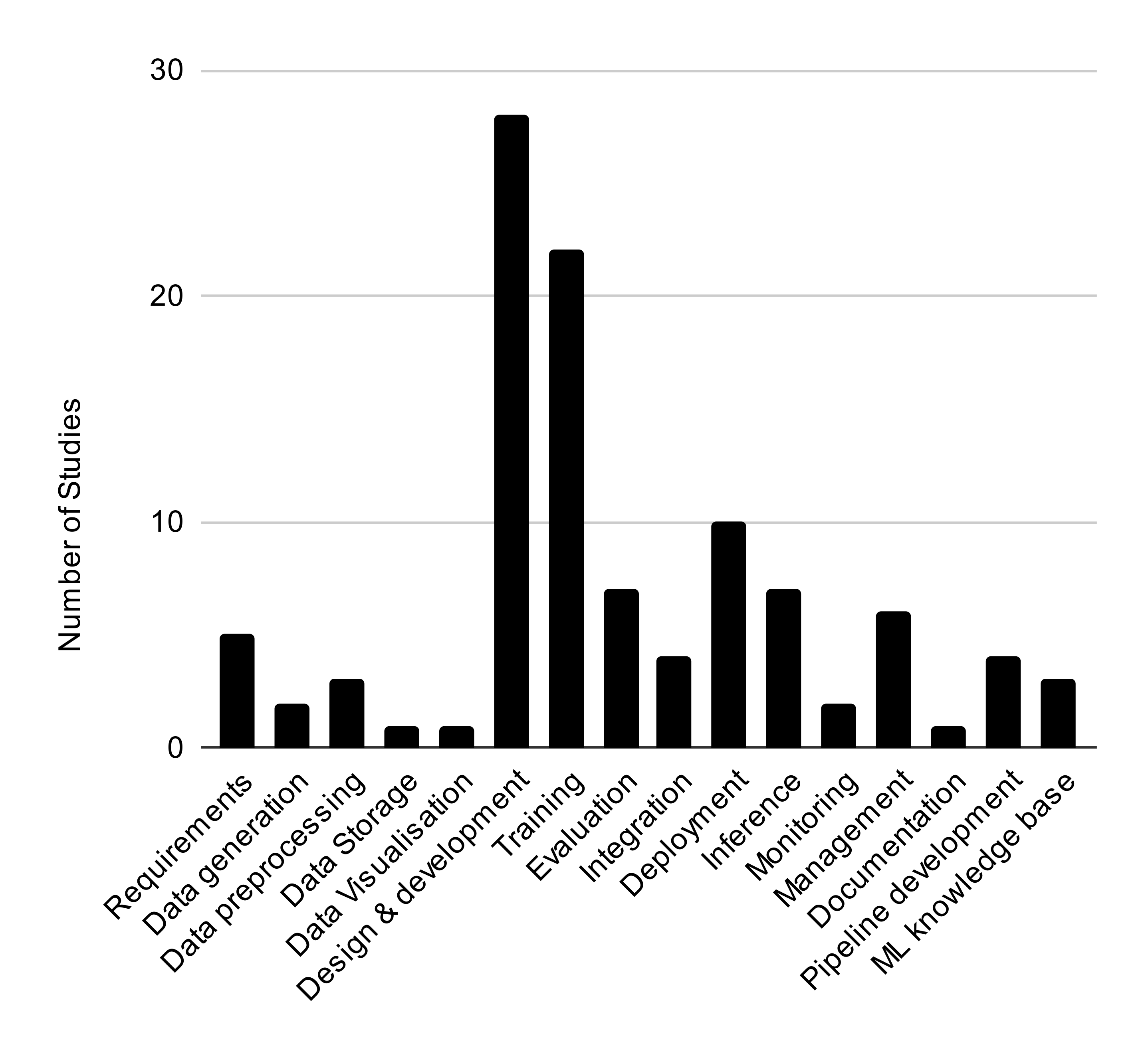}
    \caption{Machine Learning Aspects Supported in Studies}
    \label{fig:MLaspects}
    \end{subfigure}
\hfill
    \begin{subfigure}{0.49\textwidth}
    \centering
    \includegraphics[width=\textwidth]{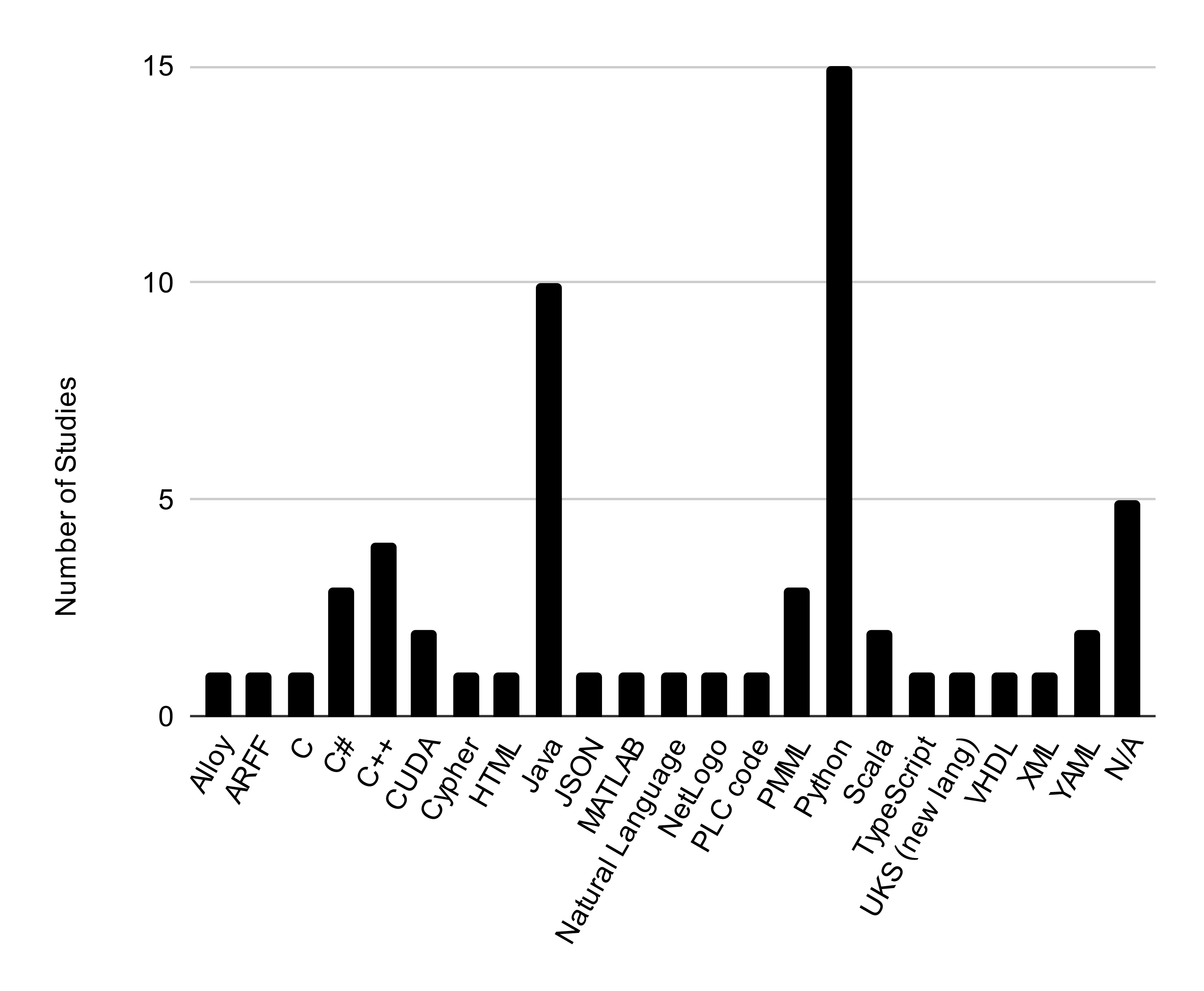}
    \caption{Generated Code and Text Languages in Studies}    
    \label{fig:generatedlanguages}
\end{subfigure}
\caption{ML aspects and Generated Text Languages}
\end{figure}

\subsubsection{Supported Machine Learning Aspects}
From the included primary studies we found 17 different ML aspects that were addressed by studies, 30 out of 46 studies focused on two or more ML aspects, whereas the remaining 16 studies only focused on one ML aspect. Figure \ref{fig:MLaspects} shows the distribution of ML aspects in studies, these include the development stages of ML components and other related aspects. The development stages consist of \textit{requirements engineering}, to gather, analyse and specify requirements for ML-based systems (e.g., P3, P35, P39); \textit{data preprocessing}, cleaning and transforming data into a suitable format for input to the ML model (e.g., P13, P14); \textit{design and development of ML models/components}, algorithm selection, feature engineering and coding for the ML component (e.g., P7, P24, P36);  \textit{training}, feeding data to a ML model for it to learn patterns and relationships (e.g., P22, P25, P42); \textit{evaluation}, evaluating the ML model's performance on unseen (test) data before deployment (e.g., P1, P27, P39); \textit{deployment}, releasing the ML component into the production environment (e.g., P9, P35, P37); \textit{integration}, integrating ML components with other software or system components (e.g., P20, P23, P31); \textit{inference}, using a trained ML model to make predictions on unseen data after deployment (e.g., P3, P6, P21); \textit{monitoring}, observing the ML component at runtime for correct behaviour (e.g., P3, P6) and \textit{management}, managing ML components after deployment through retraining, parameter adjustment, version control and maintenance (e.g., P12, P26, P31). Other related ML aspects consist of  \textit{data generation}, artificially creating data for training ML models (e.g., P39, P45); \textit{data storage}, storing and organizing datasets (e.g. P13); \textit{data visualization}, creating visual representations of data to uncover patterns, relationships, and insights (e.g., P26); \textit{documentation}, recording various aspects of the ML component such as dataset details, architecture, training setup and more (e.g., P10); \textit{ML pipeline development}, building the process for developing a ML component from data pre-processing to deployment (e.g., P4, P16, P32); and \textit{ML knowledge base development}, creating a repository of information about ML models and resources (e.g., P15, P16, P34). More than half the studies (28 out of 46 studies) were targeted toward designing and developing ML models or components, while nearly half (22 out of 46 studies) were for training the ML models. A reasonable portion of studies (10 out of 46 studies) addressed deploying ML components with a primary focus on practices such as DevOps and MLOps. The least explored ML aspects were documentation, data storage, visualization with only one study, and monitoring and data generation with two studies.

An ML framework is a comprehensive platform that offers a structured foundation to build, train, and deploy ML models. In other words, they offer various tools and functionalities that cover the entire ML model development lifecycle. An ML library is a collection of functions and methods that allow developers to perform specific ML tasks, e.g., splitting datasets into training and test sets. Libraries are generally more lightweight than frameworks and focus on particular ML pipeline aspects. We examined the ML frameworks and libraries used in the studies, our results summarised in Table \ref{table:metaTools}. Based on our review of the 46 studies, Tensorflow was the most used ML framework, followed by MXNet, whereas Weka was the most used ML library, followed by Scikit-learn and Numpy.

\begin{table}[htb]
\centering
\caption{Machine Learning Frameworks and Libraries}
\footnotesize
\begin{tabular}{ p{4cm} p{3.5cm} | p{4cm} p{3.2cm} } 
\hline
\textbf{\mbox{Machine Learning Frameworks}} & \textbf{Studies} & \textbf{Machine Learning Libraries} & \textbf{Studies}\TBstrut \\
\hline
AI-toolbox & P40  &                             Encog	& P20, P28  \TBstrut \\ 
Caffee & P8, P11, P12  &                        Keras	& P1, P22  \TBstrut \\                                                       
\mbox{Deep Learning for Java (DL4J)} &	P4 &    NetLogo for Reinforcement Learning	& P30   \TBstrut \\
Infer.NET &	P17, P21 &                          NumPy	& P35, P37, P44 \TBstrut \\
ZenML	& P37    &                              Neuroph &	P7  \TBstrut \\ 
MXNET	& P5, P8, P11, P12 &                    OpenAI Gym &	P18  \TBstrut \\ 
PyTorch	& P6, P25, P37    &                     Pandas	& P17, P35      \TBstrut \\ 
Tensorflow	 & P5, P6, P8, P11, P12, P13, P22,P24, P31 &   Scikit-learn & P13, P22, P35 \TBstrut \\ 
Tensorflow Lite	& P9 &                         Weka	& P16, P34, P36, P45 \TBstrut \\

\hline
\end{tabular}
\label{table:MLframeworks}
\end{table}

\subsubsection{Tool Support}
\sectopic{Model Transformations.} MDE involves transforming models into text or different kinds of models. We refer to these as \textit{model-to-text (M2T)} and \textit{model-to-model (M2M)} transformations respectively. M2T transformations include model transformations into code, documentation, or any kind of textual artifact whereas M2M transformations comprise model transformations into different kind of models. ~\cite{brambilla2017model}. We found the largest portion of studies (35 out of 46 studies) using solely M2T transformations (P2, P5-P8, P10-P17, P19-P24, P26-P28, P30-P32, P34-P39, P43-P46). For example, P7 generates a Predictive Model Markup Language (PMML) file from a model with ML model details. A small portion of studies (4 out of 46 studies) use only M2M transformations (P18, P29, P33, and P40). For instance, P18 converts goal models into formal specification models to generate domain simulations for reinforcement learning. We also found some studies (7 out of 46 studies) that apply both M2M and M2T transformations in their MDE solutions (P1, P3, P4, P9, P25, P41, and P42). For example, P25 applies M2T transformation to convert a model with ML prediction and dataset details into deep learning code, and M2M transformation to convert models with wireless network plan details into models with component allocation and resource usage. Another example is P35 in which models with big data analytics tasks are converted into ML code and documentation. We also found that all 46 included studies use only forward engineering.\\

\sectopic{Generated Artifacts.} Table \ref{table:genArtifacts} summarises the artifacts generated by the studies. These artifacts include \textit{ML model code or training code} in 36 studies, software or intermediate \textit{models} in 15 studies, \textit{deployment configurations} in 8 studies, and \textit{datasets} or subsets of datasets in 4 studies. The remaining were \textit{text files}, \textit{API code}, \textit{recommendation rules or queries}, and \textit{meta-models} in 2, 2, 2, and 1 studies, respectively. We further examined the languages in which text and code were generated. Our findings are shown in figure \ref{fig:generatedlanguages}, the highest number of studies generated artifacts in \textit{Python} (15 out of 46 studies), \textit{Java} (10 out of 46 studies) and \textit{C++} (4 out of 46 studies). Study P11 generates ML components in Python and C++ using artifact and reference models. \\

\sectopic{Automation Levels.} Figure \ref{fig:AutomationLevels} shows the automation categories supported by the transformations, \textit{fully automated} and \textit{partially automated}. The former category works independently while the latter requires some manual effort. Transformations in the majority of studies (38 out of 46 studies) were fully automated (P1, P2, P4-P10, P12-P18, P20, P22-P25, P27-P37, P40-P44, and P46), whereas, in a few studies (8 out of 46 studies) they were partially automated (P3, P11, P19, P21, P26, P38, P39, P45).\\

\begin{table}[htbp]
\centering
\caption{Generated Artifacts}
\label{table:genArtifacts}
\footnotesize
\begin{tabular}{ p{5cm} p{10.5cm}  } 
\hline
\textbf{Generated Artifacts} & \textbf{Studies} \TBstrut \\
\hline
ML Model/Training code	  \TBstrut  & P1, P2, P3, P5, P7, P8, P9, P11, P12, P13, P14, P16, P17, P19, P20, P21, P22, P23, P24, P25, P27, P28, P30, P31, P32, P35, P36, P37, P38, P39, P41, P42, P43, P44, P45, P46\TBstrut \\ 
Model	                     & P1, P4, P6, P7, P9, P14, P18, P20, P25, P28, P29, P33, P40, P41, P42 \TBstrut \\
Deployment configurations    &	P3, P6, P13, P16, P26, P31, P34, P37\TBstrut \\ 
Dataset	                     & P31, P36, P39, P45\TBstrut \\ 
Text files	                 & P10, P35 \TBstrut \\ 
Recommendation rules/Queries & P15, P38 \TBstrut \\ 
API code	                 &  P4, P8\TBstrut \\ 
Meta-model	                 & P26\TBstrut \\ 
\hline
\end{tabular}

\end{table}

\sectopic{Tool Availability.} We searched primary studies for details of developed tool development details, with our findings shown in Figure \ref{fig:ToolType}. An open-source tool was provided in 17 studies (P1, P2, P4, P8, P10, P13, P14, P16, P17, P22, P23, P32, P34, P35, P39, P42, and P44), a proprietary tool was mentioned in 6 studies (P7, P26, P27, P31, P33, and P45) and no tool was mentioned in 23 studies (P3, P5, P6, P9, P11, P12, P15, P18-P21, P24, P25, P28-P30, P36-P38, P40, P41, P43, and P46).\\

\sectopic{Meta Tools and Frameworks.} 
Tools and frameworks that facilitate the development of modeling languages, modeling tools and frameworks are known as \textit{meta} tools and frameworks, for example, the Eclipse Modeling Framework (EMF). Table \ref{table:metaTools} provides an overview of the modeling frameworks, meta tools, and model transformation languages found in the selected studies. Among modeling frameworks, the most frequent was the \textit{Eclipse modeling framework (EMF)} in 15 out of 46 studies. The second most frequent was the \textit{MontiAnna/MontiArc} framework in 4 out of 46 studies. Considering meta tools, \textit{Sirius} was the most common tool found in 10 out of 46 studies. The next most common tools were \textit{MontiAnna/MontiArc} and \textit{Eclipse IDE} both used in 4 studies. For model transformation languages, \textit{XTend} and \textit{Epsilon Generation Language (EGL)} were most common, found in 5 and 4 studies, respectively. The studies absent from the table did not mention the framework, meta-tool, or model transformation language.

\begin{table}[htbp]
\centering
\caption{Modeling Frameworks, Meta tools and Model Transformation Languages}
\footnotesize
        \begin{tabular}{p{2.6cm} p{2.8cm} | p{2.4cm} p{2.1cm} | p{2.5cm} p{1.5cm} }
        \hline
       \TBstrut  \textbf{Modeling Framework} \TBstrut  & \textbf{Studies} & \textbf{Meta Tool} & \textbf{Studies} & \textbf{Model Transformation Language } & \textbf{Studies}   \TBstrut \\
       
            \hline
            \TBstrut Eclipse modeling framework (EMF) \TBstrut	& P1, P2, P4, P6, P7, P9, P14, P22, P27, P28, P30, P32, P33, P34, P37 & Sirius & P1, P7, P9, P14, P25, P27, P30, P34, P37, P38 & XTend & P9, P14, P22, P28, P39\TBstrut \\

             xText \TBstrut	& P9, P22, P27, P28, P39 & Eclipse IDE  & P22, P28, P29, P39  &Epsilon Generation Language (EGL)  & P2, P6, P7, P17   \TBstrut \\
       
           \TBstrut MontiAnna/MontiArc framework \TBstrut	& P5, P8, P11, P12 & MontiArc/MontiAnna & \mbox{P5, P8, P11, P12} & MontiAnna/MontiArc generators  & \mbox{P5, P8, P11,} P12 \TBstrut \\

            \TBstrut PyEcore 	& P25, P38 & Papyrus & P17, P33 &  Acceleo & P1,P27,P34 \\

            Generic Modeling Environment (GME) 	& P13, P36 \TBstrut & Flexmi  & P2, P6 & \mbox{Atlas Transformation} language (ATL) & P25, P29, P30\TBstrut\\

            \TBstrut Meta object facility (MOF) framework  \TBstrut	& P17 &  TouchCore  & P4 & TouchCore & P4 \TBstrut \\

             \TBstrut GreyCat (extension of KMF)	& P23 & IntelliJ IDE  & P23 & Apache Velocity & P23 \TBstrut \\
                 
              Langium  \TBstrut	& P10 & \mbox{Langium Workbench} & P10 & Langium  & P10 \TBstrut \\            SyncMeta \TBstrut	& P24 &  SyncMeta  & P24 & ANTLR & P28 \TBstrut \\
                  
            JastAdd \TBstrut	& P20 & Pyro  & P32 & JastAdd  & P20 \TBstrut \\

            \TBstrut i* framework \TBstrut	& P18 & MetaEdit+ & P35 & Xpand language & P30 \TBstrut \\ 

            CINCO framework \TBstrut	& P32 & CINCO Workbench   & P32 & ENLIL & P3 \TBstrut \\

            OPC UA framework \TBstrut	& P45 &  OPC UA modeler & P45 & OPC UA code \mbox{generator} & P45 \TBstrut \\
                   
      KM3 framework \TBstrut	& P29 & DL LDM tool  & P26 &  &  \TBstrut \\
                     
                             \TBstrut	&  & WebGME  & P13 &  &  \TBstrut \\
            \hline
        \end{tabular}
    \label{table:metaTools}
\end{table}

\begin{center}
\begin{myframe}[width=45em,top=5pt,bottom=5pt,left=5pt,right=5pt,arc=10pt,auto outer arc,title=\centering\textbf{RQ2 Answer Summary}]
\footnotesize

A variety of ML aspects have been covered in studies, frequent ones being the design, development, and training of ML components with TensorFlow as the most used ML framework. Over 93\% of studies propose models at the PIM level with most being design models. The majority of studies provide new DSLs with model representation almost equally divided between graphical and textual. In transformations, 91\% of studies apply M2T transformations in their solutions, and 85\% of transformations are fully automated. However, only tools from 50\% of the studies are available. Upon examining the generated artifacts, we found that 78\% of studies generated ML model code or training code, and Python was the preferred choice of programming language. Further, we found EMF for modeling framework, Sirius for meta-tool, and XTend for model transformation language were chosen by most studies.
       
    \end{myframe}
\end{center}

\subsection{RQ3 - MDE4ML Studies Evaluation}
\subsubsection{Target Area} 
We looked at the domain examples and evaluation context in each primary study - this can be \textit{academia}, \textit{industry}, or both. We classify studies as academia if examples and evaluations occur in controlled environments like labs and as industry when conducted in real-world settings. Figure \ref{fig:TargetAreas} shows the distribution of target areas of studies. A large portion of studies (89\%) were from an academic context (P1-P11, P13, P14, P16-P30, P32, P34, and P36-P46) and a small portion (9\%) were from an industrial context (P12, P15, P31, P33). Interestingly, one study, P35 with domain-specific visual languages for Big Data Analytics and ML had evaluations in academic and industrial contexts. 

\subsubsection{Evaluation Methods} 
\begin{figure}[htbp]
    \centering
    \begin{subfigure}{0.3\textwidth}
    \includegraphics[width=\textwidth]{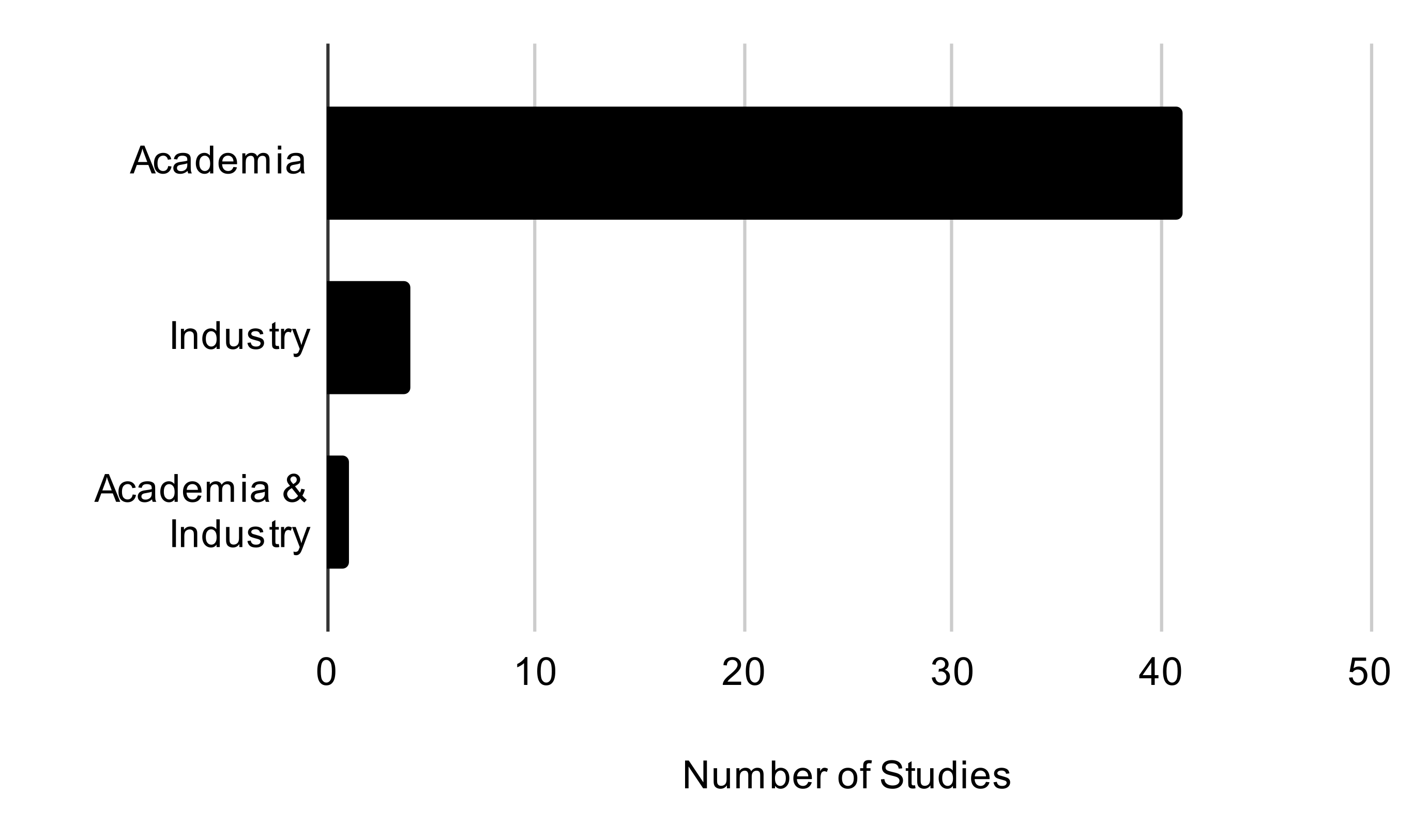}
    \caption{Target Area of Studies}
    \label{fig:TargetAreas}
    \vspace*{-1em}
    \end{subfigure}
\hfill
    \begin{subfigure}{0.3\textwidth}
    \centering
    \includegraphics[width=\textwidth]{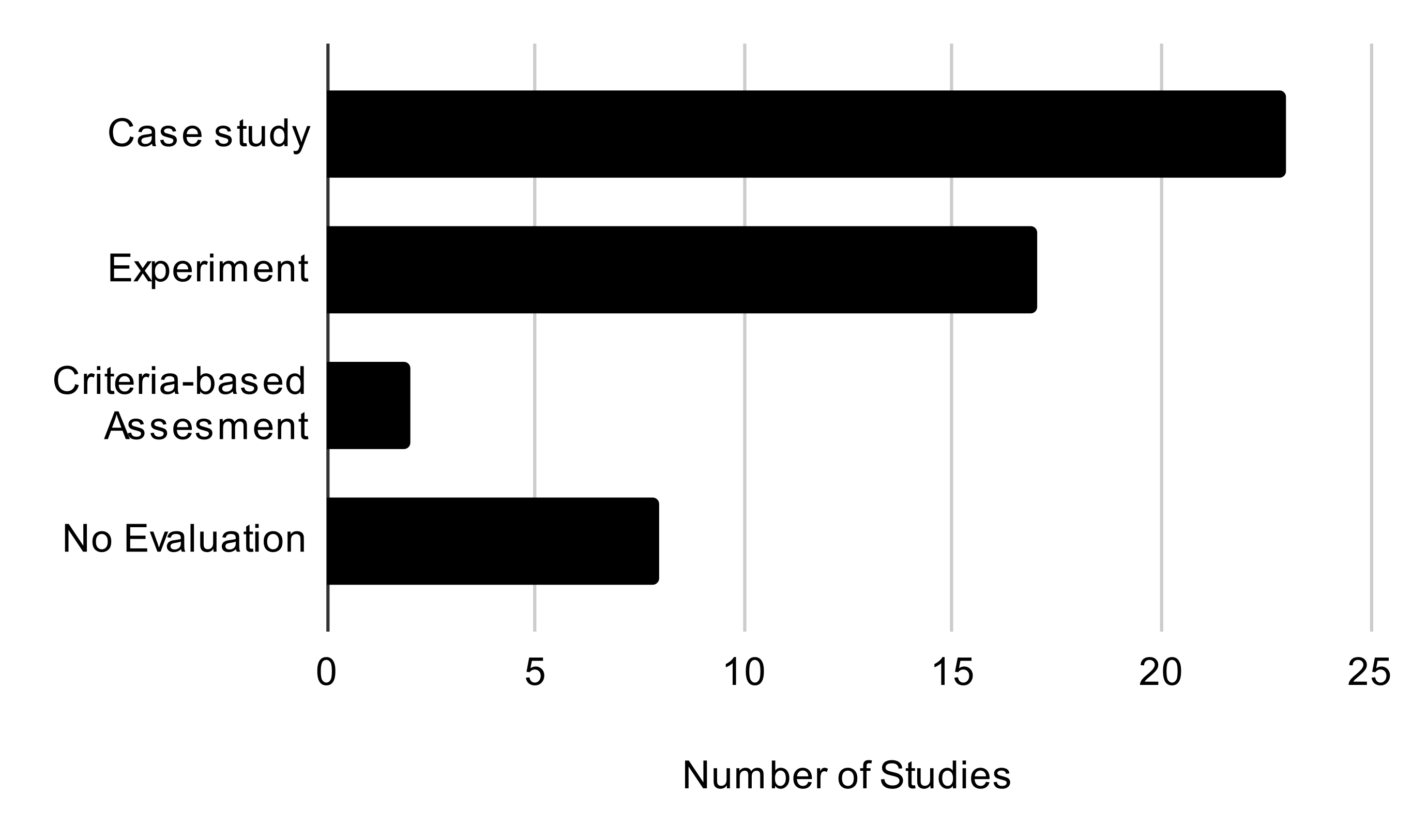}
    \caption{Evaluation Methods in Studies}
    \label{fig:evalMethods}
    \vspace*{-1em}
\end{subfigure}
\hfill
    \begin{subfigure}{0.3\textwidth}
    \centering
    \includegraphics[width=\textwidth]{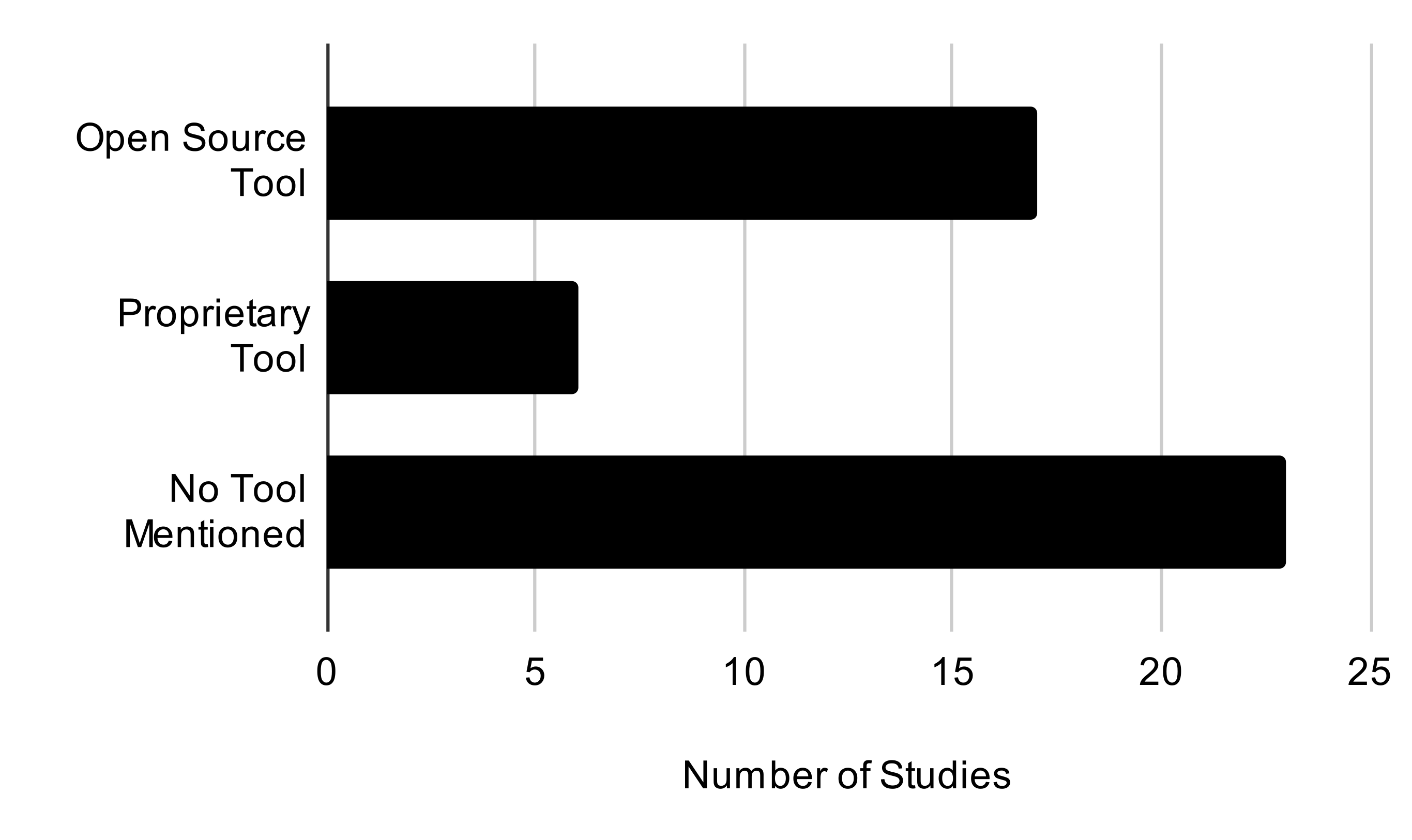}
    \caption{Types of Tools in Studies}
    \label{fig:ToolType}
    \vspace*{-1em}
\end{subfigure}
    \vspace*{1em}
 \caption{Evaluation and Tools in Studies}
\end{figure}

We analyzed the evaluation methods in studies and categorized them under the following five categories: \textit{case study}, \textit{experiment}, \textit{survey}, \textit{criteria-based assessment}; \textit{no evaluation}. 
We borrow our categorization of empirical studies from Wohlin et al.~\cite{wohlin2012experimentation}. We note that for case studies, we classify them otherwise if the primary study terms their evaluation method as a case study, albeit it does not match the SE definition~\cite{wohlin2012experimentation}. We also added a new category here ``criteria-based assessment'' where the primary studies are compared against certain well-established guidelines/criteria in the area.
Figure \ref{fig:evalMethods} shows the majority of MDE solutions (23 out of 46 studies) were evaluated on case studies (P1-P3, P9, P11-P15, P17, P19, P20, P22, P23, P27, P29, P30, P33, P35, P36, P39, P40, and P45). Out of these, a few (4 of the 23) were industrial case studies (P12, P15, P33, and P35). The second most common evaluation method in studies (17 out of a total 46 studies) was experiments (P4, P7, P10, P16, P22, P24, P25, P28, P31, P34, P35, P37, P38, P41, P42, P44, and P46). Among these, only one was an industrial experiment study (P31), and four were `user studies' (P10, P22, P24, and P35). We note that the user studies also included a post-experiment interview or questionnaire (survey) with the participants; however, the experiment was the main method of evaluation.
The criteria-based assessment was done in two studies (P28 and P35). For example, P35 performed an evaluation against the `physics of notation' guidelines in their study. 
Three primary studies (P22, P28, and P35) used multiple methods to evaluate their proposed MDE solution. For example, P22 evaluated their work with an IoT case study and a user study with four volunteers. No evaluation was found in eight studies (P5, P6, P8, P18, P21, P26, P32 and P43).

\subsubsection{Evaluation Metrics and Datasets} 

\begin{figure}[htbp]
    \centering
    \begin{subfigure}{0.49\textwidth}
    \includegraphics[width=\textwidth]{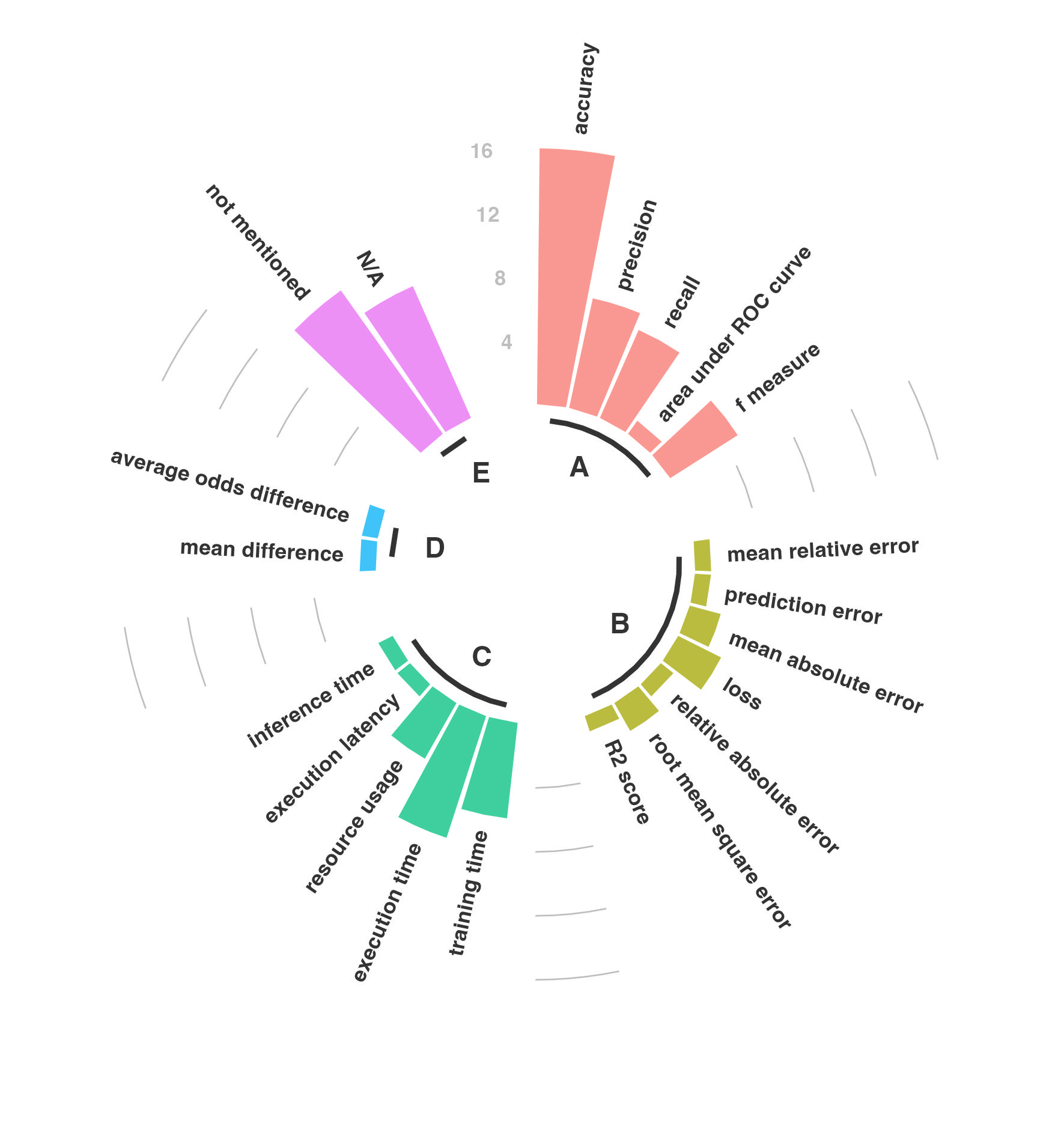}
    \caption{ML Evaluation Metrics in Studies}
          \begin{center}{\footnotesize A) Classification B) Regression C) Time and Resource D) Fairness E) No metrics} \end{center}
    \label{fig:MLmetrics}
    \vspace*{-0.3em}

    \end{subfigure}
\hfill
    \begin{subfigure}{0.49\textwidth}
    \centering
    \includegraphics[width=\textwidth]{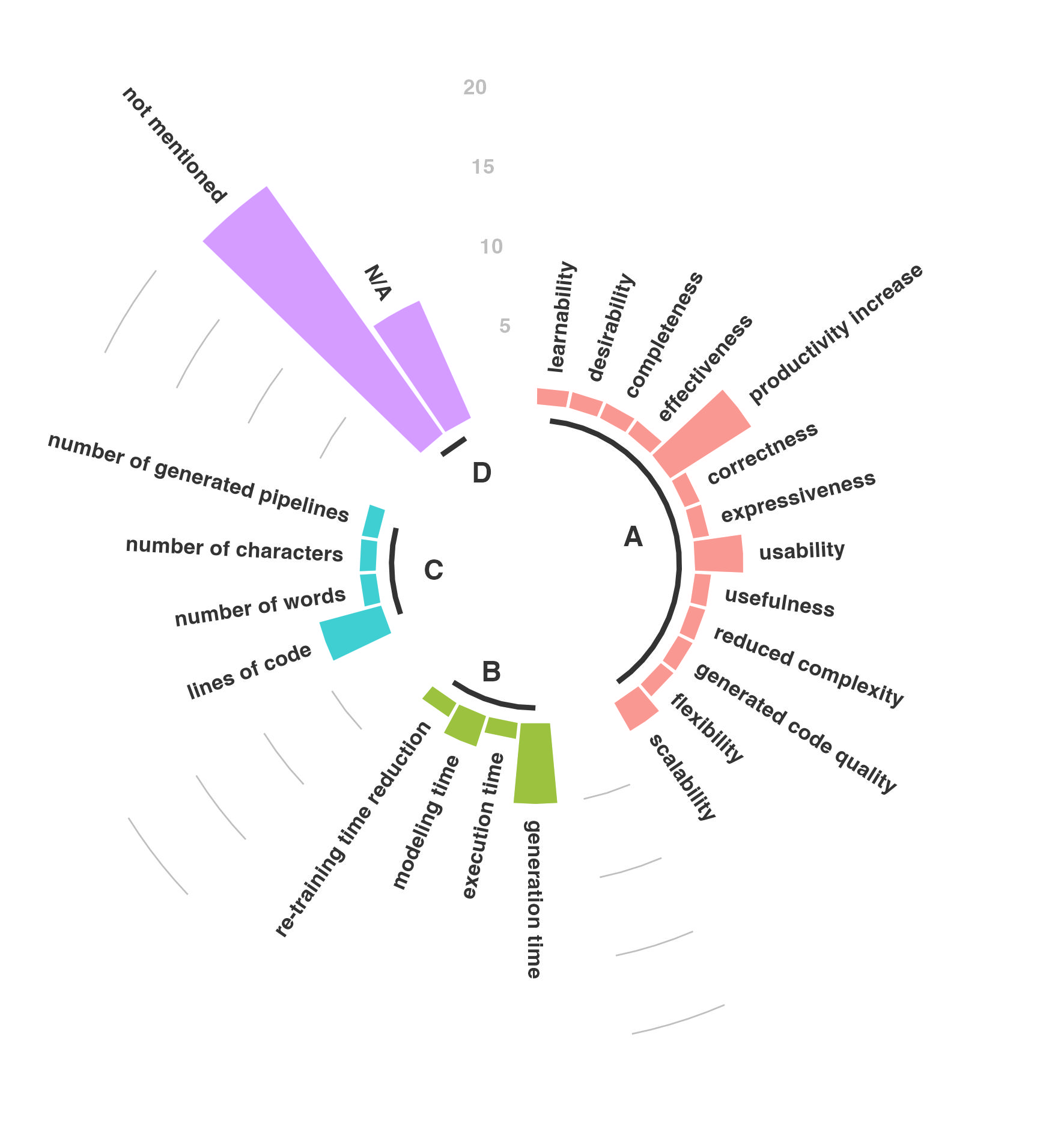}
    \caption{MDE Evaluation Metrics}
     \begin{center}{ \footnotesize A) Quality B) Time and Resource C) Code metrics D) No metrics}\end{center}
    \label{fig:MDEmetrics}
\end{subfigure}
\caption{Evaluation Metrics}
\end{figure}

Evaluation metrics in studies are divided into metrics for ML and metrics for MDE, as shown in Figure \ref{fig:MLmetrics} and Figure \ref{fig:MDEmetrics}. The category \textit{not mentioned} includes studies that perform evaluation but do not mention the metrics. We found  10 studies (P14, P15, P20, P24, P29, P30, P33, P35, P36, and P40) that do not mention any ML metrics and 18 studies (P1, P3, P7, P9, P11, P13-P15, P20, P22, P23, P27, P29, P34, P36, P39, P42, and P46) that do not mention any MDE metrics. The \textit{not applicable (N/A)} category contains studies that either have no evaluation or the solution cannot be evaluated through such metrics. For instance, P10 provides a DSL to model dataset descriptions and then transform the models into HTML documents. Since this study has no relevant ML evaluation, we categorized it as N/A in Figure \ref{fig:MLmetrics}. There are ten studies (P5, P6, P8, P10, P18, P21, P26, P32, and P43) for which ML metrics are not applicable and eight studies (P5, P6, P8, P18, P21, P26, P32, and P43) for which MDE metrics are not applicable.

We examined the primary studies and found ML evaluation metrics related to \textit{classification}, \textit{regression}, \textit{time and resource}, and \textit{fairness}. Among all studies, classification metrics were used the most, with the frequently occurring ones being \textit{accuracy} of the ML component in 16 studies' evaluation (P1, P2, P7, P9, P11, P13, P16, P17, P22, P23, P27, P34, P37-P39, and P44) and \textit{precision} of the ML component in 7 studies' evaluation (P1, P4, P9, P13, P22, P27, and P44). Other classification evaluation metrics include \textit{recall} in six studies (P1, P3, P4, P9, P22, and P27), \textit{f measure} in five studies (P13, P22, P27, P34, and P38) and \textit{area under ROC curve} in one study (P16). The second highest evaluation metrics were time and resource metrics, with \textit{execution time} in eight studies (P13, P19, P23, P25, P41, P42, P44, and P46) and \textit{training time} in six studies (P9, P12, P22, P37, P38, and P44). Other time and resource evaluation metrics such as \textit{resource usage} were found in four studies (P19, P25, P42, and P44), \textit{execution latency} in one study (P13) and \textit{inference time} in one study (P44). In comparison, regression and fairness metrics were found in much fewer studies' evaluations. Among evaluation metrics for regression tasks, \textit{loss} was found in three studies (P1, P13, and P39), \textit{root mean square error} in two studies (P31 and P45), \textit{mean absolute error} in two studies (P12 and P31), \textit{mean relative error} in one study (P37), \textit{prediction error} in one study (P31),  \textit{relative absolute error} in one study (P45), and \textit{R\textsuperscript{2} score} in one study (P13). Among evaluation metrics for fairness tasks, \textit{mean difference} and \textit{average odds difference} were both found in one study (P2).

MDE metrics found in the primary studies were related to \textit{quality}, \textit{time and resource}, and \textit{code}. Quality metrics were often used to evaluate MDE approaches, especially \textit{productivity increase} when developing an ML solution found in six studies (P25, P30, P31, P37, P38, and P41), and \textit{usability} of the MDE approach found in three studies (P10, P24, and P35).  Other quality-related evaluation metrics such as \textit{scalability} were found in two studies (P19 and P44), \textit{learnability} in one study (P35), \textit{desirability} in one study (P35), \textit{completeness} in one study (P33), \textit{effectiveness} in one study (P30), \textit{correctness} in one study (P4), \textit{expressiveness} in one study (P10), \textit{usefulness} in one study (P10), \textit{reduced complexity}	in one study (P10), \textit{generated code quality} in one study (P17), and \textit{flexibility} in one study (P19). The second highest was time and resource metrics, with \textit{generation time} and \textit{modeling time} used the most in five studies (P2, P25, P37, P38, and P45) and two studies (P25 and P45), respectively. Other time and resource-related evaluation metrics found in the primary studies were \textit{execution time} in one study (P2) and \textit{re-training time reduction} in one study (P12). Code-related evaluation metrics were found in relatively fewer studies. Among these metrics, we found \textit{lines of code} in four studies (P2, P28, P30, and P40), \textit{number of words} in one study (P28), \textit{number of characters} in one study (P28), and \textit{number of generated pipelines} in one study (P16).

During the analysis of the evaluations described in the primary studies, 33 datasets were identified. The datasets most frequently used in the evaluations were the \textit{MNIST} handwritten digits dataset (7 out of 46 studies) and the \textit{Iris} flowers dataset (3 out of 46 studies) -- we note that both these datasets are widely used for ML classification.

\begin{center}
\begin{myframe}[width=45em,top=5pt,bottom=5pt,left=5pt,right=5pt,arc=10pt,auto outer arc,title=\centering\textbf{RQ3 Answer Summary}]
\footnotesize
Out of 46 primary studies, only five evaluate their MDE solution in an industrial context while the remaining studies provide examples or evaluations in academic contexts. The evaluation method identified in 23 out of 46 studies is a case study. The occurrence of other evaluation methods such as experiments and user studies was relatively low. Analysis of ML and MDE evaluation metrics shows that MDE metrics were mentioned in a few studies or MDE aspects were evaluated in a few studies. The dataset most often used for evaluations in the primary studies was MNIST. The results from the studies show that MDE approaches for ML are seldom evaluated in industrial contexts and during evaluation more emphasis is placed on ML aspects.
    \end{myframe}
\end{center}

\subsection{RQ4 - Limitations and future work of existing MDE4ML studies}

\subsubsection{Limitations in the Primary Studies}
In our analysis of the selected primary studies we found several limitations and we classified these into three high-level categories: approach, evaluation, and solution quality. Among the selected 46 studies, the following 19 studies, P3, P5, P7, P8, P11, P16, P18, P26, P27, P29, P31, P33, P37, P40-P44, and P46, did not mention any limitations.\\

\sectopic{Limitations in Approach.} A key limitation found in the approach taken by several of the selected primary studies is the manual effort required to configure the generated artifacts (studies P19, P20, P22, P24, P36, and P39). Whereas considerable manual effort goes into modeling for P15 and code generator implementation for P21. Limited ML models are supported in studies P7, P22, P23, P30, P33, P34, and P37. This may restrict the applicability of these approaches to a broader range of ML techniques. P4 has the limitation that a single error in the model can defeat the purpose of the entire solution. The approach in P21 is not generic and is only relevant to one ML framework (Infer.NET).  

\sectopic{Limitations in Evaluation.} We discovered evaluation-related limitations in several primary studies. However, only a few of them were actually discussed in the studies. P17, P28, P30 mention the absence of a user study to evaluate the approach with real-world users. This raises concerns about the practical usability of the approach, as user feedback is never obtained. Another limitation found in two studies (P39 and P40) was the absence of an industrial evaluation. This signifies an important gap regarding the application and usefulness of MDE solutions for ML-based systems in the industry. Interestingly, among the studies that do provide an evaluation, ten (10) studies (P9, P12, P14, P19, P20, P23, P24, P38, P42 and P45) describe their evaluations as being limited to simple scenarios or single case studies. Furthermore, P18 highlights the lack of an evaluation of the proposed solution.

\sectopic{Limitations in Quality.} The quality of a proposed solution is of high importance; while analyzing the studies we found quality limitations related to scalability and accessibility. P14 and P39 propose approaches that are difficult to scale, making them less suitable for large-scale applications. The study P35 describes accessibility issues during evaluation due to the MetaEdit+ meta-tool. These issues include the lack of a web interface and the need for a tool license. 

\subsubsection{Future Work Suggested in the Primary Studies.} 

While analyzing our selected primary studies, we were interested in examining the key future work and challenges described. We have classified these future works into three high-level categories related to enhancements in the approach, solution quality, and evaluation. P5, P11, P13, P26, P29, P41, and P44 do not mention any future work.

\sectopic{Improvement or Extension of Approach.} Many kinds of improvements to the approach or extensions to add new features have been grouped under this category. The addition of new features and improvements like a recommender and new modeling concepts were proposed in many studies, such as P6, P9, P10, P12, P14-P16, P18-P20, P22, P24, P25, P28, P34-P36, P39, and P42. For example, the authors of P9 intend to add support for future collaborative training of ML models. Studies P1, P21, P27, P32, and P35 plan to support additional platforms like embedded systems and websites to make their solutions compatible with a wider range of platforms.  Similarly, in P7, P8, P22, and P36 it is suggested that the code generators will support more programming languages. Currently, limited ML models are supported in P7, P22, P23, P30, P33, P34, P37, and P45; future work involves supporting a wider range of ML models to enhance system capabilities. In P3, P6, P15, and P36, the future goal involves considering more complex scenarios in the approach. Some studies like P9, P13, P14, P17, P22, and P34 aim to add training data processing and preparation as a part of their solutions. A future goal described in P16 is the development of a DSL, whereas the authors of P14 and P27 suggest the creation of a textual DSL to support the existing graphical one. P15 states tool implementation of their approach as a future work, although we note that 50\% of studies do not specify if they have developed a tool for their approach.

\sectopic{Further Evaluation.} As discussed in the primary study limitations section, the evaluations performed in many of the selected primary studies have limitations. To address this gap, thirteen studies P9, P12, P14, P17, P19, P20, P23, P24, P38-P40, P42, and P45, intend to perform additional evaluations. For example, in P23 multiple use cases for ML models were planned to be evaluated on high-power computers and P40 considered industrial evaluation an important next step. Surprisingly, from the eight studies that do not provide any sort of evaluation, only P18 states evaluation as a future task. Similarly, out of the 42 studies that do not conduct a user study, only three studies, P17, P28, and P30, report it as work to be done in the future. 

\sectopic{Quality Enhancement.} A small portion of the primary studies mention quality enhancements as a future target. The two more frequent quality improvements include integration with other languages and tools (P1, P8, P24, P35, P43) and interoperability support (P24, P32, P45). For instance, in P1 a future goal is to integrate with EAST-ADL an architectural language for automotive embedded systems. Other enhancements mentioned are optimization of generated code, in P2 and P33, optimal resource allocation in P37, support for model checking in P32, and improved scalability, reusability, and adaptability in P40, P20, and P25 respectively. \\

\begin{center}
\begin{myframe}[width=45em,top=5pt,bottom=5pt,left=5pt,right=5pt,arc=10pt,auto outer arc,title=\centering\textbf{RQ4 Answer Summary}]
\footnotesize
The key limitations in the selected primary studies are related to the evaluation with over 88\% of studies having no industrial evaluation and user study. Furthermore, 48\% of studies evaluate only one aspect of their approach (either MDE or ML), and 17\% of studies do not provide any kind of evaluation. In terms of solution quality, some limitations mentioned in the selected primary studies are relevant to scalability and accessibility. For future work, the majority of papers (46\%) state additional features and enhancements in the proposed approach along with further evaluations (28\%).
    \end{myframe}
\end{center}
\clearpage

\section{Threats to Validity}~\label{sec:Threats}

\subsection{Internal Validity}
To mitigate threats to internal validity, an SLR protocol was developed by the first author and reviewed by the other authors before conducting the study. The search string was modified and executed several times on multiple scientific databases to optimize the results. Since the Science Direct database does not allow searching with long strings, we created multiple smaller combinations of our search string and executed those. The studies were filtered in various rounds by the first author and validated by the other authors. The first round of filtering was based on the title and abstract. The second round was based on a brief reading of the paper, and the third round on a detailed reading. These measures ensure minimal selection bias in our study. After selecting the final pool of studies, a data extraction form was created, and all the authors participated in pilot tests for extracting data from these papers. 

\subsection{Construct Validity}
 We attempted to reduce the threat to construct validity by searching seven relevant scientific databases and employing two search strategies (automated and manual). The selected primary studies were highly relevant to MDE for ML and our RQs. After several rounds of discussions, we refined our inclusion and exclusion criteria to ensure that our criteria support selecting the most suitable studies for this SLR. Some of the chosen studies use inconsistent terminology for ML, which is a potential threat to our study. However, all ambiguities were discussed with the second and the third authors to reach a consensus.
 
\subsection{Conclusion Validity}
We aimed to minimize threats to conclusion validity through a well-planned and validated search and data extraction process. A data extraction form was created with questions based on our RQs, ensuring the selected data was relevant to the study. The first author extracted data using a data extraction form for a small subset of studies. All other authors followed the same method and extracted data for the same subset of studies. We compared the data extracted by the first author and other authors and found a close match between them, after which the first author proceeded with data extraction of the remaining studies. To reduce bias during data analysis and synthesis all authors had several rounds of discussion on how to best categorize and represent data.

\subsection{External Validity}
To mitigate threats to external validity, we employed a systematic search process combining automatic search and manual search (snowballing) from the widely accepted guidelines in \cite{kitchenham2009systematic} and \cite{wohlin2014guidelines}. For both searches, we had clearly defined inclusion and exclusion criteria. To ensure the quality of studies considered in our SLR, we only included peer-reviewed academic studies, excluding grey literature, book chapters, opinion-, vision-, and comparison papers. 
We only included studies in the English language since it is the most widely used language for reporting research studies. While we acknowledge that this may have led to the exclusion of some potentially relevant studies, we deem the impact of this bias on our research is minimal. We did not exclude any study based on publication quality to eliminate publication bias in our study. Additionally, our search was not restricted to any time frame to capture all developments in the area of MDE for ML.

\section{Discussion and Research Roadmap}~\label{sec:Discussion}
From the analysis of our selected primary studies on MDE for ML, we present several interesting insights and recommendations for future research.
\subsection{MDE solutions for ML}
\subsubsection{Data for ML}
ML is a data-driven technique~\cite{braun2018open}, but most studies focus on other aspects instead. We found a surprisingly small number of studies related to data generation, pre-processing, storage, and visualization. Studies often assume that data has been pre-processed and is ready to use, whereas in reality, cleaning, wrangling, and transforming raw data is a tedious and time-consuming process~\cite{duong2021review}. Hence, the MDE techniques should consider the cases where real data is unavailable or insufficient and make data processing a part of the MDE solution.
We also identified six studies (P3, P19, P29, P31, P41, and P43) that do not consider the training data or training process in their MDE solutions for ML.

We recommend researchers consider data as a first-class citizen when devising MDE solutions for ML-based systems. MDE can be particularly useful for generating data or simulators from models ~\cite{jahic2023semkis}, creating meaningful visualizations ~\cite{barzdins2022metamodel}, modeling data pre-processing workflows, and generating code to prepare data for training ~\cite{bhattacharjee2019stratum, ries2021mde}. 

\subsubsection{Expressiveness of Models}
A key issue we identified in a few studies (P17, P20, and P21) was that ML concepts were not adequately expressed in models. For example, P17 and P21 rely on probabilistic graphical models (PGMs) to represent ML models and software models~\cite{moin2022model}. However, PGMs are not expressive enough~\cite{moin2022model} to sufficiently represent complex functions, software structures, and connections between ML components and traditional software components.   

We suggest researchers interested in using PGMs for MDE use them with software models to comprehensively capture both the statistical and software aspects of ML components and systems. 

\subsubsection{Solution Focus}

\sectopic{Development Aspect.} One of our major findings from this study was the high volume of MDE solutions for the design, development, and training of ML components. This narrow solution focus leads to the issue of many other important development aspects for ML being neglected, such as requirements engineering and integration. Requirements engineering for ML and integration of ML components with traditional software components is particularly challenging~\cite{ahmad2023requirements, atouani2021artifact} due to the inherent under-specification of ML~\cite{d2022underspecification} and the unique differences between ML components and traditional software components~\cite{atouani2021artifact,kusmenko2019modeling}. However, we only found five papers focusing on requirements engineering and four on integration. Furthermore, we found a lack of MDE solutions for ML pipelines~\cite{raedler2023model}, automated deployment, and monitoring of ML components, also known as MLOps. Despite the importance of runtime monitoring for early detection of unwanted behavior~\cite{nigenda2022amazon}, we only found two papers relevant to this. Another key area that has not received enough attention is documentation~\cite{giner2023domain}; documentation of datasets, ML models, training parameters, deployment configurations, and ML pipelines is extremely important for maintenance, reusability, and scalability. 

We recommend that researchers broaden the focus of their studies and leverage MDE capabilities to address the challenges in requirements engineering, integration, deployment, monitoring, and documentation of ML components. MDE can be beneficial in several ways, such as DSLs to support requirements engineering for ML, automated generation of runtime monitors from models, and models to automate and ease integration.

\sectopic{Machine Learning Type.} We identified a large fraction of studies focused on supervised and deep learning, while unsupervised and reinforcement learning were not as widely covered in the literature. Unsupervised learning is a powerful technique to identify hidden patterns in large unlabeled datasets. To this end, we did not find any paper providing an MDE solution focusing solely on unsupervised learning. Reinforcement learning has strengths in learning from experience, sequential decision-making, and handling complex state spaces, which are highly useful in robotics, personalized learning, and gaming. Only four studies proposed MDE solutions for reinforcement learning. 

We recommend researchers should explore developing MDE solutions that cater to unsupervised and reinforcement learning. Given the wealth of unlabeled data in various domains, MDE tools for unsupervised learning could pave the way for more efficient data analysis and knowledge extraction.
Similarly, reinforcement learning, with its expansive applications, presents a significant potential for applying MDE.

\sectopic{Model-driven Engineering Details.} We found a lack of MDE details in studies published in ML venues. These studies were focused on textual DSLs for specifying ML operations and transforming them into high-performance code for various hardware platforms like CPUs and GPUs. For instance, P41 provides a textual DSL for ML, and the model is automatically converted into optimized CUDA code for GPUs. Other similar studies include P40, P42, P43, P44 and P46. Although there are significant mathematical details about ML in these studies, the explanation of MDE aspects, such as the meta-models and transformations, is often overlooked. Additionally, we also found some studies in the MDE domain that lacked MDE details, such as studies P15, P18, P19, P24, P26, P29, P32, and P38.
 
We recommend authors, when specifying MDE approaches, add some level of detail with regard to the MDE steps taken in their solution. This would allow researchers and practitioners from the SE and MDE domains to better understand and apply these solutions.

\subsubsection{Solution Maturity}
Most reported MDE solutions for ML in our selected primary studies are still in their early stages, e.g., based on simple cases, and do not support an end-to-end ML lifecycle. This is not surprising since active research in the area dates only to a few years back (e.g., in our 46 primary studies, the earliest paper is from 2008, but 39/46 papers are from 2018 and later). Upon examination of the literature, we found that tools are available for 23 studies. Out of these, only 17 studies provided open-source tools. Some notable ones are P2, P10, P22, and P23. Moreover, existing solutions often overlook complex scenarios and focus only on simple models, for example, P3, P6, P15, and P26. Our analysis also revealed that manual configurations are required for the artifacts generated by ten studies (P3, P14, P19, P20, P22, P24, P35, P36, P39, and P45). The lack of available tool support, automation, and consideration of complex scenarios hinders the adoption, extension and reproducibility of MDE solutions for ML. 

We suggest researchers and practitioners further develop their solutions to consider the entire ML lifecycle and develop research prototypes that cater to end users. In the interest of lowering costs, fostering collaboration, and creating more opportunities for innovation, we recommend researchers to open-source their solutions, data and tools. Open-source software hosting platforms like GitHub, Zenodo, SourceForge, and Gitlab can be leveraged. Accessible and mature solutions are also more likely to be adopted in the industry. 

\subsubsection{MDE Solutions for Domain Experts}
While there were a substantial number of studies providing MDE solutions for ML engineers and software engineers, we found a lack of solutions for domain experts. With the rise of new trends like \textit{no code} and \textit{low code}~\cite{cabot2020positioning}, the complexity of developing ML-based systems can be significantly reduced for domain experts. No code and low code approaches allow users (mostly domain experts) to develop and deploy applications without writing much code~\cite{cabot2020positioning}. Additionally, ML experts, engineers, and analysts can also benefit from such low-code platforms.

We recommend researchers and practitioners develop low-code platforms for systems with ML components so that non-ML experts can benefit from the capabilities of ML. Low code platforms can significantly reduce development complexity and time to deployment.

\subsubsection{Machine Learning Algorithms and Terminology}
We came across varied ML terminology and granularity levels used in studies to describe ML. For instance, P9 and P32 mention at a very high level that their MDE solution is for ML, but studies like P8 and P30 are more specific and specify the exact nature of ML techniques, e.g., neural networks and reinforcement learning. The inconsistent terminologies made it challenging to analyze studies and draw conclusions. This issue was further exacerbated when studies did not explicitly mention the supported ML algorithms; these include P12, P24, P25, P27, P35, P38, and P40. 

We encourage the MDE community to build a consensus on the terminology used for devising ML-based solutions to facilitate understanding and comparisons with other studies. We further highlight the need for clearly specifying the details of the ML algorithms supported in the study.

\begin{figure}[htbp]
    \centering
    \begin{subfigure}{0.3\textwidth}
    \includegraphics[width=\textwidth]{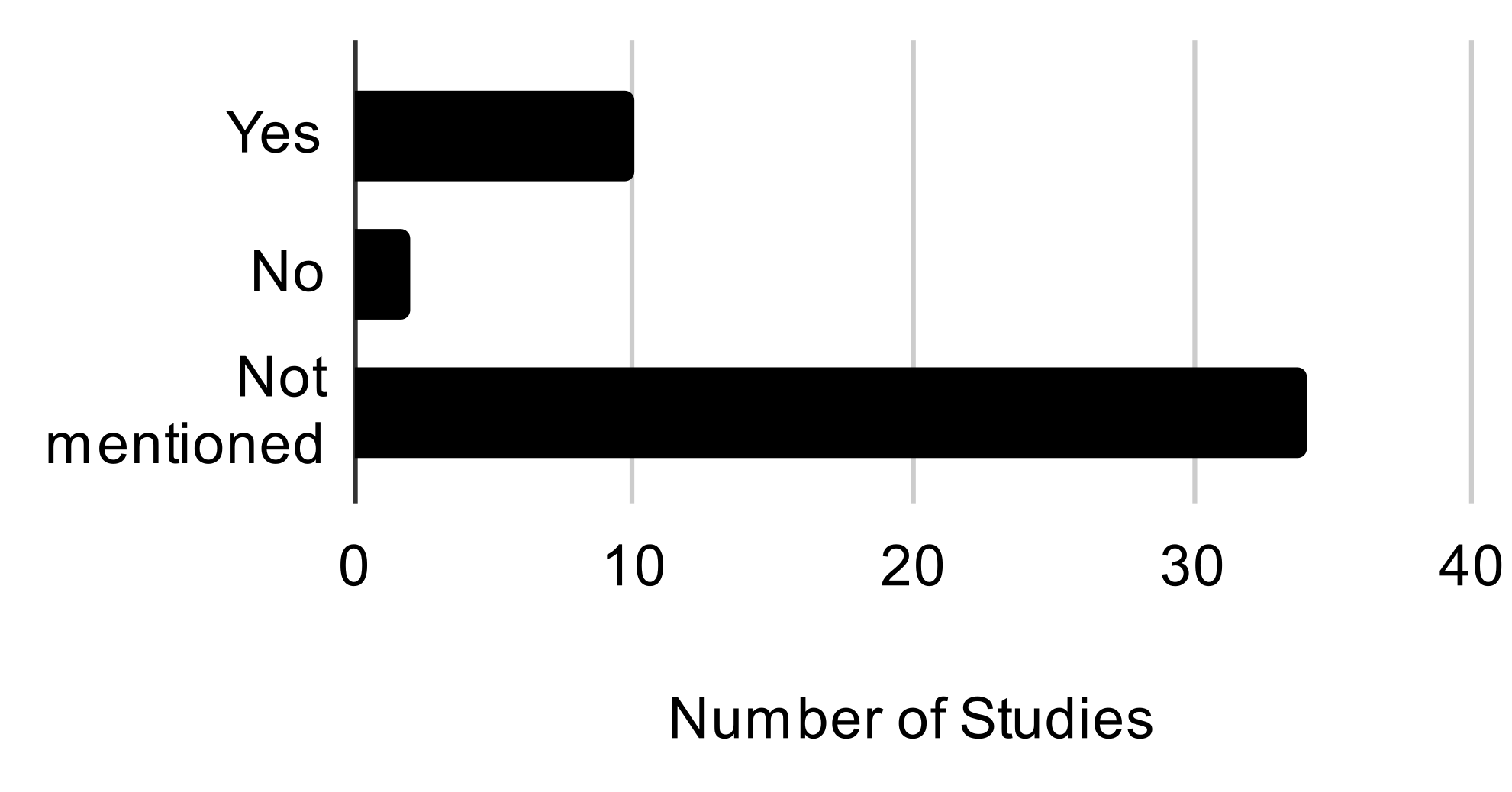}
    \caption{Scalability Support in Studies}
    \label{fig:scalabilityLimits}
    \vspace{-0.8ex}

    \end{subfigure}
\hfill
    \begin{subfigure}{0.3\textwidth}
    \centering
    \includegraphics[width=\textwidth]{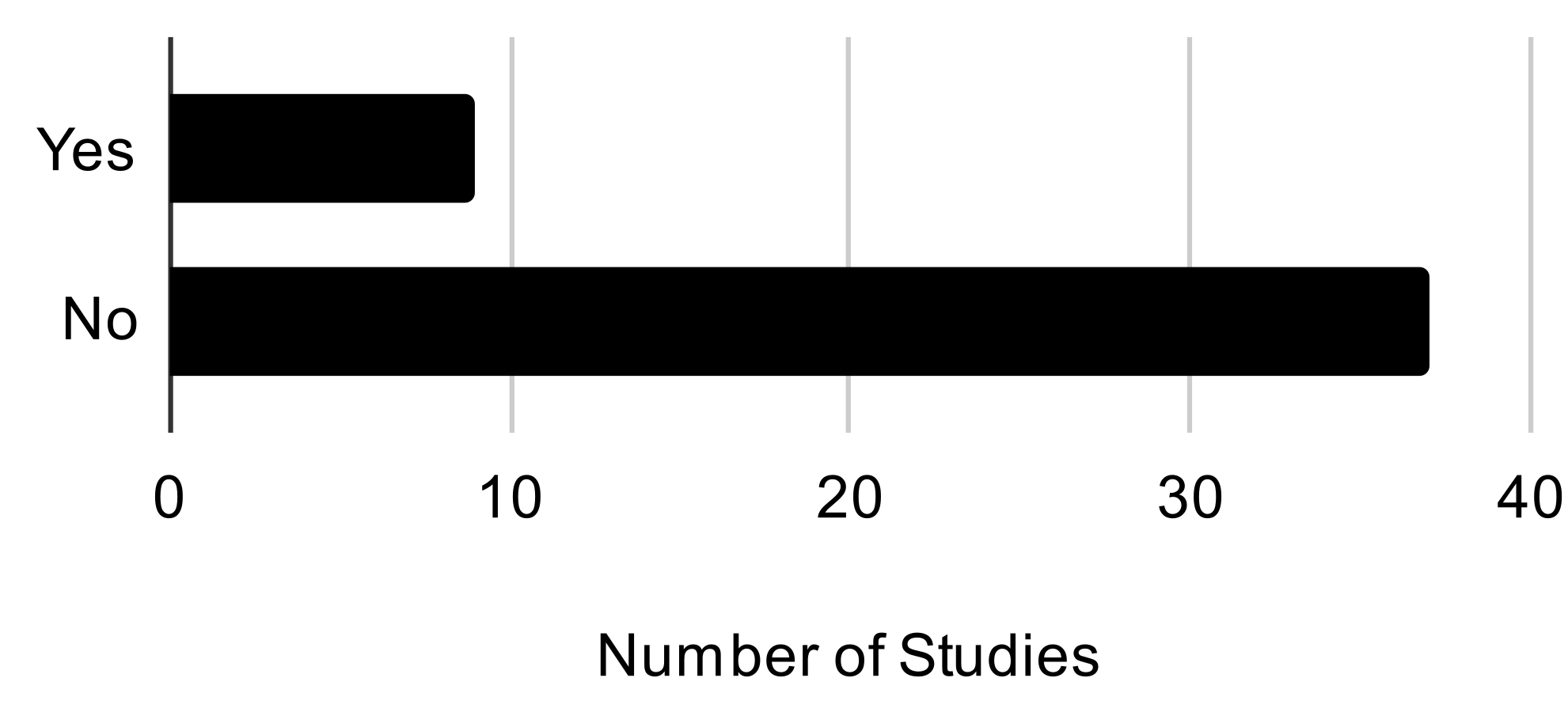}
    \caption{Responsible ML in Studies}
    \label{fig:responsibleLimits}
    \vspace{-0.3ex}
\end{subfigure}
\hfill
    \begin{subfigure}{0.3\textwidth}
    \centering
    \includegraphics[width=\textwidth]{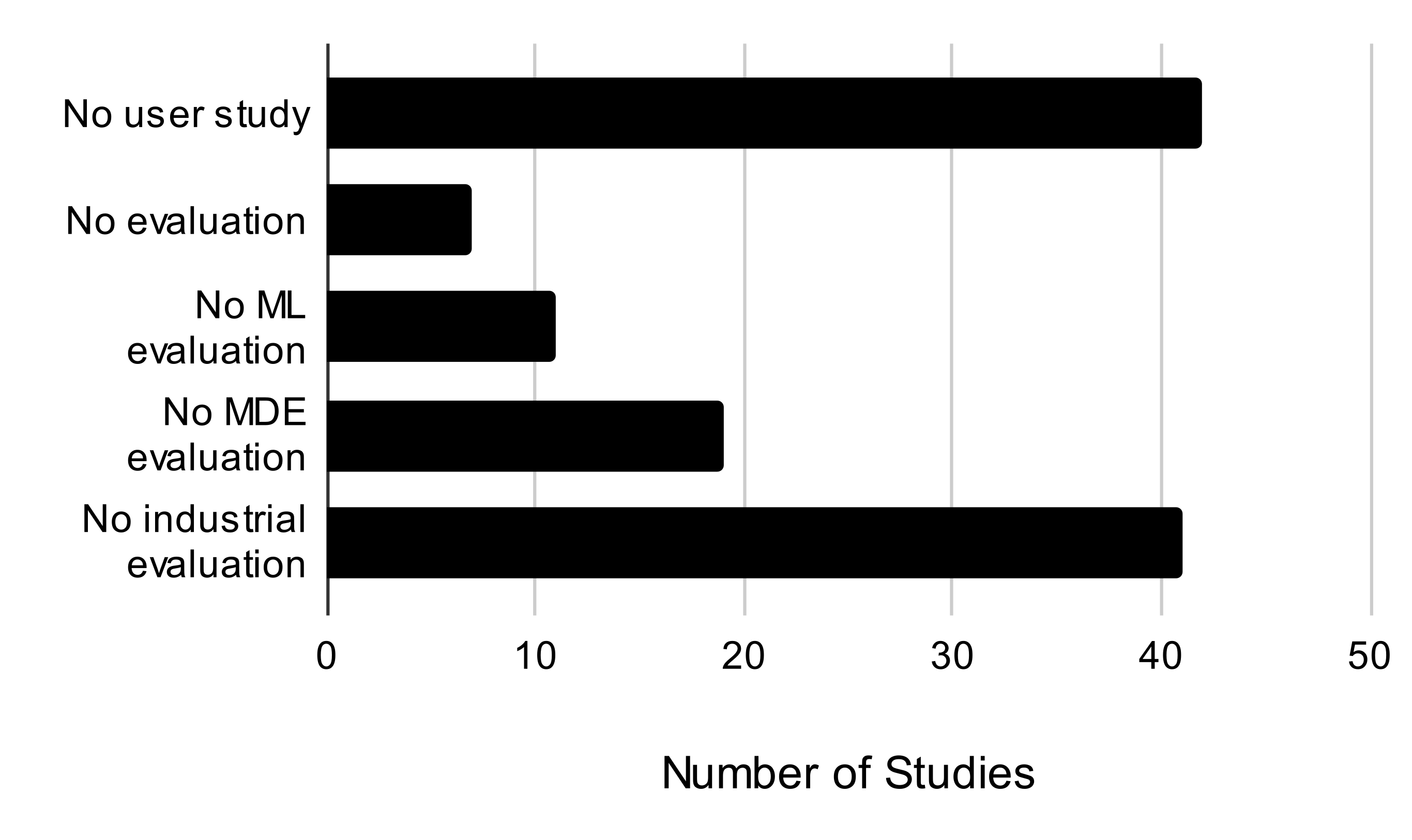}
    \caption{Evaluation in Studies}
    \label{fig:evalLimits}
\end{subfigure}
\caption{Gaps Identified in Studies}
\end{figure}

\subsubsection{Solution Scalability}
Scalability in MDE is a well-known challenge~\cite{bucchiarone2020grand, kolovos2013research}. Little to no focus is placed on scalability in the selected studies on MDE for ML. As shown in Figure \ref{fig:scalabilityLimits}, almost 75\% of the studies (P1-P8, P10-P12, P15, P17, P18, P20-P22, P24-P29, P31, P32, P34-P36, P38, P40, P42, P43, P45, and
P46) do not discuss the scalability of their solution. One possible reason behind this could be 
that often, the primary goal is to develop a proof of concept, leading to the development of basic MDE tools that only work for simple projects. Scalability is often an afterthought in such solutions, rendering them of little or no practical use.

We recommend researchers consider the importance of scalability in MDE solutions and report their scalability results to facilitate practical adoption and comparison. 

\subsubsection{Responsible ML}
Responsible use of ML refers to applying ML to maximize the benefits for the end users and society while minimizing harm. This consists of developing and managing ML components prioritising human-centric needs such as fairness, trust, safety, explainability, privacy, and human values~\cite{zhu2022ai}. Despite the growing awareness of responsible ML, only nine of the selected primary studies (P2, P3, P7, P10, P27, P31, P33, P35, and P45) out of the total 46 considered human-centric aspects and/or the responsible use of ML in their MDE solutions as shown in Figure \ref{fig:responsibleLimits}. This is concerning since neglecting these critical human-centric requirements can have serious consequences, such as ML-based systems that are not trustworthy, biased, and violate ethical principles and legal policies~\cite{zhu2022ai}.

We suggest researchers prioritize the responsible applications of ML in their MDE solutions. As a result, the ML artifacts generated will better meet the needs of end-users and foster trust~\cite{zhu2022ai}. Some potential MDE approaches to achieve this include DSLs tailored for modeling human-centric requirements in ML components. These automated code generators of ML components conform to responsible ML practices and runtime monitoring for responsible ML. 

\subsection{Evaluations of MDE solutions for ML}
Figure \ref{fig:evalLimits} shows various limitations related to evaluation found in the selected primary studies.
\subsubsection{Real-world Evaluation}
Real-world evaluations in an industrial context demonstrate the usefulness of MDE solutions in the industry. Such evaluations also impact the likelihood of research being applied to the industry. Yet, only five of the selected studies (P12, P15, P31, P33, P35) perform an industrial evaluation. User studies (or surveys) are another useful method for evaluating the practical usability of the approach with actual end users. However, we found only four studies (P10, P22, P24, P35) with user study as an evaluation method.

We encourage researchers to evaluate their proposed MDE solutions on industrial case studies and with real-world users to get realistic results and feedback. More value can be added to user studies by evaluating with diverse groups of users representing a wide range of demographics and perspectives. We realize this may not always be possible, but it remains an ideal goal to strive in order to achieve a more comprehensive and inclusive evaluation.

\subsubsection{Evaluation Rigor}
 
One of the recurring issues we observed in studies was the lack of rigor in evaluation. For instance, eight studies (P5, P6, P8, P18, P21, P26, P32, and P43) did not report on any evaluation, 22 studies performed a partial evaluation (i.e., focused on either MDE or ML but not both), 14 studies (P7, P14, P15, P20, P29, P36-P42, P45, P46) had insufficient evaluation details, no clear rationale behind the evaluation settings (e.g., the choice of evaluation metrics and parameters selection) or provided only limited discussion on the implication of results. In studies P14, P15, P20, P24, P29, P30, P33, P35, and P40, there is no evaluation of the ML aspect. A greater number of studies P1, P3, P7, P9, P11, P13, P22, P23, P27, P34, P39, P42, and P46, provide no evaluation of the MDE aspect. These limitations impact the quality of the resulting evaluation and the ability to draw meaningful and reliable conclusions from the results.


We propose researchers perform and report empirical evaluations with respect to both aspects of MDE and ML. Clearly justify their choices of evaluation methods and metrics and provide comprehensive details of the result, based on the guidelines for reporting empirical software engineering research (e.g., ~\cite{runeson2009guidelines,ralph2020empirical}). Rigorous evaluation methods increase the credibility of research findings, making them more trustworthy for academia and industrial applications.

\section{Conclusion}~\label{sec:Conclusion}
Software engineering for ML-based systems is in many ways more complicated and challenging for developers compared to traditional software systems~\cite{atouani2021artifact}. In this article, we report on an SLR of MDE solutions for software systems with ML components. The goal was to explore the potential of MDE for ML-based systems and identify trends, benefits, and limitations of existing work. For the SLR, we followed the systematic process described by Kitchenham et al.~\cite{kitchenham2009systematic} and selected 46 highly relevant primary studies from an initial pool of 3,496 papers. We have explored many interesting aspects of the MDE solutions and reported our findings along with gaps and potential directions for future research. Our key findings suggest that over the last five years, there has been a significant increase in the studies on MDE4ML; with the rapidly growing popularity of ML and AI, we expect this trend to continue in the future. 

Our examination of selected studies suggests that: 1) there are few studies on the data required for ML; 2) proposed solutions are limited to the design, development, and training of ML components (less studies on requirements engineering, integration, ML pipelines, automated deployment, monitoring and documentation); 3) there are limited studies for unsupervised learning and reinforcement learning; 4) MDE steps are not comprehensively explained; 5) MDE solutions for ML-based systems require more maturity and better tool support; 6) there is a lack of MDE4ML solutions for domain experts; 7) studies use inconsistent terminology for describing ML; 8) there is a need for more focus on solution scalability; 9) there is a lack of responsible ML and human-centric development practices in MDE approaches for ML-based systems; 10) most evaluations lack rigor and are conducted in an academic setting.

\section*{Acknowledgements}

Naveed is supported by a Faculty of IT PhD scholarship. Grundy is supported by ARC Laureate Fellowship FL190100035. This work has been partially supported by ARC Discovery Project DP200100020.

\appendix
\section{Selected Primary Studies}

\begin{footnotesize}
\textbf{P1:} Safdar, Aon, Farooque Azam, Muhammad Waseem Anwar, Usman Akram, and Yawar Rasheed. "MoDLF: a model-driven deep learning framework for autonomous vehicle perception (AVP)." In Proceedings of the 25th International Conference on Model Driven Engineering Languages and Systems, pp. 187-198. 2022, doi: 10.1145/3550355.3552453

\textbf{P2:} Yohannis, Alfa, and Dimitris Kolovos. "Towards model-based bias mitigation in machine learning." In Proceedings of the 25th International Conference on Model Driven Engineering Languages and Systems, pp. 143-153. 2022, doi: 10.1145/3550355.3552401

\textbf{P3:} Langford, Michael Austin, Kenneth H. Chan, Jonathon Emil Fleck, Philip K. McKinley, and Betty HC Cheng. "Modalas: Model-driven assurance for learning-enabled autonomous systems." In 2021 ACM/IEEE 24th International Conference on Model Driven Engineering Languages and Systems (MODELS), pp. 182-193. IEEE, 2021, doi: 10.1109/MODELS50736.2021.00027

\textbf{P4:} Shi, Yechuan, Jörg Kienzle, and Jin LC Guo. "Feature-oriented modularization of deep learning APIs." In Proceedings of the 25th International Conference on Model Driven Engineering Languages and Systems: Companion Proceedings, pp. 367-374. 2022, doi: 10.1145/3550356.3561575

\textbf{P5:} Gatto, Nicola, Evgeny Kusmenko, and Bernhard Rumpe. "Modeling deep reinforcement learning based architectures for cyber-physical systems." In 2019 ACM/IEEE 22nd International Conference on Model Driven Engineering Languages and Systems Companion (MODELS-C), pp. 196-202. IEEE, 2019, doi: 10.1109/MODELS-C.2019.00033

\textbf{P6:} Kourouklidis, Panagiotis, Dimitris Kolovos, Joost Noppen, and Nicholas Matragkas. "A model-driven engineering approach for monitoring machine learning models." In 2021 ACM/IEEE International Conference on Model Driven Engineering Languages and Systems Companion (MODELS-C), pp. 160-164. IEEE, 2021, doi: 10.1109/MODELS-C53483.2021.00028

\textbf{P7:} Al-Azzoni, Issam. "Model Driven Approach for Neural Networks." In 2020 International Conference on Intelligent Data Science Technologies and Applications (IDSTA), pp. 87-94. IEEE, 2020, doi: 10.1109/IDSTA50958.2020.9264067

\textbf{P8:} Kusmenko, Evgeny, Sebastian Nickels, Svetlana Pavlitskaya, Bernhard Rumpe, and Thomas Timmermanns. "Modeling and training of neural processing systems." In 2019 ACM/IEEE 22nd International Conference on Model Driven Engineering Languages and Systems (MODELS), pp. 283-293. IEEE, 2019, doi: 10.1109/MODELS.2019.00012

\textbf{P9:} Moin, Armin, Moharram Challenger, Atta Badii, and Stephan Günnemann. "Supporting AI engineering on the IoT edge through model-driven TinyML." In 2022 IEEE 46th Annual Computers, Software, and Applications Conference (COMPSAC), pp. 884-893. IEEE, 2022, doi: 10.1109/COMPSAC54236.2022.00140

\textbf{P10: } Giner-Miguelez, Joan, Abel Gómez, and Jordi Cabot. "A domain-specific language for describing machine learning datasets." Journal of Computer Languages 76 (2023): 101209, doi: 10.1016/j.cola.2023.101209

\textbf{P11: } Atouani, Abdallah, Jörg Christian Kirchhof, Evgeny Kusmenko, and Bernhard Rumpe. "Artifact and reference models for generative machine learning frameworks and build systems." In Proceedings of the 20th ACM SIGPLAN International Conference on Generative Programming: Concepts and Experiences, pp. 55-68. 2021, doi: 10.1145/3486609.3487199

\textbf{P12: } Baumann, Nils, Evgeny Kusmenko, Jonas Ritz, Bernhard Rumpe, and Moritz Benedikt Weber. "Dynamic data management for continuous retraining." In Proceedings of the 25th International Conference on Model Driven Engineering Languages and Systems: Companion Proceedings, pp. 359-366. 2022, doi: 10.1145/3550356.3561568

\textbf{P13: } Bhattacharjee, Anirban, Yogesh Barve, Shweta Khare, Shunxing Bao, Zhuangwei Kang, Aniruddha Gokhale, and Thomas Damiano. "Stratum: A bigdata-as-a-service for lifecycle management of iot analytics applications." In 2019 IEEE International Conference on Big Data (Big Data), pp. 1607-1612. IEEE, 2019, doi: 10.1109/BigData47090.2019.9006518

\textbf{P14: } Ries, Benoit, Nicolas Guelfi, and Benjamin Jahic. "An mde method for improving deep learning dataset requirements engineering using alloy and uml." In Proceedings of the 9th International Conference on Model-Driven Engineering and Software Development, pp. 41-52. SCITEPRESS, 2021, doi: 10.5220/0010216600410052

\textbf{P15: } Enríquez, José Gonzalez, Antonio Martínez-Rojas, David Lizcano, and Andrés Jiménez-Ramírez. "A unified model representation of machine learning knowledge." Journal of web engineering (2020): 319-340, doi: 10.13052/jwe1540-9589.1929

\textbf{P16: } Benni, Benjamin, Mireille Blay-Fornarino, Sébastien Mosser, Frederic Precisio, and Günther Jungbluth. "When DevOps meets meta-learning: A portfolio to rule them all." In 2019 ACM/IEEE 22nd International Conference on Model Driven Engineering Languages and Systems Companion (MODELS-C), pp. 605-612. IEEE, 2019, doi: 10.1109/MODELS-C.2019.00092

\textbf{P17: } Koseler, Kaan, Kelsea McGraw, and Matthew Stephan. "Realization of a Machine Learning Domain Specific Modeling Language: A Baseball Analytics Case Study." In Proceedings of the 7th International Conference on Model-Driven Engineering and Software Development, pp. 13-24. 2019., doi: 10.5220/0007245800130024

\textbf{P18: } Liaskos, Sotirios, Shakil M. Khan, Reza Golipour, and John Mylopoulos. "Towards Goal-based Generation of Reinforcement Learning Domain Simulations."  In Proceedings of 15th International i* Workshop (2022)

\textbf{P19: } Arif, Madeha, Farooque Azam, Muhammad Waseem Anwar, and Yawar Rasheed. "A model-driven framework for optimum application placement in fog computing using a machine learning based approach." In Information and Software Technologies: 26th International Conference, ICIST 2020, Kaunas, Lithuania, October 15–17, 2020, Proceedings 26, pp. 102-112. Springer International Publishing, 2020, doi: 10.1007/978-3-030-59506-7\_9 

\textbf{P20: } Schöne, René, Johannes Mey, Boqi Ren, and Uwe Aßmann. "Bridging the gap between smart home platforms and machine learning using relational reference attribute grammars." In 2019 ACM/IEEE 22nd International Conference on Model Driven Engineering Languages and Systems Companion (MODELS-C), pp. 533-542. IEEE, 2019, doi: 10.1109/MODELS-C.2019.00083

\textbf{P21: } Breuker, Dominic. "Towards Model-Driven Engineering for Big Data Analytics--An Exploratory Analysis of Domain-Specific Languages for Machine Learning." In 2014 47th Hawaii International Conference on System Sciences, pp. 758-767. IEEE, 2014, doi: 10.1109/HICSS.2014.101

\textbf{P22: } Moin, Armin, Moharram Challenger, Atta Badii, and Stephan Günnemann. "A model-driven approach to machine learning and software modeling for the IoT: Generating full source code for smart Internet of Things (IoT) services and cyber-physical systems (CPS)." Software and Systems Modeling 21, no. 3 (2022): 987-1014, doi: 10.1007/s10270-021-00967-x

\textbf{P23: } Hartmann, T., A. Moawad, F. Fouquet, and Y. Le Traon. "The Next Evolution of MDE: A Seamless Integration of Machine Learning into Domain Modeling. In 2017 ACM/IEEE 20th International Conference on Model Driven Engineering Languages and Systems (MODELS). 180–180." (2017), doi: 10.1007/s10270-017-0600-2

\textbf{P24: } Neumann, Alexander Tobias, Peter de Lange, and Ralf Klamma. "Collaborative creation and training of social bots in learning communities." In 2019 IEEE 5th International Conference on Collaboration and Internet Computing (CIC), pp. 11-19. IEEE, 2019, doi: 10.1109/CIC48465.2019.00011

\textbf{P25: } Krstić, Dragana, Nenad Petrović, and Issam Al-Azzoni. "Model-driven approach to fading-aware wireless network planning leveraging multiobjective optimization and deep learning." Mathematical Problems in Engineering 2022 (2022), doi: 10.1155/2022/4140522

\textbf{P26: } Barzdins, Paulis, Audris Kalnins, Edgars Celms, Janis Barzdins, Arturs Sprogis, Mikus Grasmanis, Sergejs Rikacovs, and Guntis Barzdins. "Metamodel Specialisation based Tool Extension." Baltic Journal of Modern Computing 10, no. 1 (2022): 17-35, doi: 10.22364/bjmc.2022.10.1.02

\textbf{P27: } Tabbiche, Mohammed Nadjib, Mohammed Fethi Khalfi, and Reda Adjoudj. "Applying Machine Learning and Model-Driven Approach for the Identification and Diagnosis Of Covid-19." International Journal of Distributed Systems and Technologies (IJDST) 14, no. 1 (2023): 1-27, doi: 10.4018/IJDST.321648

\textbf{P28: } García-Díaz, Vicente, Jordán Pascual Espada, B. Cristina Pelayo G-Bustelo, and Juan Manuel Cueva Lovelle. "Towards a standard-based domain-specific platform to solve machine learning-based problems." International Journal of Interactive Multimedia and Artificial Intelligence 3, no. 5 (2015), doi:  10.9781/ijimai.2015.351 

\textbf{P29: } Mili, and Djamel Meslati. "Modeling bio inspired systems: Towards a new view based on MDE." Journal of Theoretical and Applied Information Technology 38, no. 1 (2012): 63-73

\textbf{P30: } Santos, Fernando, Ingrid Nunes, and Ana LC Bazzan. "Model-driven agent-based simulation development: A modeling language and empirical evaluation in the adaptive traffic signal control domain." Simulation Modelling Practice and Theory 83 (2018): 162-187, doi: 10.1016/j.simpat.2017.11.006

\textbf{P31: } Pilarski, Sebastian, Martin Staniszewski, Matthew Bryan, Frederic Villeneuve, and Dániel Varró. "Predictions-on-chip: model-based training and automated deployment of machine learning models at runtime: For multi-disciplinary design and operation of gas turbines." Software and Systems Modeling 20 (2021): 685-709, doi: 10.1007/s10270-020-00856-9

\textbf{P32: } Zweihoff, Philip, and Bernhard Steffen. "Pyrus: an online modeling environment for no-code data-analytics service composition." In Leveraging Applications of Formal Methods, Verification and Validation: 10th International Symposium on Leveraging Applications of Formal Methods, ISoLA 2021, Rhodes, Greece, October 17–29, 2021, Proceedings 10, pp. 18-40. Springer International Publishing, 2021, doi: 10.1007/978-3-030-89159-6\_2

\textbf{P33: } Yang, Zhibin, Yang Bao, Yongqiang Yang, Zhiqiu Huang, Jean-Paul Bodeveix, Mamoun Filali, and Zonghua Gu. "Exploiting augmented intelligence in the modeling of safety-critical autonomous systems." Formal Aspects of Computing 33, no. 3 (2021): 343-384, doi: 10.1007/s00165-021-00543-6

\textbf{P34: } Espinosa, Roberto, Diego García-Saiz, Marta Zorrilla, José Jacobo Zubcoff, and Jose-Norberto Mazón. "S3Mining: A model-driven engineering approach for supporting novice data miners in selecting suitable classifiers." Computer Standards and Interfaces 65 (2019): 143-158, doi: 10.1016/j.csi.2019.03.004

\textbf{P35: } Khalajzadeh, Hourieh, Andrew J. Simmons, Mohamed Abdelrazek, John Grundy, John Hosking, and Qiang He. "An end-to-end model-based approach to support big data analytics development." Journal of Computer Languages 58 (2020): 100964, doi: 10.1016/j.cola.2020.100964

\textbf{P36: } Lechevalier, David, Anantha Narayanan, Sudarsan Rachuri, Sebti Foufou, and Y. Tina Lee. "Model-based engineering for the integration of manufacturing systems with advanced analytics." In Product Lifecycle Management for Digital Transformation of Industries: 13th IFIP WG 5.1 International Conference, PLM 2016, Columbia, SC, USA, July 11-13, 2016, Revised Selected Papers 13, pp. 146-157. Springer International Publishing, 2016, doi: 10.1007/978-3-319-54660-5\_14

\textbf{P37: } Petrović, Nenad. "Model-Driven Approach to Blockchain-Enabled MLOps." In Proceedings of 9th International Conference IcETRAN (2022): 533-538

\textbf{P38: } Petrovic, Nenad, Issam Al-Azzoni, Dragana Krstic, and Abdullah Alqahtani. "Base station anomaly prediction leveraging model-driven framework for classification in Neo4j." In 2022 International Conference on Broadband Communications for Next Generation Networks and Multimedia Applications (CoBCom), pp. 1-5. IEEE, 2022, doi: 10.1109/CoBCom55489.2022.9880776

\textbf{P39: } Jahić, Benjamin, Nicolas Guelfi, and Benoît Ries. "SEMKIS-DSL: A Domain-Specific Language to Support Requirements Engineering of Datasets and Neural Network Recognition." Information 14, no. 4 (2023): 213, doi: 10.3390/info14040213

\textbf{P40: } Molderez, Tim, Bjarno Oeyen, Coen De Roover, and Wolfgang De Meuter. "Marlon: a domain-specific language for multi-agent reinforcement learning on networks." In Proceedings of the 34th ACM/SIGAPP Symposium on Applied Computing, pp. 1322-1329. 2019, doi: 10.1145/3297280.3297413

\textbf{P41: } Sujeeth, Arvind, HyoukJoong Lee, Kevin Brown, Tiark Rompf, Hassan Chafi, Michael Wu, Anand Atreya, Martin Odersky, and Kunle Olukotun. "OptiML: an implicitly parallel domain-specific language for machine learning." In Proceedings of the 28th International Conference on Machine Learning (ICML-11), pp. 609-616. 2011

\textbf{P42: } Zhao, Tian, and Xiaobing Huang. "Design and implementation of DeepDSL: A DSL for deep learning." Computer Languages, Systems and Structures 54 (2018): 39-70, doi: 10.1016/j.cl.2018.04.004

\textbf{P43: } Weimer, Markus, Tyson Condie, and Raghu Ramakrishnan. "Machine learning in ScalOps, a higher order cloud computing language." In NIPS 2011 Workshop on parallel and large-scale machine learning (BigLearn), vol. 9, pp. 389-396. 2011

\textbf{P44: } Podobas, Artur, Martin Svedin, Steven WD Chien, Ivy B. Peng, Naresh Balaji Ravichandran, Pawel Herman, Anders Lansner, and Stefano Markidis. "Streambrain: an hpc framework for brain-like neural networks on cpus, gpus and fpgas." In Proceedings of the 11th International Symposium on Highly Efficient Accelerators and Reconfigurable Technologies, pp. 1-6. 2021, doi: 10.1145/3468044.3468052

\textbf{P45: } Shin, Seung-Jun. "An opc ua-compliant interface of data analytics models for interoperable manufacturing intelligence." IEEE Transactions on Industrial Informatics 17, no. 5 (2020): 3588-3598, doi: 10.1109/TII.2020.3024628

\textbf{P46: } Elango, Venmugil, Norm Rubin, Mahesh Ravishankar, Hariharan Sandanagobalane, and Vinod Grover. "Diesel: DSL for linear algebra and neural net computations on GPUs." In Proceedings of the 2nd ACM SIGPLAN International Workshop on Machine Learning and Programming Languages, pp. 42-51. 2018, doi: 10.1145/3211346.3211354

\end{footnotesize}



\printcredits

\bibliographystyle{elsarticle-num}

\bibliography{refs}

\end{document}